\documentclass[11pt,a4paper]{article} 
\pdfoutput=1
\usepackage{jcappub}
\usepackage{enumerate}
\usepackage{epsfig}
\usepackage{wasysym} 
\usepackage{mathrsfs} 
\usepackage{graphicx} 
\usepackage{amsfonts} 
\usepackage{amsbsy} 
\usepackage{amscd}
\usepackage{slashed} 
\usepackage{multirow} 
\usepackage{gensymb}
\usepackage{subcaption}
\usepackage{natbib}
\allowdisplaybreaks
\usepackage[section]{placeins}
\usepackage{color}
\definecolor{myred}{rgb}{0.6,0,0} 
\definecolor{myblue}{rgb}{0,0.2,0.4}
\definecolor{mygreen}{rgb}{0,0.9,0.1}
\definecolor{hc}{rgb}{.9,0.1,0.7}
\definecolor{hcout}{rgb}{.9,0.7,0.9}
\definecolor{Orange}{rgb}{1.,0.65,0.}

\makeatletter

\newcommand{\fmslash}[2][0mu]{%
  \mathchoice
    {\fmsl@sh\displaystyle{#1}{#2}}%
    {\fmsl@sh\textstyle{#1}{#2}}%
    {\fmsl@sh\scriptstyle{#1}{#2}}%
    {\fmsl@sh\scriptscriptstyle{#1}{#2}}}
\newcommand{\fmsl@sh}[3]{%
  \m@th\ooalign{$\hfil#1\mkern#2/\hfil$\crcr$#1#3$}}
\makeatother

\newcommand{\lsim}{{\;\raise0.3ex\hbox{$<$\kern-0.75em\raise-1.1ex\hbox{$\sim$}}\;}}
\newcommand{\gsim}{{\;\raise0.3ex\hbox{$>$\kern-0.75em\raise-1.1ex\hbox{$\sim$}}\;}}

\usepackage{array}
\newcolumntype{C}[1]{>{\centering\arraybackslash$}p{#1}<{$}}

\usepackage{bm} 
\newcommand{\be}{\begin{equation}}
\newcommand{\ee}{\end{equation}}
\newcommand{\bes}{\begin{equation*}}
\newcommand{\ees}{\end{equation*}}
\newcommand{\bea}{\begin{eqnarray}}
\newcommand{\eea}{\end{eqnarray}}
\newcommand{\beas}{\begin{eqnarray*}}
\newcommand{\eeas}{\end{eqnarray*}}
\title{Dark matter, extra-terrestrial gamma-rays and the MSSM: a
  viability study}

\author[a]{Arpan Kar,} \author[b]{Sourav Mitra,} \author[a]{Biswarup 
Mukhopadhyaya,}  \author [c]{Tirthankar  Roy Choudhury}
\affiliation[a]{Regional Centre for Accelerator-based Particle
  Physics, Harish-Chandra Research Institute, HBNI, Chhatnag Road,
  Jhunsi, Allahabad - 211 019, India} \affiliation[b]{Surendranath
  College, 24/2 M. G. ROAD, Kolkata, West Bengal 700009}
\affiliation[c]{National Centre for Radio Astrophysics, TIFR, Post Bag
  3, Ganeshkhind, Pune 411007, India} \emailAdd{arpankar@hri.res.in}
\emailAdd{hisourav@gmail.com} \emailAdd{biswarup@hri.res.in}\emailAdd{tirth@ncra.tifr.res.in}

\abstract{ We fit the $\gamma$-ray excess from the galactic centre
  (GC) in terms of parameters of the minimal supersymmetric standard
  model (MSSM).  Consistency with other $\gamma$-ray observation, such
  as those from dwarf spheroidal galaxies, is also ensured, in
  addition to the constraints from direct dark matter
  search. Furthermore, we expect the contribution to the relic density
  from the MSSM dark mater candidate, namely, the lightest neutralino,
  should not go below the stipulated value; otherwise it will amount
  to going beyond the MSSM by including some additional dark matter
  source. After a detailed scan of the parameter space in terms of
  four representative types of particle spectra, we identify the ones
  that are best fit to the observed data. However, these two are
  somewhat unsatisfactory in terms of $\chi^2_{min}$ as well as
  $p$-values. In some case(s), the unacceptability of low-$\chi^2_{min}$
  regions due to direct search constraint is responsible for this. In others,
  the observed shape of the $\gamma$-ray spectrum makes the fits unsatisfactory.
  The imposed lower limit on relic density, too, has a role to play all along.
  On the whole, the conclusion is that the MSSM is not a very satisfactory
  fit for the GC $\gamma$-ray compounded with other cosmological observations 
  and direct search limits. }

\subheader{HRI-RECAPP-2017-017}


\begin{document}
\maketitle

\newpage

\section{Introduction} \label{introduction}
The observation that more than 26\% of the energy density of our
universe is `cold dark matter' poses a big challenge to fundamental
physics. It is considered very likely, especially in the light of
various astrophysical and cosmological observations (for example,
gravitational lensing effects around galaxy clusters like the bullet
cluster, Big Bang nucleosynthesis, cosmological large-scale
structures, the cosmic microwave background radiation), that dark
matter (DM) is constituted out of hitherto unknown massive elementary
particles \cite{Feng:2010gw}. The very existence of any such
particle(s) together with its dynamics implies physics beyond the
standard model (SM) which otherwise describes so well most aspects of
strong, weak and electromagnetic interactions.

The particle interpretation of dark matter has spawned twofold
activities. On the one hand, direct DM search experiments are being
carried out widely \cite{Aprile:2017iyp, PhysRevLett.118.251302,
  Akerib:2015rjg, Akerib:2016lao}.  On the other, various
extra-terrestrial signals, in the form of electromagnetic waves at
various frequency ranges \cite{Colafrancesco:2005ji, Chan:2017aup} and
also (anti)protons, positrons etc. \cite{Beck:2015rna}, are being
thoroughly probed for excesses, for which DM annihilation can be
responsible.  Even in the absence of positive signals, both the above
kinds of efforts play a very useful role in constraining or ruling out
some among the plethora of theoretical frameworks proposed to
accommodate the existence of DM. One of course obtains useful guidance
in this respect from the observed relic density of the universe, as
revealed by the WMAP and subsequently Planck data
\cite{Ade:2015xua}. The present work is an attempt in this direction,
mainly using extra-terrestrial $\gamma$-rays, and stringently applying
the relic range and direct search constraints. Our investigation is
around a scenario where the minimal supersymmetric standard model
(MSSM) is the only source of dark matter.

Supersymmetry (SUSY), a scenario that naturally stabilises the Higgs
mass through the postulate of a boson-fermion symmetric action, also
offers a dark matter candidate in the form of the lightest SUSY
particle (LSP) if baryon number (B) is conserved while lepton number
(L) is not violated by odd units. The thus conserved quantum number
called R-parity, with $R~=~(-1)^{(3B+L+2J)}$, emerges as the $Z_2$ symmetry
lending stability to the LSP, thus making the latter a viable DM candidate. 
The lightest neutralino ($\chi^0_1$) is the most common choice, since
the other possibility, namely, a sneutrino ($\tilde{\nu}$), is disfavoured 
in the MSSM by direct searches. The rest of our discussion pertains to a
$\chi^0_1$ LSP. 

The Large Hadron Collider (LHC) has set lower limits close to or around
2 TeV on strongly interacting superparticles in the MSSM over most of
its parameter space. However, the electroweak sector can still in
general be considerably lighter, thus retaining the candidature of the
$\chi^0_1$ DM. The relic density resulting from it should lie in the
range $\Omega h^2 = 0.1199 \pm 0.0022$, according to the Planck data
\cite{Ade:2015xua}. However, when any scenario is to be matched with
observation, theoretical errors are non-negligible. Consequently, it
has been argued in the context of the MSSM that theoretical estimates
yielding relic densities in the range $\Omega h^2 = 0.12 \pm 0.012$
may be treated as consistent \cite{Harz:2016dql, Klasen:2016qyz,
  Badziak:2017uto}, once such errors are factored in. One should
however note that, {\em once MSSM is accepted as the new physics
  prevailing at low-energy, there is just one DM candidate, and thus
  one has to take the lower limit on $\Omega h^2$ as binding}.  This
fact, often ignored, strengthens the constraints on a particular MSSM
spectrum, and will be applied in our analysis more stringently than in
most recent studies \cite{Achterberg:2017emt, Caron:2015wda,
  Bertone:2015tza, Calore:2014nla, Butter:2016tjc}.

   It is true that not saturating the relic density does not create an
   impossible situation for neutralino DM. However, this may mean {\em some new physics beyond MSSM}. Alternatively, having
   a modified cosmological history before BBN 
   is an explanation (see for
example \cite{Profumo}), as is additional ways of entropy injection.  However, this again amounts to {\em going beyond
     standard cosmology}.  Non-thermal production of neutralino dark
   matter, on the other hand, could also account for under-abundance \cite{Profumo}
   when calculated without taking such production into account.
   Again, this would mean substantial interaction strength of the
   neutralino DM with hitherto unknown long-lived superheavy fields. This
   is a pointer in some way to physics beyond MSSM.  Thus our
   contention is that one has to go {\em either beyond MSSM or
     beyond standard cosmology if the relic density calculated is
     well below the measured band, even after factoring in all
     uncertainties.}  Here we examine the
   suitability of the MSSM as the source of DM if neither of the above
   possibilities hold. This, if turned around, would also act as a
   pointer to the need of non-standard cosmological situation, or of
   SUSY beyond MSSM.

The pair-annihilation of the neutralino DM leads, among other things,
to photons in various frequency bands. Of these, a large volume of
data has accumulated in the radio and $\gamma$-ray ranges (with some
observations of X-ray data as well \cite{Chan:2017aup}), either in
the form of actual excess(es) \cite{TheFermi-LAT:2015kwa,
  Calore:2014xka, TheFermi-LAT:2017vmf, Karwin:2016tsw, Thierbach:2002rs, Geringer-Sameth:2015lua} or as
upper limits \cite{Beck:2015rna, Zhao:2017pcz, Drlica-Wagner:2015xua,
  Ackermann:2013yva, Griffin:2014bra} on the flux from some specific
source. A systematic and comprehensive way of using them to probe DM
scenarios essentially should comprise all or some of the following
steps:

\begin{enumerate}
\item Identify the free parameters of the theoretical scenario
      under consideration, and make sure that one stays within
      their values admissible from terrestrial/accelerator experiments.

\item Scan over the admissible range of these parameters and
      fit the data on excess in flux.

\item Check during the scan that one is satisfying the upper limits 
      on the flux, for sources where excess is not yet seen.

\item Keep the relic density $\Omega h^2$ within the allowed range,
      satisfying the upper as well as lower limit, if the scenario
      under investigation aims to be the only source of cold DM.

\item Check consistency with updated direct search results \cite{Aprile:2017iyp, PhysRevLett.118.251302}.

\item Thus identify the allowed region of the parameter space at, say,
  95.6\% confidence level (C.L.). Take note of the value of $\chi^2$
  per degree of freedom (DOF) and find out how good the fit is.

\item Predict the signals corresponding to the best fit region(s)
      at other frequency ranges for various celestial objects.    

\end{enumerate}

It is not, of course, an entirely straightforward process.  For
example, one needs to factor in all experimental uncertainties as well
as the dependence of the emitted flux on DM density profiles. It is
also necessary to include possible theoretical errors which
effectively modify/expand the allowed limits/range.  It has been
already mentioned that, though the latest Planck results require one
to stay within $\Omega h^2~=~0.1199 \pm 0.0022$ at, say, the $1\sigma$ level,
the calculation of $\Omega h^2$ in the MSSM has theoretical
uncertainties \cite{Harz:2016dql}. Thus a bigger margin needs to be
treated as `allowed' in a realistic estimate, where we try to be
as conservative and accommodative of various uncertainties
as possible, keeping in mind what has been said above in connection
with the relic density lower limit. Furthermore, there may
be issues to address in the implementation and interpretation of step
$6$ above, especially when the fit is not so good, as will be
discussed in detail later.

As the title of this paper shows, we primarily focus on $\gamma$-ray
data.  The excess from our galactic centre (GC) has been a central
point in such analysis. This excess is inferred from the observations
carried out using the Fermi Large Area Telescope which seem to imply
that the GC region emits $\gamma$-rays more than what is expected from
existing models of the diffuse emission and catalogues of known
astrophysical sources \cite{TheFermi-LAT:2017vmf}. The origin of this
excess is still unknown. 

There are astrophysical explanations where the excess is believed to be arising from, for example, an unresolved population of millisecond pulsars (MSPs) \cite{Abazajian:2012pn,Yuan:2014rca,Mirabal:2013rba,Bartels:2015aea, Ploeg:2017vai} or cosmic ray particles injected near the GC region about $\sim 10^6$ years back \cite{Petrovic:2014uda,Carlson:2014cwa}. Among these, the explanation based on the MSPs has been studied in great detail in the literature mainly because the spectral shape of the excess has been found to be consistent with the spectra of MSPs \cite{Abazajian:2010zy}. In addition, since the GC region has a high density, it is an ideal location for hosting star formation activities which in turn can lead to MSP formation \cite{Ponti}. However, there exist arguments based on the observed population of bright low-mass X-ray binaries which claim that the unresolved MSPs in the GC region can at best account for only $\sim 5\%$ of the excess \cite{Cholis:2014lta,
  Linden:2015qha}. Given that there are uncertainties in the astrophysical explanations of the excess, it has been suggested that the GCE is a
direct result of annihilation of the dark matter particles \cite{Calore:2014xka,Daylan:2014rsa,Huang:2015rlu,TheFermi-LAT:2017vmf}. The
resolution of this issue will undoubtedly depend on more observations of the GC expected in
the future. With this caveat, it may make sense to study explanations based on dark matter annihilation, even when one does not feel fully committed to it. This is especially useful keeping its particle physics explanations is mind.

In addition, we take into account the
observation of $\gamma$-ray signals from dwarf spheroidal galaxies. Of
these, an excess over backgrounds was reported earlier for Reticulum II some
time ago \cite{Geringer-Sameth:2015lua}, when the so-called Pass 7
data were used. Subsequent analyses, using the Pass 8 data, found no
such excess but instead bib-by-bin upper limits on the $\gamma$-ray
energy distribution \cite{Zhao:2017pcz, Drlica-Wagner:2015xua}, which was
matching well with similar limits in the case of other dwarf
spheroidal galaxies.

Several earlier studies have concentrated on finding the region of the
MSSM parameter space taking the GC excess alone, together with relic
density upper limits and collider constraints
\cite{Achterberg:2017emt, Caron:2015wda, Bertone:2015tza,
  Calore:2014nla, Butter:2016tjc}. Side by side, investigations have
been carried out to find the best fit for the GC excess and the Pass 7
claim \cite{Achterbeg:2015dca}, thereby pointing towards the optimum
DM profile encapsulated in the J-factor that determines the flux. It
has however been assumed there that the Reticulum II excess is real,
something that has been contradicted by the Pass 8 data
released later \cite{Zhao:2017pcz, Drlica-Wagner:2015xua}. At the same time,
attempts have been made to simultaneously fit the radio synchrotron
data from the Coma cluster and the Reticulum II excess as per the Pass
7 claim \cite{Beck:2017wxu}. It has also been shown that they can be
fitted consistently with the $\gamma$-ray data from M31, if one
confines the fit to the energy range 1 - 12 GeV, where the errors are
not inordinately large \cite{Ackermann:2017nya}. In addition, the
thermal average $\langle \sigma v \rangle$, where $\sigma$ is the DM
pair-annihilation cross-section and $v$, the relative velocity of the
annihilating pair, has been treated just as a free parameter
\cite{Beck:2017wxu}, extracted by demanding saturation of the 
observed radio data for a particular DM
mass. Such an analysis, also done in several other related works
\cite{Beck:2015rna}, raises some questions, since the annihilation
cross-section involves details of the MSSM spectrum and the resulting
dynamics, where consistency with all existing constraints needs to be 
ensured. To demand specific values of $\langle \sigma v \rangle$ for 
any DM mass, therefore,   may not always be justified. 
Finally, {\em while most investigations of the
  above kind have applied the upper limit of the relic density
  stringently, the lower limit has not been used in an equally serious
  manner}. This allows cases where the MSSM DM (the lightest
neutralino in all such studies) may `underclose' the universe,
something that necessitates other DM candidates and takes us beyond a
scenario where the MSSM is the only new physics around, and just
above, the TeV scale.

In this backdrop, the study presented here has the following 
distinctive features:

\begin{itemize}
\item We carry out a Markov Chain Monte Carlo (MCMC) 
based maximum likelihood analysis of the GC excess, {\em constrained by} the Pass 8 upper limits
  \cite{Zhao:2017pcz, Drlica-Wagner:2015xua}. The 95.6 \% C.L. 
 region is identified. The results are consistent with the M31 $\gamma$-ray
  data \cite{Ackermann:2017nya} as well.

\item The lower limit on the relic density is used as a constraint as
  stringent as the upper limit, modulo theoretical uncertainties.
  This alone keeps one within a `strictly MSSM' explanation of dark matter.

\item The latest constraints from colliders as well as low-energy experiments 
   are used in identifying the allowed regions of the MSSM parameter space.

\item Each of four different kinds of MSSM spectra is used for
  identification of the 95.6\% C.L. This includes the co-annihilation
  region, in which the dominant mode of DM annihilation during the
  freeze-out process need not be the same as that at the GC, a dwarf
  galaxy or a galactic cluster.

\item The updated direct search constraints, as obtained from
LUX as well as XENON1T \cite{Aprile:2017iyp}, are used. 
Regions that do not satisfy these constraints are left out during 
the likelihood analysis itself. 
\end{itemize} 

While the results presented here are mostly based on the above kind of
fit, we also compare it with another approach, where the likelihood analysis
is carried out with the GC, Reticulum II, M31 and accelerator
data/limits, over and above the relic density constraints. The
95.6\%C.L. region thus obtained is then subjected to direct search
constraints. We have shown in a recent work that most of the above region then becomes
disallowed for MSSM spectra which otherwise yield the most favourable fits
to extra-terrestrial data \cite{Kar:2017dmg}. In this paper, we extend the
analysis presented in \cite{Kar:2017dmg} by including the details of various
steps used in the calculations. Contrasting this with the
correspondingly complementary fits of the Higgs boson mass
\cite{Baak:2011ze, Flacher:2008zq}, involving electroweak precision
data and collider searches, one is guided to the conclusion that {\em
  the MSSM is perhaps not a good fit for all data connected with dark
  matter.}

In section \ref{astrophysics}, we outline the procedure of obtaining the $\gamma$-ray flux
from
the annihilation cross-section for DM particles. The overall scheme of our
analysis,
including the constraints applied, benchmark spectra  and the method of
obtaining
the fits, is summarised in section \ref{scheme}. Section \ref{Results} contains a detailed
discussions of our results.
In section \ref{radio} we comment on the implications of our analysis based on
$\gamma$-ray
observations for radio synchrotron flux from the Coma cluster. A further
set of remarks
arising out of  our investigation, questioning the appropriateness of the
MSSM as a
candidate DM scenario in the light of the $\gamma$-ray data, are
incorporated in section \ref{alternative_analysis}.
We summarise and conclude in section \ref{summary}.

\section{$\gamma$-rays from  galactic centre and Reticulum II} \label{astrophysics}

As has been mentioned in the introduction, the main astronomical
inputs in our analysis come from $\gamma$-ray data from the GC
\cite{TheFermi-LAT:2017vmf, Calore:2014xka} as well as the dwarf
spheroidal galaxy Reticulum II \cite{Geringer-Sameth:2015lua,
  Zhao:2017pcz, Drlica-Wagner:2015xua} .\footnote{As will be
  re-iterated in the next section, consistency with $\gamma$-ray
  observations from other sources has been ensured at every step of
  the analysis. This includes the M31 data as well as the upper limits
  on the flux from several other dwarf spheroidal galaxies.}

For any source, the $\gamma$-ray flux due to DM annihilation 
(in units of $\rm{GeV^{-1}} \rm{cm^{-2} s^{-1}}$) at photon energy $E$
is given by \cite{Calore:2014xka, Ackermann:2013yva, Agrawal:2014oha}
\begin{equation}  \label{eq:flux1}
\phi(E) = \frac{\left\langle \sigma v \right\rangle}{8 \pi
  m^{2}_{\chi}} \frac{dN_{\gamma}}{dE}(E) \int_{l.o.s}
\rho^{2}(r(s,\theta)) ds d\Omega
\end{equation}
where $\left\langle \sigma v \right\rangle$ is the velocity averaged
annihilation rate of DM particles to SM particles at zero temperature
\cite{Colafrancesco:2005ji}. $m_{\chi}$ is the DM particle mass. $\frac{dN_{\gamma}}{dE}$ denotes the photon energy distribution per annihilation. The DM density squared is integrated
along the line-of-sight ($ds$) as well as the solid angle (d$\Omega$),
assuming azimuthal symmetry and denoting by $\theta$ the angular
distance with respect to the direction of the centre of the
source. Azimuthal symmetry enables us to express the DM density as
$\rho = \rho^{2}(r(s,\theta)$.  When one is considering the flux per
steradian, one further divides by $\int d\Omega$, over the angular
width of the source, which defines the region of interest
(ROI). Finally, the distance along the line of sight ($s$) is related
to the angle $\theta$ and the distance $r$ from the central point of
the source, by \cite{Calore:2014xka, Cirelli:2010xx}
\begin{equation}
r(s,\theta)= \sqrt{r_{0}^{2} + s^{2} - 2 r_{0} s cos\theta}
\end{equation}
where $r_0$ is the distance of the observer from the centre of the source.
(When the GC is observed, $r_0 = r_{\odot}$, the distance of the sun from 
centre of the galaxy.)

For GC, we have used a general Navarro-Frenk-White (NFW) profile with
slope $\gamma$ \cite{Daylan:2014rsa}:
\begin{equation}
\rho(r) = \frac{\rho_{0}}{(\frac{r}{r_{s}})^{\gamma} (1 +
  \frac{r}{r_{s}})^{3-\gamma}}
\end{equation}
the slope $\gamma$ and the `scale radius' $r_s$ being free parameters
\cite{Iocco:2011jz}.  The normalisation $\rho_{0}$ is determined by
setting $\rho = \rho_{\odot}$ at the position of sun $r_{\odot}$. The
J-factor, defined as
\begin{equation}
J(\theta) = \int_{l.o.s} \rho^{2}(r(s,\theta)) ds
\end{equation}
is a measure of the total mass of annihilating DM along the
line-of-sight in any direction. One further defines $J_{av}$, the
angular average of $J$ as \cite{Cirelli:2010xx, Agrawal:2014oha}
\begin{equation}
J_{av} = \frac{\int J(\theta) d\Omega}{\int d\Omega}
\end{equation}
so that the flux per unit solid angle is expressed as

\begin{equation}  \label{eq:flux2}
\frac{d\phi}{d\Omega} = \frac{\left\langle \sigma v \right\rangle}{8 \pi
  m^{2}_{\chi}} \frac{dN_{\gamma}}{dE}(E) J_{av}
\end{equation}

The average J-factor for GC is routinely calculated over the ROI
$2\degree < |b| < 20\degree$ and $|l| < 20\degree$
\cite{Calore:2014nla, Calore:2014xka, Daylan:2014rsa}, where $b$ and
$l$ are respectively the galactic latitude and longitude.  The best
fit J-factor, as obtained from galactic rotation curve data
\cite{Iocco:2011jz}, corresponds to the profile parameters $\gamma =
1.26$, $r_{s} = 20$ Kpc, $\rho_{\odot} = 0.4$ $\rm {GeV} \rm {cm^{-3}}$
\cite{Catena:2009mf, Salucci:2010qr}. In Table \ref{Table_Jfactor}, we
present the best fit value of J-factor for GC with $2\sigma$
uncertainty. In our analysis we have used the $2\sigma$ maximum value
for the J-factor, since this leads to the most optimistic fit of the
GC $\gamma$-ray excess in terms of MSSM.

\begin{table}[h]
\begin{center}
\begin{tabular}{|c|c|}
\hline  & averaged J-factor for GC ($\rm {GeV^{2} cm^{-5} sr^{-1}}$)\\ \hline
minimum & $3.51 \times 10^{22}$ \\ \hline
best-fit & $2.06 \times 10^{23}$ \\ \hline
maximum & $1.09 \times 10^{24}$ \\ \hline

\end{tabular}
\caption{Uncertainty in the averaged J-factor for GC. Information in
  this table taken from \cite{thesis, Bertone:2015tza}}
\label{Table_Jfactor}
\end{center}
\end{table}
\vspace{0.5cm} 

\begin{figure}[h]
\begin{center}
\includegraphics[height=0.37\textwidth, angle=0]{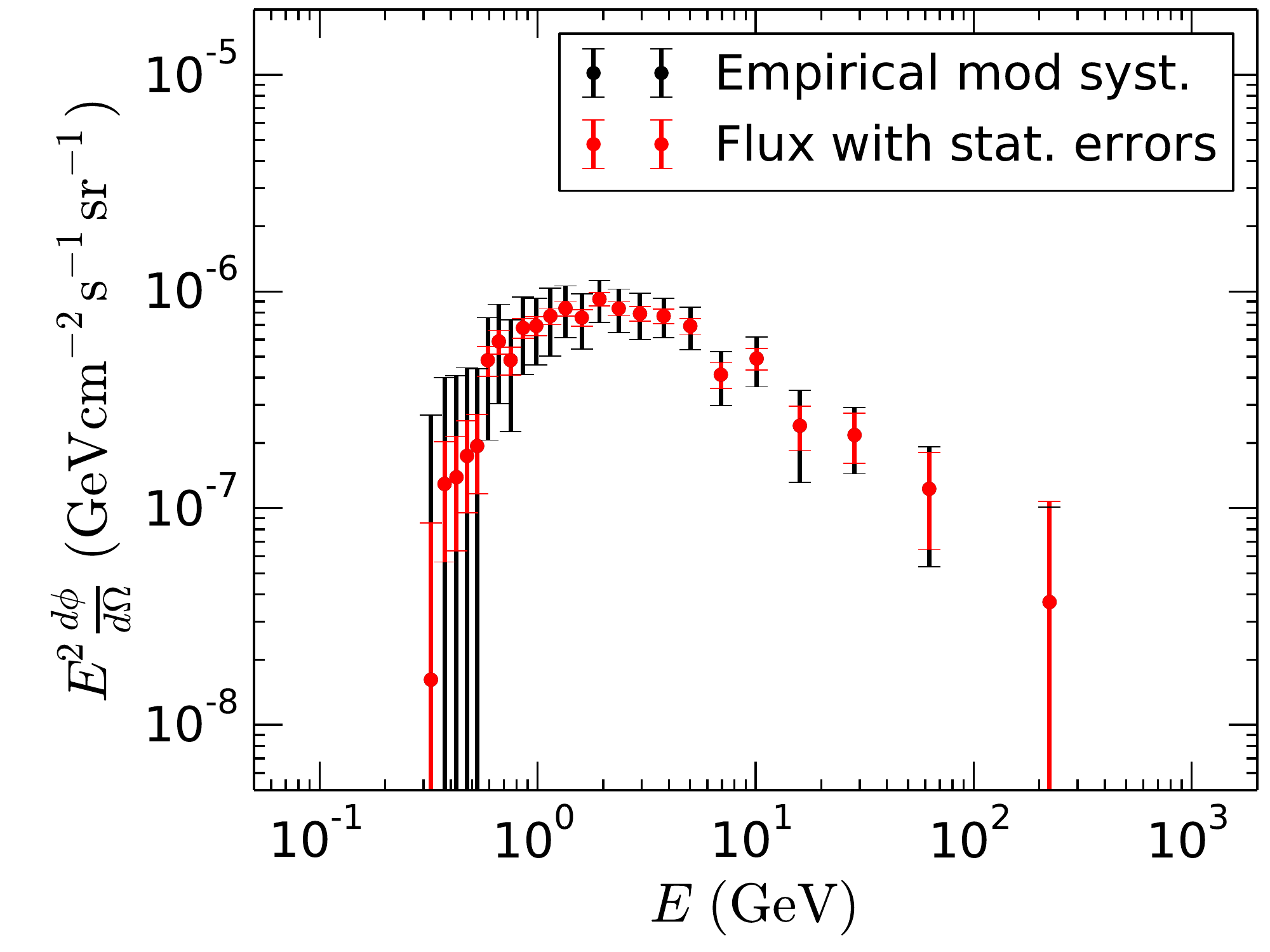}
\caption{Galactic Centre Excess spectrum, together with statistical
  and systematical errors, averaged over the ROI $2\degree < |b| <
  20\degree$ and $|l| < 20\degree$. Data and errors are taken from \cite{GC_matrix}.}
\label{GC_data}
\end{center}	
\end{figure}

In Fig. \ref{GC_data} we have shown the $\gamma$-ray excess spectrum
from the Fermi Large Area Telescope (Fermi-LAT) observations, together
with systematic and statistical uncertainties, corresponding to the
aforementioned ROI \cite{Calore:2014xka, TheFermi-LAT:2017vmf}. The
statistical uncertainties for different bins are mutually
uncorrelated. The systematics include instrumental uncertainties as
well as  those in background modeling, the latter having
correlation among various energy bins and thus introducing additional
diagonal and off-diagonal entries in the covariance
matrix. Fig. \ref{GC_data} takes into account the diagonal
entries. The off-diagonal entries have been included in the likelihood analyses
presented later. The analysis takes further note of theoretical
uncertainties in the flux estimate, coming mainly from the
fragmentation functions that determine production rates for pions and
other sources of $\gamma$-ray photons. In our subsequent analysis,
this uncertainty is taken to be 10\%, following earlier works
\cite{Caron:2015wda, Butter:2016tjc}.

For calculating the $\gamma$-ray flux for Reticulum II, we use the
value of the `total' J-factor, defined as $J_{tot} = \int J(\theta) d\Omega$,
found from the Jeans analysis \cite{Bonnivard}. The value used is
$J_{tot} = 10^{19.6}$ $\rm {GeV^{2}.cm^{-5}}$, obtained on
integration over the ROI $\Delta\Omega = 2.4 \times 10^{-4} \rm {sr}$ 
corresponds to the angular radius $0.5\degree$.

Till very recently, excess in the 1 - 10 GeV range was claimed by
observations using the so-called Pass 7 data from Reticulum II
\cite{Geringer-Sameth:2015lua}.  Beyond about 10 GeV, the error is too
large for any meaningful fit.  While such announcements generated
considerable activity towards fitting such excess in terms of DM
models including the MSSM \cite{Beck:2015rna, Achterbeg:2015dca}, the
later analyses, in terms of the Pass 8 data, disavowed such claims
\cite{Zhao:2017pcz, Drlica-Wagner:2015xua}. Upper limits on the flux
in the aforesaid range continued to exist, side by side with
comparable limits from other dwarf spheroidal galaxies.

Fig. 2 shows the $\gamma$-ray excess upper-limit according to the Pass
8 data. In our subsequent fits, carried out for the GC excess,
consistency with the upper limits from Reticulum II Pass 8 data is
ensured throughout. \footnote {The Pass 8 data, as compared to the
  Pass 7 excess, make little difference to one's conclusion regarding
  the candidature of the MSSM as a DM model for explaining the GC
  excess. More non-trivial constraints emerge when one takes into
  consideration the lower limit on the relic density along with the
  direct search constraints from XENON1T. This is because our analysis
  yields a slight shortfall upon fitting the GC $\gamma$-ray excess,
  especially in the low-frequency bins. The effort to minimise such
  shortfall, together with the commitment to obeying the Reticulum II
  Pass 8 upper limits, pushes one towards a best fit which is not
  substantially different from that obtained when the Pass 7 excess is
  recognised.}

\begin{figure*}
\centering
  \includegraphics[height=0.37\textwidth, angle=0]{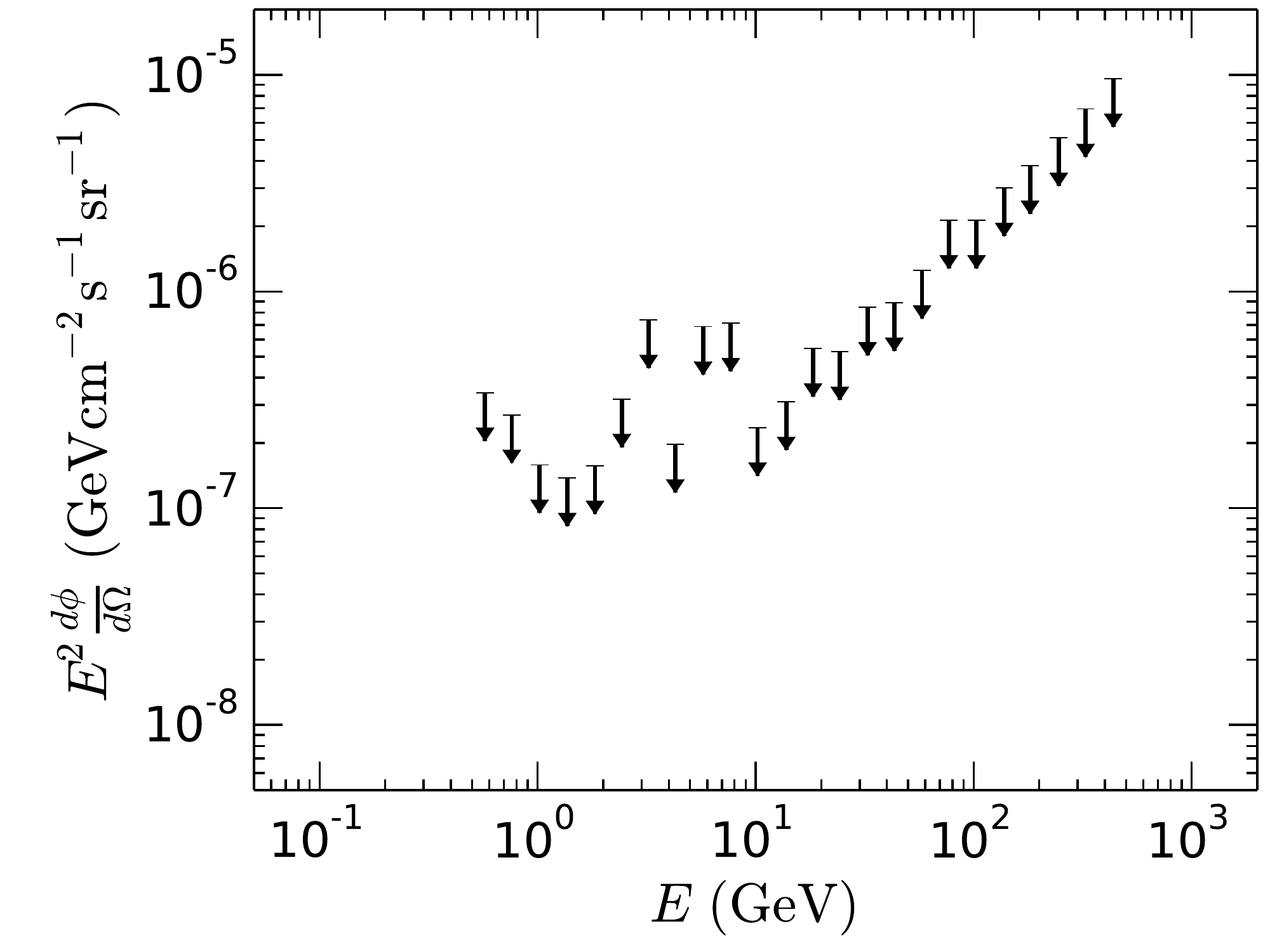}
    
  \caption{Reticulum II $\gamma$-ray excess upper limit
    (Pass 8 data) \cite{Zhao:2017pcz, Drlica-Wagner:2015xua}}.

\label{RetII_data}

\end{figure*}

\section{Scheme of analysis} \label{scheme}
\subsection{Constraints on the MSSM parameter space} 

In order to examine the candidature of the lightest neutralino
($\chi^0_1$) as DM constituent, one has to scan over a reasonable
region of the MSSM parameter space, spanned by quantities which can
have a bearing on  signals related to its annihilation, direct detection
etc. The MSSM spectrum has been generated using SuSpect 2.41 \cite{Djouadi:2002ze}.
The viability of the region thus examined 
depend on consistency with the following issues.

\begin{itemize}

\item {\bf Higgs mass}:\\ As per observation, we almost certainly have
  a spin-zero neutral object with mass close to 125 GeV
  \cite{Aad:2014aba, Khachatryan:2014jba}.  Keeping this in mind, we
  have confined ourselves to those regions which yield the mass of
  lightest neutral scalar $h$ in the range $122~\rm{GeV} \le m_h \le
  128~\rm{GeV}$. This is conservative range, accounting for theoretical
  uncertainties \cite{Allanach:2004rh}.

\item {\bf Constraints from accelerator data:}\\ The LEP lower limit
  on the lighter chargino ($\chi^{\pm}_1$) \cite{chargino,
    Lipniacka:2002sw} has been used to start with.  Furthermore,
  consistency with Higgs search results at LEP leads to allowed
  regions in the $\mu - M_2$ space \cite{chargino_atlas}, where $\mu$
  and $M_2$ are the Higgsino mass parameter and the SUSY-breaking $SU(2)$
  gaugino mass parameter respectively. We have only scanned regions
  that are consistent with this constraint. The masses of sleptons are
  similarly constrained by LEP \cite{slepton}. Thus, while a part of
  our analysis holds all sleptons at high values, we have also
  included the light-$\tilde{\tau}_1$ scenario which may lead to
  co-annihilation of the DM candidate.

LHC data impose model-independent lower bounds most markedly on
strongly interacting superparticles. Taking these into account, we
have kept the gluino and first two family squark masses above 2 TeV
\cite{Aaboud:2017hrg, Sirunyan:2017cwe, Aaboud:2017bac,
  Sirunyan:2017kqq}, and chosen benchmark values accordingly. In any
case, the snapshot values thus used do not affect DM-related issues
significantly.  An exception is the stop mass(es), which can be as low
as 150 GeV according to the current data \cite{Sirunyan:2017cwe,
  Aaboud:2017ayj, Aaboud:2017nfd}. Consistently with this, and also
with the observed Higgs mass, we have included  sample scenarios where
one of the stops is lighter than the rest of the strongly interacting
superparticles, but not so light as to reduce the relic
density below admissible limits.

Light sbottom scenarios, on the other hand, mostly require the
$\chi^0_1$ mass to be at least about 350 GeV \cite{Sirunyan:2017kiw, Sirunyan:2017cwe}. However, this shifts astrophysical $\gamma$-ray
peaks to higher frequency regions than what is, for example, observed
from the galactic centre.  Therefore, such a scenario is not taken
account here.

In addition, low-energy and flavour constraints have been followed at
the level of spectrum generation. This includes the contributions to
rare B-decays, muon anomalous magnetic moment etc.

\item {\bf Relic density constraint}\\
The Planck observations suggest that the relic density 
lies in the range \cite{Ade:2015xua}

\begin{align}
\Omega h^{2} = 0.1199 \pm 0.0022
\end{align}
Accordingly,
the contribution from the $\chi^0_1$ DM candidate
can be computed and compared, for which we have used the
package micrOMEGAs 4.3.1 \cite{micromega, Belanger:2001fz} .

However, there are also  theoretical uncertainties, typically on the
order of 10\%, in computing the MSSM DM relic density, which is
approximately six times the uncertainty in the observed data
\cite{Harz:2016dql, Klasen:2016qyz, Badziak:2017uto}. The main source
of the former is strong corrections to the annihilation rate, where
the dependence on the renormalisation and factorisation scales can be
significant. Accordingly, a larger window of the computed value of
$\Omega h^{2}$ may be taken as `allowed', leading to the range
$\Omega h^{2} = 0.12 \pm 0.012$.

Most recent studies have treated only the upper limit on the relic
density as a serious constraint. However, {\em when one is examining
  the viability of the MSSM, it makes sense to assume that the
  $\chi^0_1$ the only DM candidate, and that there is no additional
  new physics around the SUSY breaking scale.} Regions of the
parameter space that leads to values of $\Omega h^{2}$ considerably
smaller than 0.108 are therefore difficult to accept as {\em
  consistent with the MSSM.} We find that this requirement worsens the
status of the MSSM as a fit to the $\gamma$-ray observations.

\item {\bf The co-annihilation region}\\ 
  As has been mentioned above,
  one may have a phenomenologically consistent MSSM spectrum with, for
  example, the lighter stau \cite{Aad:2015baa} close in mass to the lightest
  neutralino. This leads to the well-studied co-annihilation region so
  long as the mass difference is within 4\% \cite{Belanger:2001fz}.

In our context, when one is in the co-annihilation region, {\em the
dominant channel of DM annihilation, which leads to its freeze-out,  
is different from the pair-annihilation channels which give rise to
astrophysical $\gamma$-rays.} Therefore, the relic density constraint
used in the fit of the $\gamma$-ray data becomes different for
such regions. This difference has not been taken into account in
erstwhile studies. We have, therefore, separately included co-annihilation 
regions with the associated relic density constraints in our analysis.

\item{\bf Direct search constraints}\\ The most stringent limits on
  the spin-independent cross-section of $\chi^0_1$-nucleon scattering
  can be derived from the XENON1T results \cite{Aprile:2017iyp}. In
  most of what follows, regions of the parameter space scanned to fit
  the $\gamma$-ray data have been filtered through this constraint.
  As before, micrOMEGAs 4.3.1 with most parameters at default values has
  been used to compute the spin-independent cross-sections.

Later in the paper, we have taken the following approach, too:
the best-fit regions from the analysis of $\gamma$-ray signals/limits 
from the GC and also dwarf spheroidal galaxies have been obtained
without any bias from direct search experiments. The 95.6\% C..L. 
regions  have then been subjected to the constraints by the  XENON1T
data. We interpret the difference of the two kinds of results
in section \ref{alternative_analysis}
\end{itemize}

\subsection{Benchmarks} 

As has been stated above, we have fixed the strongly interacting
superparticle masses (except that of the stop) at values above 2
TeV. The masses of sleptons in the first two families are similarly
fixed \cite{ATLAS:2017uun}. Keeping with the regions that reproduce the lighter neutral
scalar mass in the right band, we have varied the following quantities
as free parameters:\\ 
\vspace{0.2cm}

$M_1, M_2, \mu, m_A$,\\ 

\vspace{0.2cm}
\noindent
where $M_1$ and $M_2$
are the $U(1)$ and $SU(2)$ gaugino masses, $\mu$ is the Higgsino mass
parameter and $m_A$ is the neutral pseudoscalar mass.  $\tan\beta$,
the ratio of the vacuum expectation values (vev) of the two Higgs
doublets, have been fixed at `snapshot values' 5, 20 and 50.  The
remaining electroweak parameters, it has been checked, do not affect
our conclusions significantly.

Four representative MSSM spectra have been considered, which
broadly capture various features of DM annihilation, be it in
$\gamma$-ray sources or in the early universe. These are

\begin{enumerate}
\item All squarks and sleptons above 2 TeV. We call
this the {\em heavy squark and slepton} (HSS) scenario.
\item Only the lighter stop with a lower mass ($\approx$ 300 GeV),
consistently with the Higgs mass and the admissible relic density band.
This is the {\em light stop} (LST) scenario.
\item Only the lighter stau close enough to the neutralino dark matter
candidate, so as to cause co-annihilation\footnote{In principle, there can also
be a region answering to co-annihilation of the $\chi^0_1$ and
the  $\chi^{\pm}_1 / \chi^0_2$. This region is included in all the
scenarios under consideration, since $M_1, M_2$ and $\mu$ are varied
continuously, including regions where one of them is
close to another. }. This is called the stau co-annihilation (STC) scenario.
\item The lighter stop with low mass and also a stau in the
co-annihilation region. We call this {\em the light stop and stau co-annihilation}
(LSTSTC) scenario. 
\end{enumerate}

The four free parameters mentioned above have been varied over
the following ranges: -1000 GeV $ < M_1 < $ 1500 GeV, -1000 GeV $ < M_2 < $ 1500 GeV, -1000
GeV $ < \mu < $ 1500 GeV.  Depending on the LHC constraints we varied
$m_A$ in the range 350 GeV - 4000 GeV for $\tan\beta = 5$, 450 GeV -
4000 GeV for $\tan\beta = 20$, 850 GeV - 4000 GeV for $\tan\beta = 50$
\cite{Aaboud:2016cre, Aaboud:2017sjh}.

In Table \ref{Table_benchmark}, we list the benchmark values for the
fixed parameters for the various scenarios listed above.  Most of the
entries can be justified from the above discussion.  In the case of
the stop, the SUSY-breaking mass parameters for the $SU(2)$ doublet
and singlet components have been separately shown, since they have a
bearing of the lighter neutral scalar mass. The corresponding mass
eigenstates have been appropriately used, since they are relevant 
for the $t\bar{t}$ annihilation channel  in the LST and LSTSTC
scenarios. The trilinear SUSY-breaking parameter $A_t$ is
fixed at such values as to correctly reproduce the observed Higgs mass, 
but ensuring consistency with a charge-and
color-preserving vacuum, and  a potential bounded from
below. Lower values of $A_t$ necessitate larger $\mu$ or stop mass
parameters, which, for these two scenarios, worsen the fits to the
$\gamma$-ray data, thereby reducing the regions of good MSSM fit.

\begin{table}[h]
\begin{center}
\begin{tabular}{|c|c|c|c|c|c|}
\hline 
Case no & $tan\beta$ & $m_{\widetilde{t_{R}}}$(GeV) & $m_{\widetilde{Q_{3}}}$(GeV) & $A_{t}$(GeV) & $m_{\widetilde{\tau_{1}}}$(GeV)\\ 
\hline
1a & 20 & 2000 GeV & 3000 & -3000 & 2500 \\ 
\hline
1b & 50 & 2000 GeV & 3000 & -3000 & 2500 \\ 
\hline 
1c & 5 & 4000 GeV & 4000 & -4000 & 2500 \\ 
\hline 
2a & 20 & 300 GeV & 4000 & -4000 & 2500 \\ 
\hline
2b & 50 & 300 GeV & 4000 & -4000 & 2500 \\ 
\hline
2c & 5 & 300 GeV & 4000 & -4000 & 2500 \\ 
\hline
3a & 20 & 2000 GeV & 3000 & -3000 & 1.03 $m_{\chi}$ \\ 
\hline
3b & 50 & 2000 GeV & 3000 & -3000 & 1.03 $m_{\chi}$ \\ 
\hline 
3c & 5 & 4000 GeV & 4000 & -4000 & 1.03 $m_{\chi}$ \\ 
\hline
4a & 20 & 300 GeV & 4000 & -4000 & 1.03 $m_{\chi}$ \\ 
\hline 
4b & 5 & 300 GeV & 4000 & -4000 & 1.03 $m_{\chi}$ \\ 
\hline

\end{tabular}
\caption{List of various cases we have studied, along with the values of the
  fixed parameters. All other slepton masses are at 2.5 TeV, All other
  squark masses are at 3 Tev, gluino mass is at 3 TeV, $A_{b}$ = 4 TeV}
\label{Table_benchmark}
\end{center}
\end{table}

\subsection{Methodology}

We go in the following steps in our analysis:

\begin{enumerate}

\item For each of the four scenarios listed in the previous
section, we scan over the free parameters, holding the
others at their benchmark values as listed in Table \ref{Table_benchmark}.

We implement a Markov Chain Monte Carlo (MCMC) technique to estimate
the posterior distribution of our model parameters. We use a publicly
available {\it affine-invariant} MCMC
code \begin{small}EMCEE\end{small} (Foreman-Mackey et al. 2012) \cite{MCMC}, 
which has  considerable performance advantage over other
  standard MCMC sampling algorithms (Goodman \& Weare 2010).

We start with an ensemble of random walkers ($N_{\rm walker}$), each
with a chain length of $N_{\rm step}$, scanning over the model
parameter space in their aforementioned ranges. As each walker
progresses through the available parameter space, we compute the new
posterior probability density and compare it with the old one. Based
on their acceptance ratio, the Markov chains gradually converge to a
stationary distribution with highest probability (i.e. minimum
$\chi^2$). The mean and corresponding uncertainty of each parameter
are finally computed from the posterior distribution. To ensure the
convergence of the chain, we have used $N_{\rm walker}=400$ and
$N_{\rm step}=1000 - 2000$ i.e. a total of $400000 - 800000$ independent
samplings, which is large enough considering the usual mean
autocorrelation time.

\item Each point in the parameter space scanned is subjected to the constraints
from accelerator experiments, relic density (both lower and upper limits) and
direct searches, as discussed above.

\item A likelihood analysis is carried out using the GC excess, consistently with
the Reticulum II Pass 8 limits.  The likelihood function used in
  our analyses is given by ${\cal L} \propto \exp(-\chi^2 / 2)$. 
 The expression for  $\chi^{2}$ is
\begin{equation}  \label{eq:chisq}
\chi^{2} = \sum_{i,j = 1}^{24} (\phi_{i} -
\overline{\phi_{i}})(\Sigma^{-1})_{ij} (\phi_{j} -
\overline{\phi_{j}})  
\end{equation}
where $\overline{\phi_{i}}$ and is the observed flux from GC,
$\phi_{i}$ being the corresponding calculated $\gamma$-ray flux.  For
GC, the indices $i,j$ denote the energy bin numbers that run from 1 to
24. Once more, the calculation of ${\phi_{i}}$ has been done in
micrOMEGAs 4.3.1, interpolating the numerical results contained in
files generated from Pythia \cite{Sjostrand:2006za}.

The covariance matrix $\Sigma$ which has a non-diagonal nature in the
case of GC excess\footnote{The non-diagonal nature supposedly creeps
  in the process of background subtraction, where model-dependence
  introduces a correlation among data in the various bins.}, is
defined as (following eq.5.2 and corresponding description in
\cite{Calore:2014xka})

\begin{equation}  \label{covariance_matrix}
\Sigma_{ij} = \Sigma^{emperical}_{ij} \quad + \quad
\delta_{ij}(\sigma^{stat}_{i})^{2} \quad + \quad
\delta_{ij}(\overline{\phi_{i}}  \sigma)^{2}
\end{equation}

As defined in \cite{Calore:2014xka}, $\Sigma^{emperical}_{ij}$ is the
covariance matrix for empirical model systematics and
$\sigma^{stat}_{i}$ is the statistical error. We have used \cite{GC_matrix} 
for obtaining the covariance matrices, statistical errors and the
excess data.
The entries in $\Sigma^{empirical}_{ij}$, diagonal as well as non-diagonal, 
arise due the correlation among excesses in different bins, induced in
the process of background elimination. Following the argument of
\cite{Caron:2015wda}, we have also added a theoretical uncertainty at
the level of $\sigma = 10\%$ (last part of
eq. \ref{covariance_matrix}).

As has been already mentioned, only the upper limits on the flux from Reticulum II 
following the Pass 8 data \cite{Zhao:2017pcz, Drlica-Wagner:2015xua}
are used to constrain regions in the MSSM parameter space. 
In addition, the consistency with the M31 data \cite{Beck:2017wxu} is
ensured, especially in the frequency range mentioned in the
introduction.  We also check that the upper limits on the flux from
the other dwarf spheroidal galaxies are satisfied. The last-mentioned
constraint is not in general difficult to satisfy in our
treatment of the Reticulum data, as the J-factor in each such case is
smaller than that for Reticulum II \cite{Drlica-Wagner:2015xua}.

\item The aforementioned procedure yields not only the
best-fit value of each of our free parameters, but also
the 95.6\% C.L. region in the hyperspace spanned by them. 
From this one obtains various two-parameter marginalised
plots, for the basic parameters as well as derived quantities
such as physical masses.

\item The overall frequency distributions corresponding to our
95.6\% C.L. bands are compared with the observed GC excess
distribution. A comparison with the constraints from Reticulum II is also made.
\end{enumerate}

Finally, we compare the above results with the alternative analysis
as outlined at the section \ref{alternative_analysis}.


\section{Results}  \label{Results}

we now present the results of our analyses for the various benchmark
scenarios listed in (Table \ref{Table_benchmark}).  All Higgs and MSSM
particle masses, widths, cross sections etc. have been calculated at
the one-loop level.  While most relevant parameters are scanned over,
the benchmark spectra correspond to 'snapshot values' of $\tan\beta$,
so that the results have a generic nature, representative of all
relevant kinds of MSSM spectra that can have distinctive roles in DM
annihilation. The values of $\chi^2$ per degree of freedom for all the
benchmarks have been shown in Table \ref{Table_chisq}.

\subsection{Case 1: HSS}

This is the scenario where all sleptons and squarks masses have
high values ($>$ 2 TeV). \footnote{In practice, one can check that lowering
  them would not make the corresponding models better candidates, as
  (a) they would not greatly facilitate production of $\gamma$-rays in
  the right frequency range, and (b) if the masses are considerably
  low, the relic density falls below the stipulated level due to
  annihilation in other channels.}

\begin{figure*}
\centering
  \includegraphics[height=0.37\textwidth, angle=0]{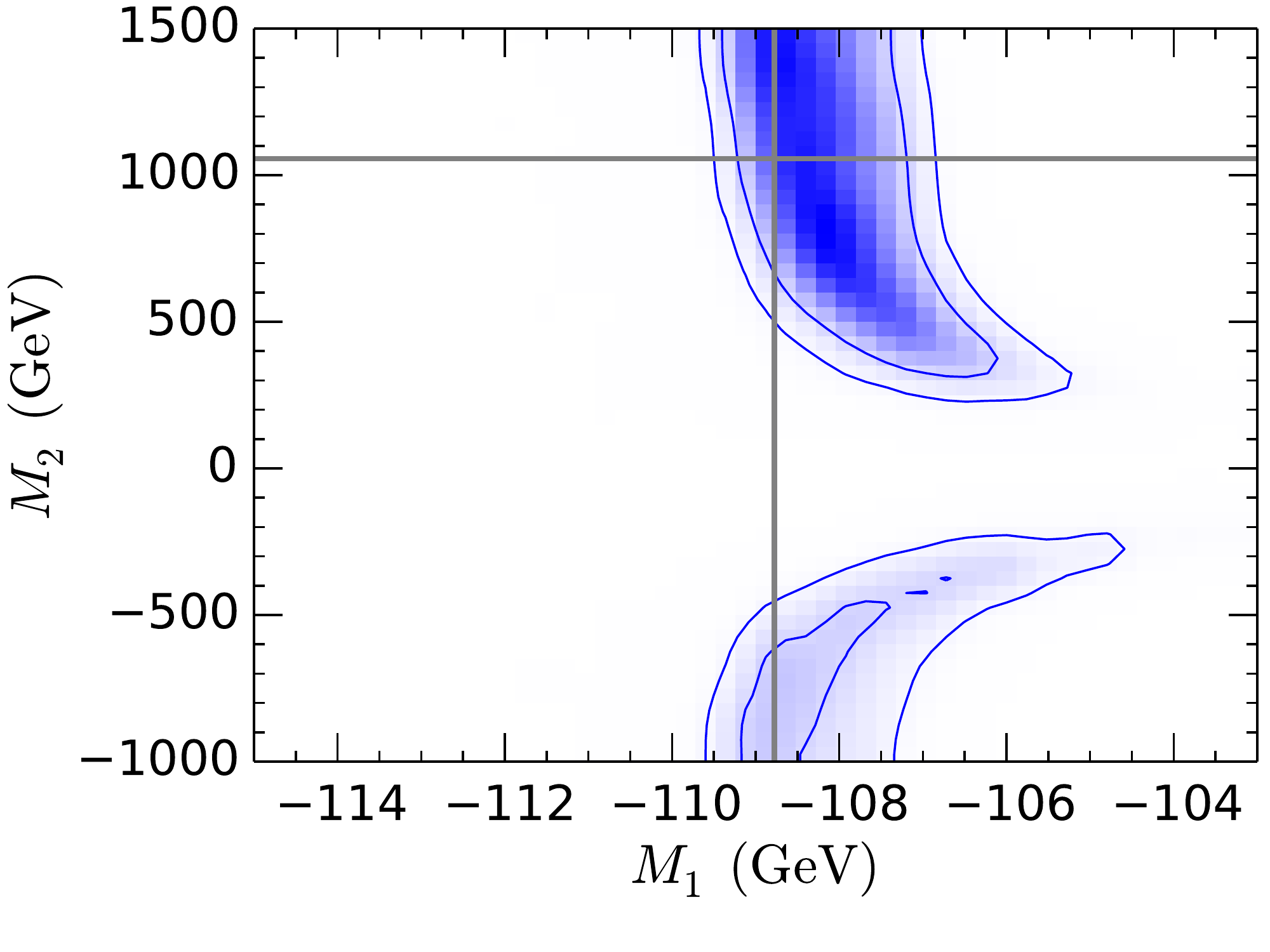}
  \includegraphics[height=0.37\textwidth, angle=0]{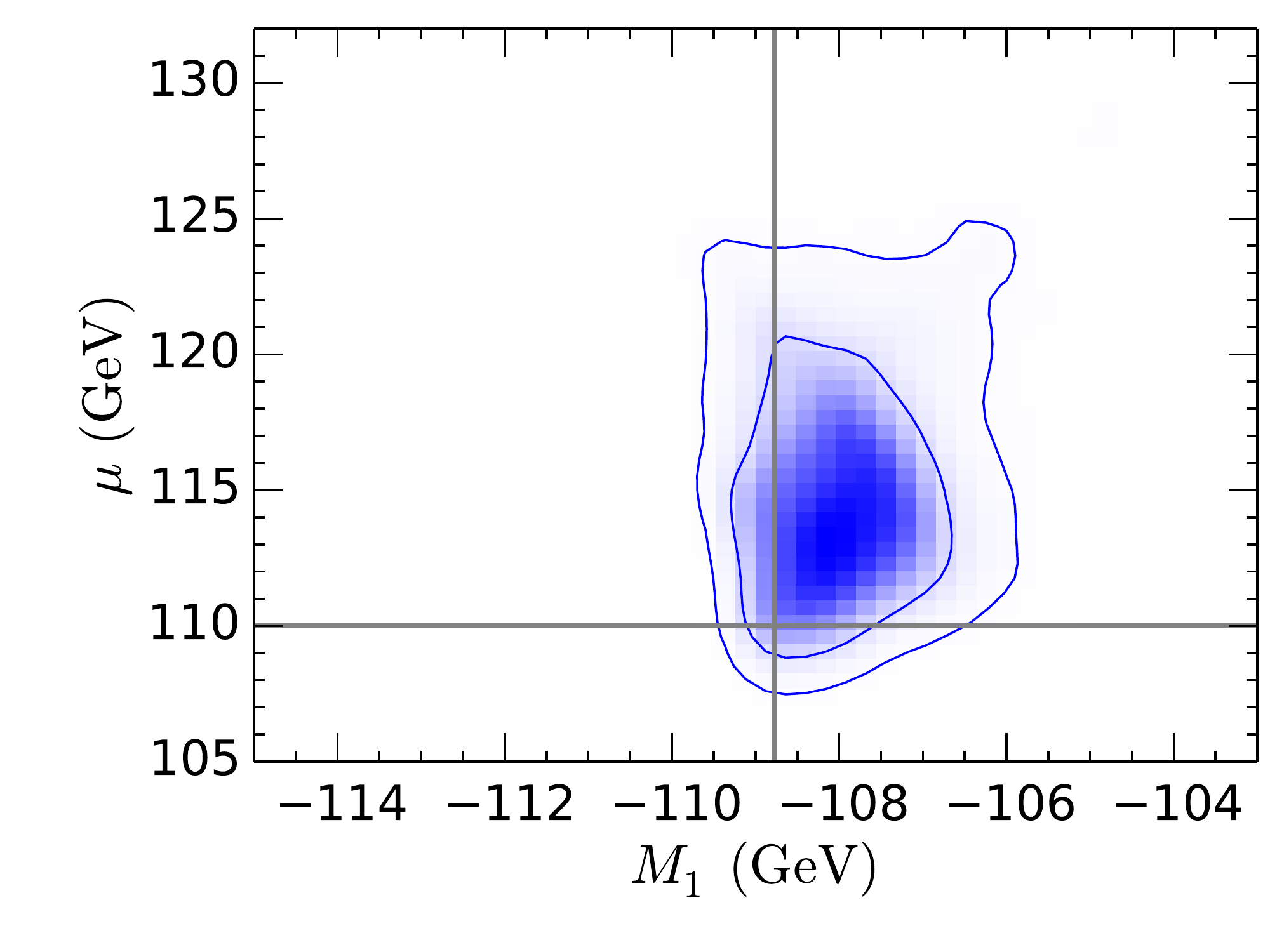}
  \includegraphics[height=0.37\textwidth, angle=0]{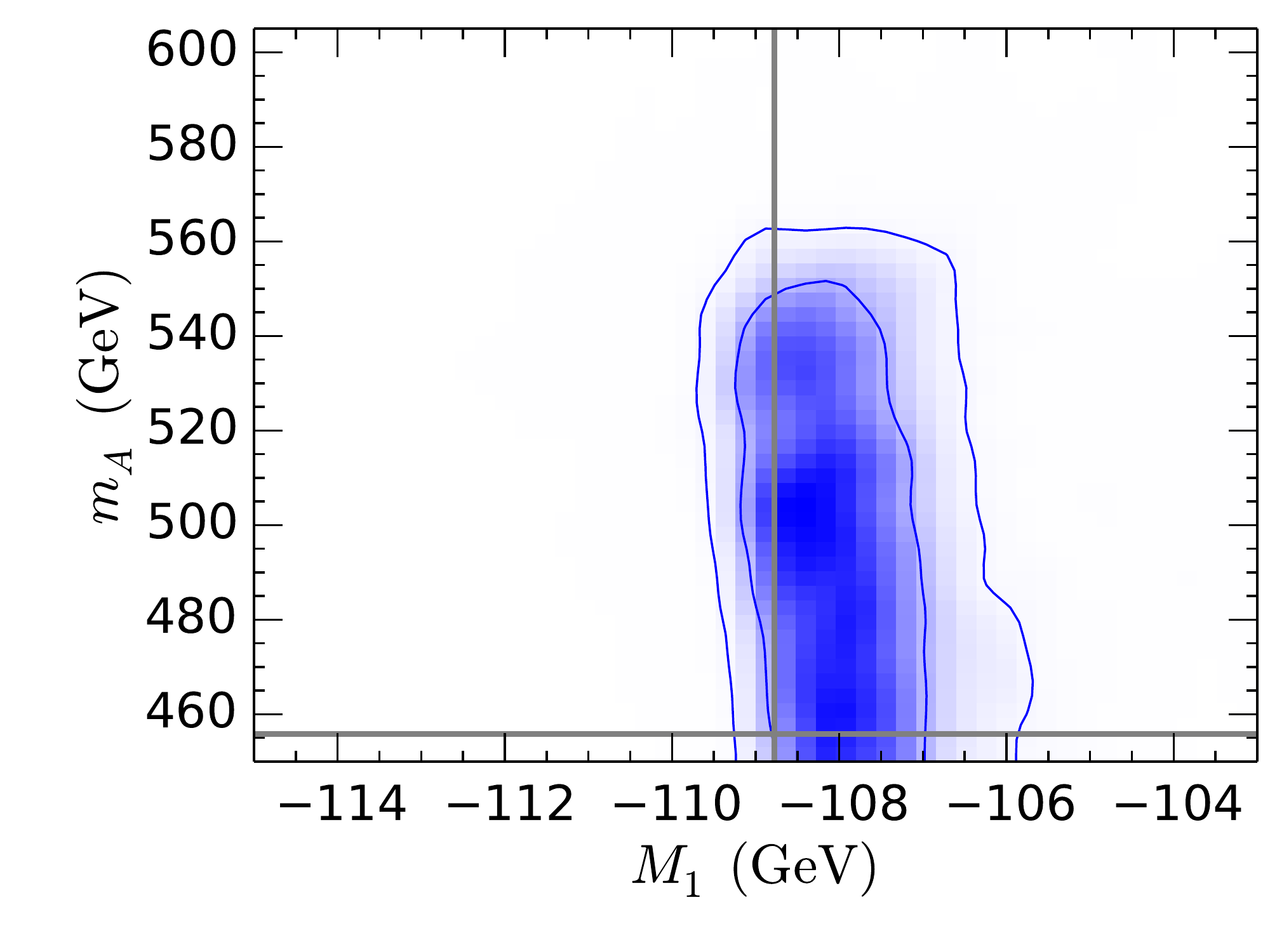}
  \includegraphics[height=0.37\textwidth, angle=0]{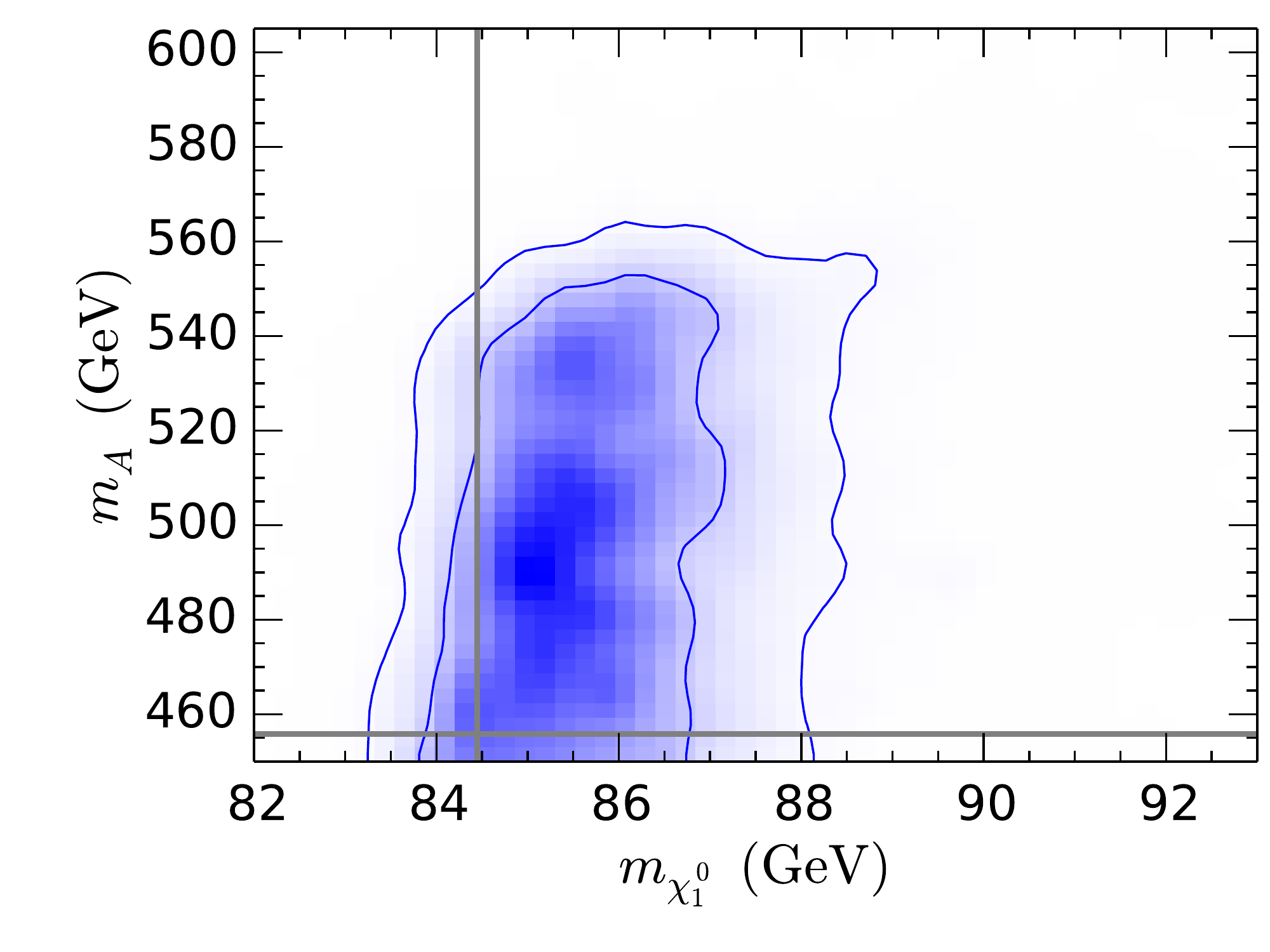}
  \includegraphics[height=0.37\textwidth, angle=0]{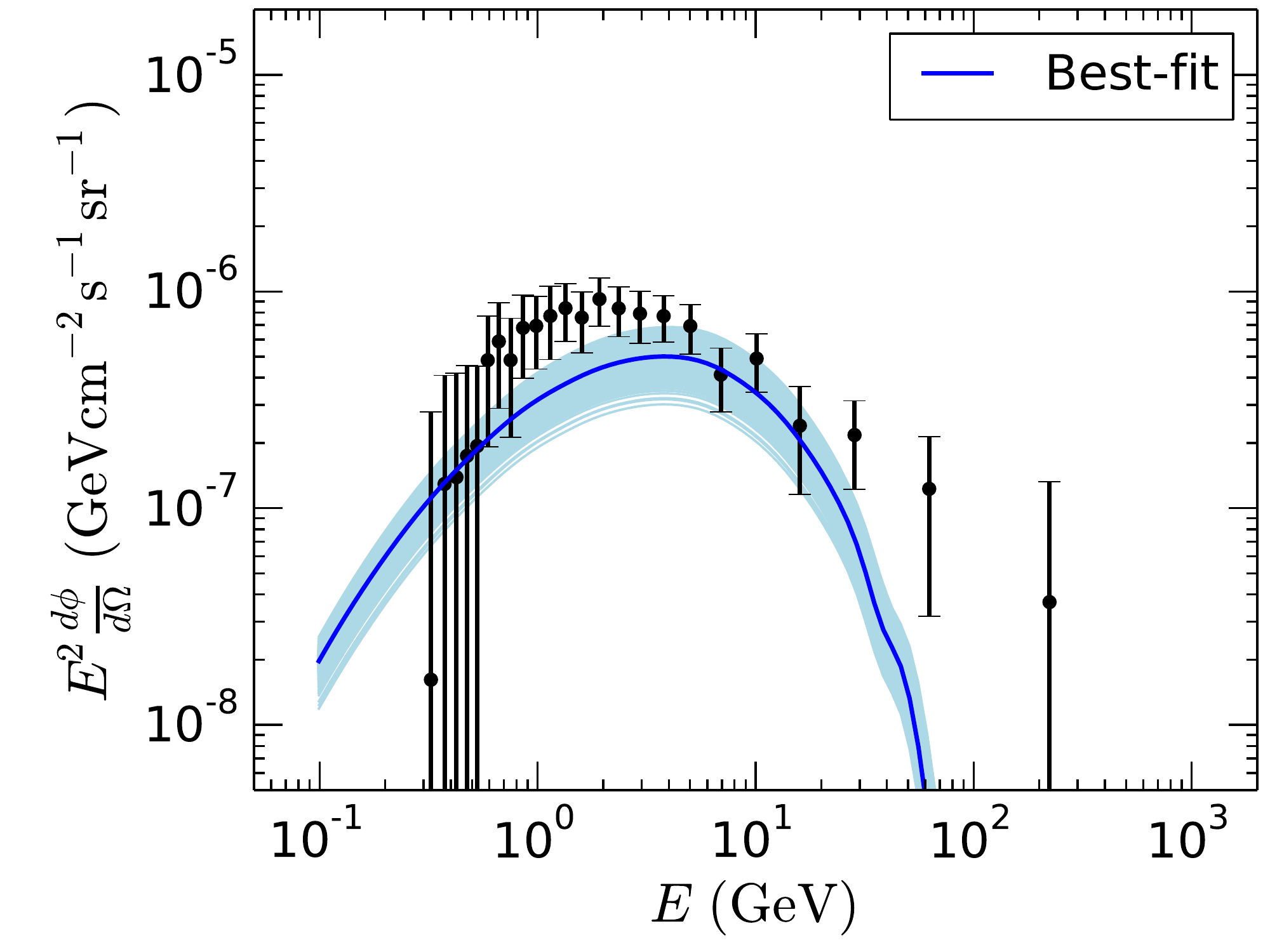}
  \includegraphics[height=0.37\textwidth, angle=0]{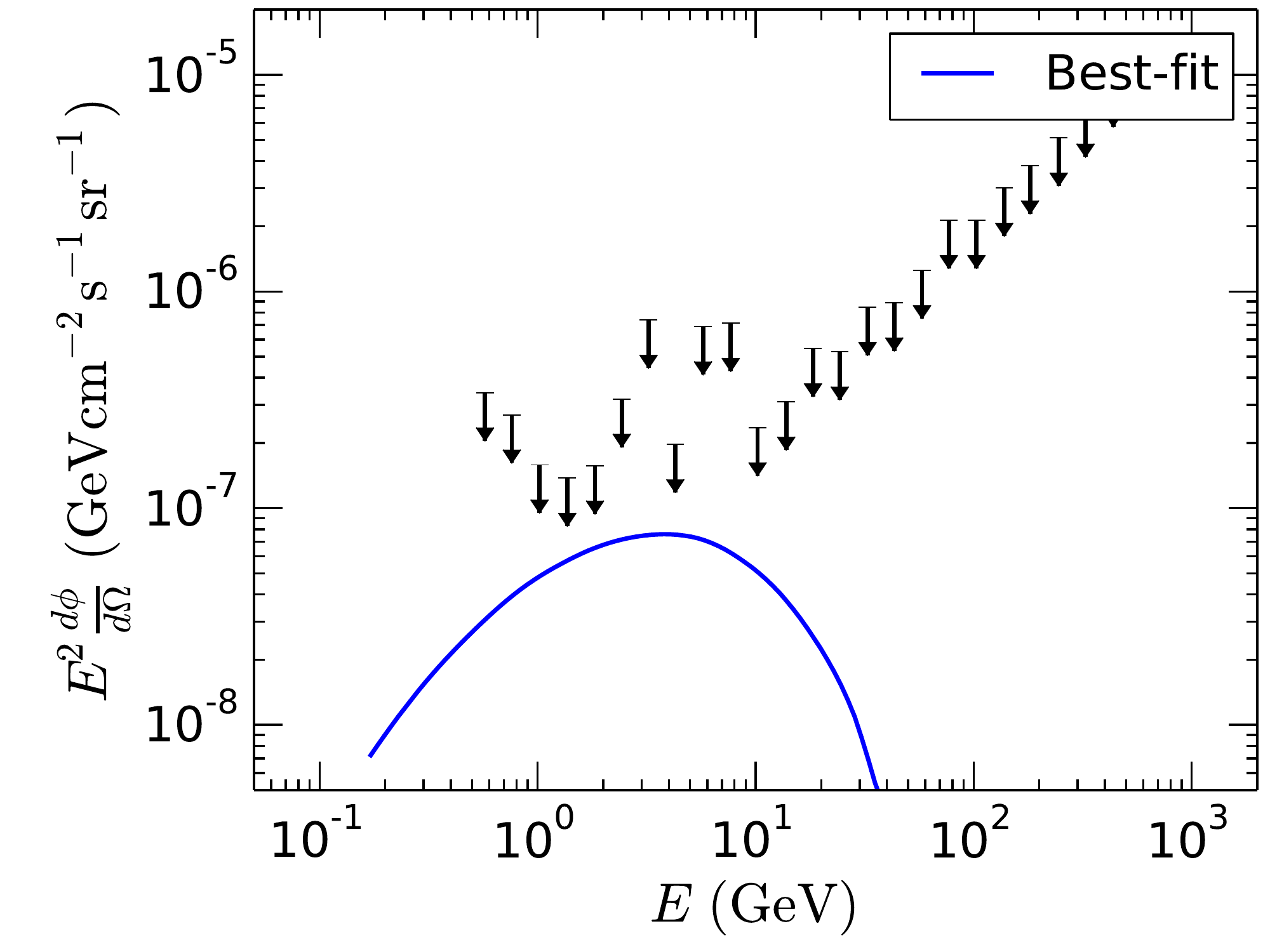}

  \caption{1$\sigma$ and 2$\sigma$ contours plots
    in the plane of $M_1 - M_2$ (top left), $M_1 - \mu$ (top right), $M_1 - m_A$ (middle left), $m_{\chi^0_1} - m_A$ (middle right) for Case 1a. Solid gray lines indicate the best fit values. Bottom left:
    2$\sigma$ bands of GC excess spectrum (light blue region)
    correspond to this case along with Fermi-LAT GC excess data and
    error bars (diagonal part of the covariance matrix). Deep blue
    line is the spectrum for best-fit points. Bottom right:
    Reticulum II $\gamma$-ray spectrum for the best-fit point (blue
    curve) along with the upper-limit on flux from Pass 8 analysis.}
\label{1a}
\end{figure*}

\underline{Case 1a}: Among all the scenarios sampled, this case fits
the GC excess spectrum best. The best-fit point corresponds to $M_{1}=
-108.75, M_{2}= 1115.85, \mu = 110.30, m_A = 451.92$, with
$\chi^{2}_{min} = 51.3$ for DOF = 24. The 2D projection of the $1\sigma$
and $2\sigma$ contour plots in the parameter hyperspace, corresponding
to the fit to the GC flux constrained by Reticulum II data, are
shown in Fig. \ref{1a}. This yields marginalised plots in various
pairs of parameters, as will also be seen in the cases to follow. Here
$M_{2}$ is less tightly constrained. The area thus marked out yields
the lightest neutralino mass ($m_{\chi^0_1}$) in the range $\simeq$ 83
- 88 GeV, and it is largely Bino, with some Higssino admixture. In any
case, since we are imposing the direct DM search constraint, the
lightest neutralino cannot have a very large Higgsino component. The
dominant channel of annihilation in this case is mainly $W^+ W^-$,
along with a small but perceptible branching fraction for
$b\bar{b}$. For $tan\beta = 20$, the LHC lower limit on $m_A$is about
450 GeV. The favoured range emerging from our $2\sigma$ fit is $m_A \simeq$ 
450 - 560 GeV.  This basically weakens the resonant annihilation channel
into third family fermion pairs, namely, $\chi^0_1 \chi^0_1 \rightarrow
f\overline{f}$, for low neutralino masses, as required for matching
the GC $\gamma$-ray spectrum.

The  limit on $m_A$ for $\tan\beta = 5$ is lower, but the $b\bar{b}$ 
coupling is weaker when $m_h$ is fixed at the observed value. For 
$\tan\beta = 50$, the corresponding limit goes up to $\simeq$ 850 GeV.

The annihilation in the $W^+ W^-$ channel is facilitated for such spectrum
by the fact that $\chi_1^{\pm}$, the lighter chargino has a
substantial Higgsino component, since $\mu$ is on the smaller side.
The close proximity of $\mu$ and $M_1$, while complying with the
lowers limits on the chargino mass, enhances the t-and u-channels of
annihilation, which interfere constructively. However, in the context 
of the early universe,  a boost in such
annihilation rate implies
correspondingly high $\left\langle \sigma v \right\rangle$, averaged
thermally. The lower bound on the relic density, modulo the
aforementioned uncertainties, therefore restricts  the
annihilation rate. This cannot be ameliorated by higher neutralino and
chargino masses, since that shifts the peak of the GC $\gamma$-ray
distribution to relatively high frequencies where the observed rate is
exceeded. A strict adherence to the lower limit on the relic density
(as one must do if the MSSM is the only new physics) thus implies
that, with the required low $m_{\chi^0_1}$, one cannot achieve as much
DM annihilation rate as is required to fully explain the GC
$\gamma$-ray excess with the observed frequency distribution. In
Fig. \ref{1a} we have presented the $2\sigma$ band of the predicted
GC excess spectrum along with the data points\footnote{Reference \cite{Achterberg:2017emt}
presents an otherwise sound study, including the latest direct search results.
However, the fact that we impose the requirement of a minimum $\Omega h^2$
worsens the fit compared to what is presented there.}.  
We re-iterate that
the band does not give sufficient saturation to the data as
$\left\langle \sigma v \right\rangle$ is constrained by the relic density 
lower bound. Fig. \ref{1a} shows the Reticulum II $\gamma$-ray
spectrum for the best-fit model in this case, where consistency with the 
Pass 8 limits is obvious.

\begin{figure*}[h]
\centering
  \includegraphics[height=0.37\textwidth, angle=0]{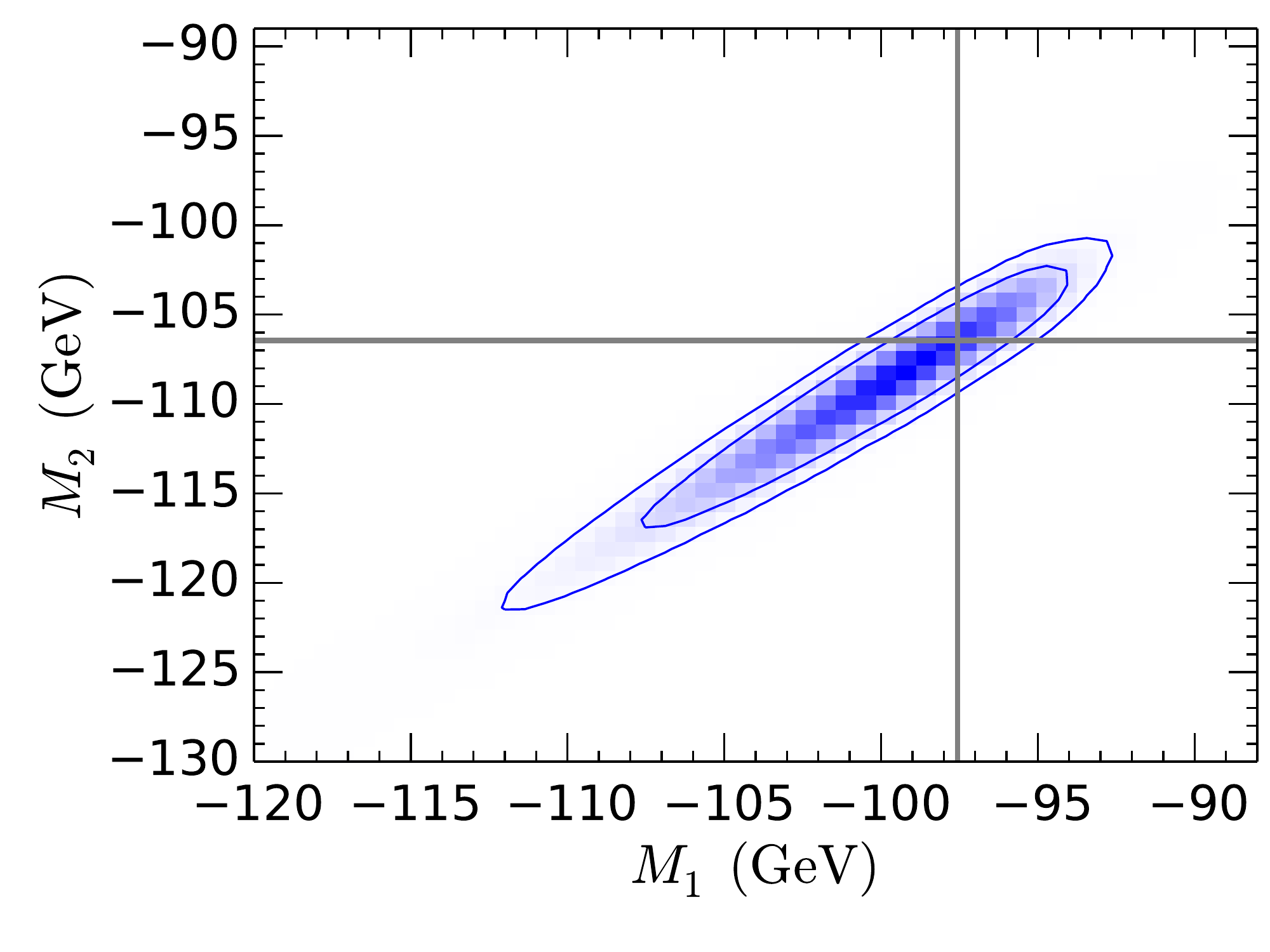}
  \includegraphics[height=0.37\textwidth, angle=0]{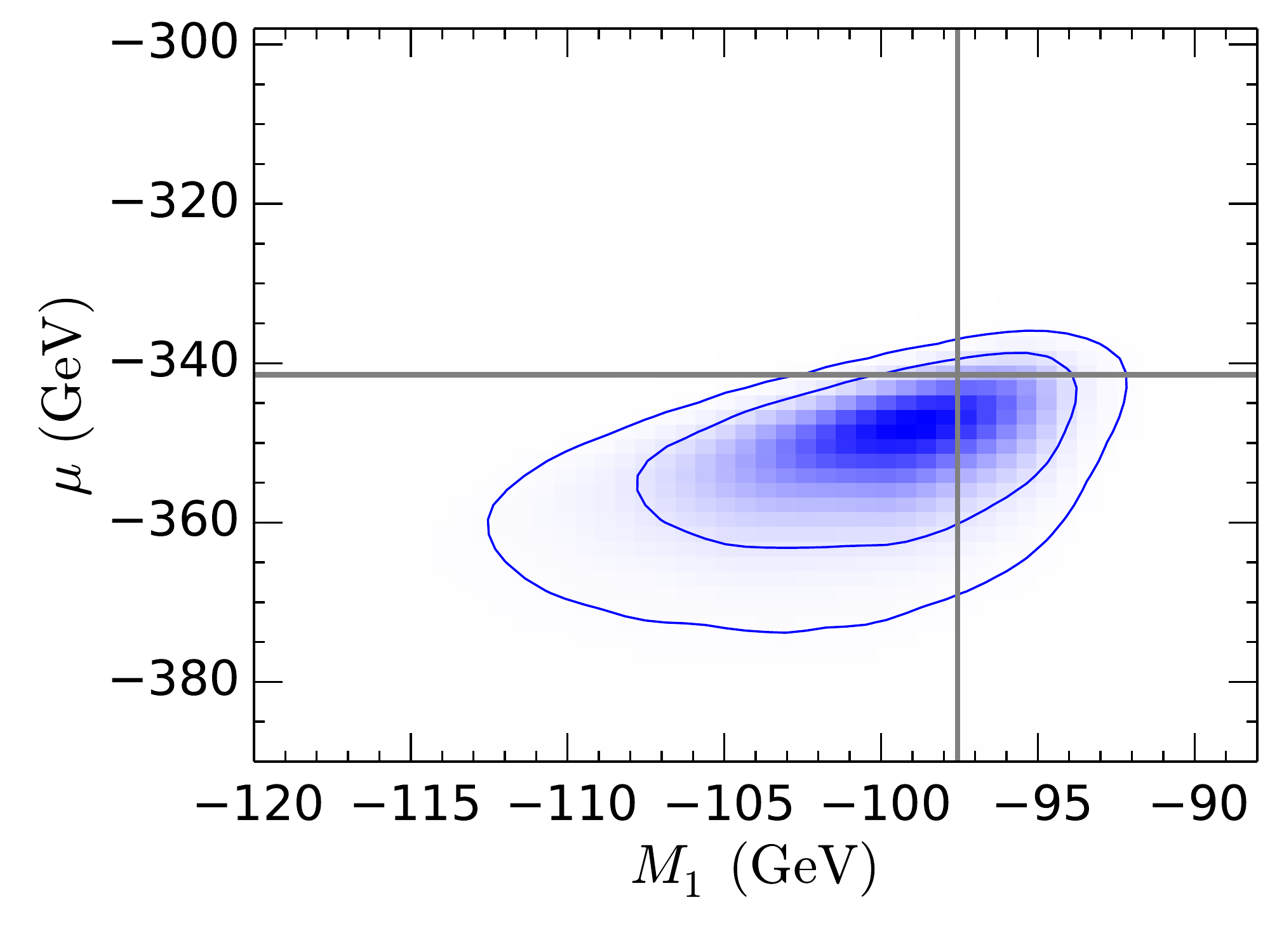}
  \includegraphics[height=0.37\textwidth, angle=0]{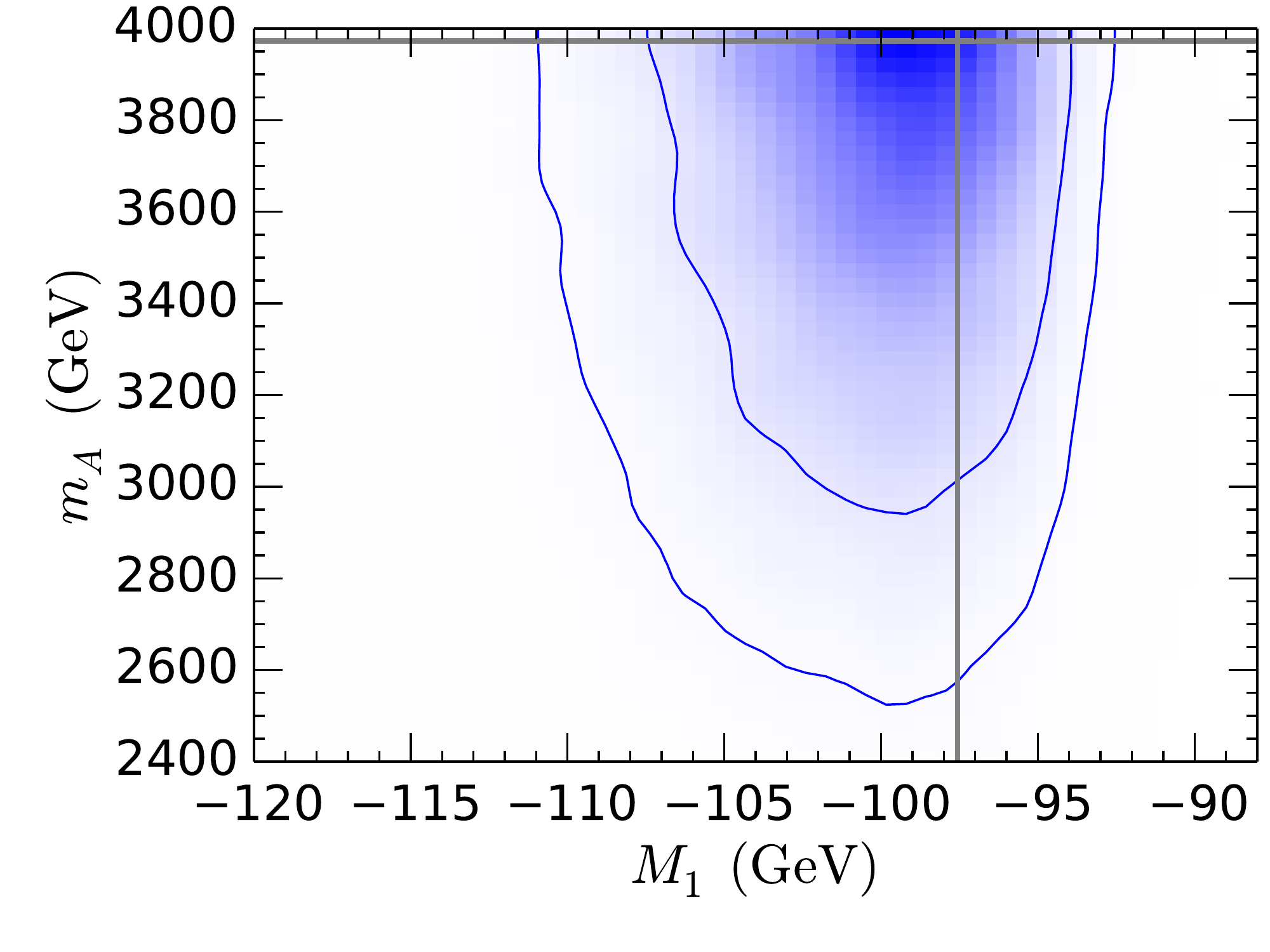}
  \includegraphics[height=0.37\textwidth, angle=0]{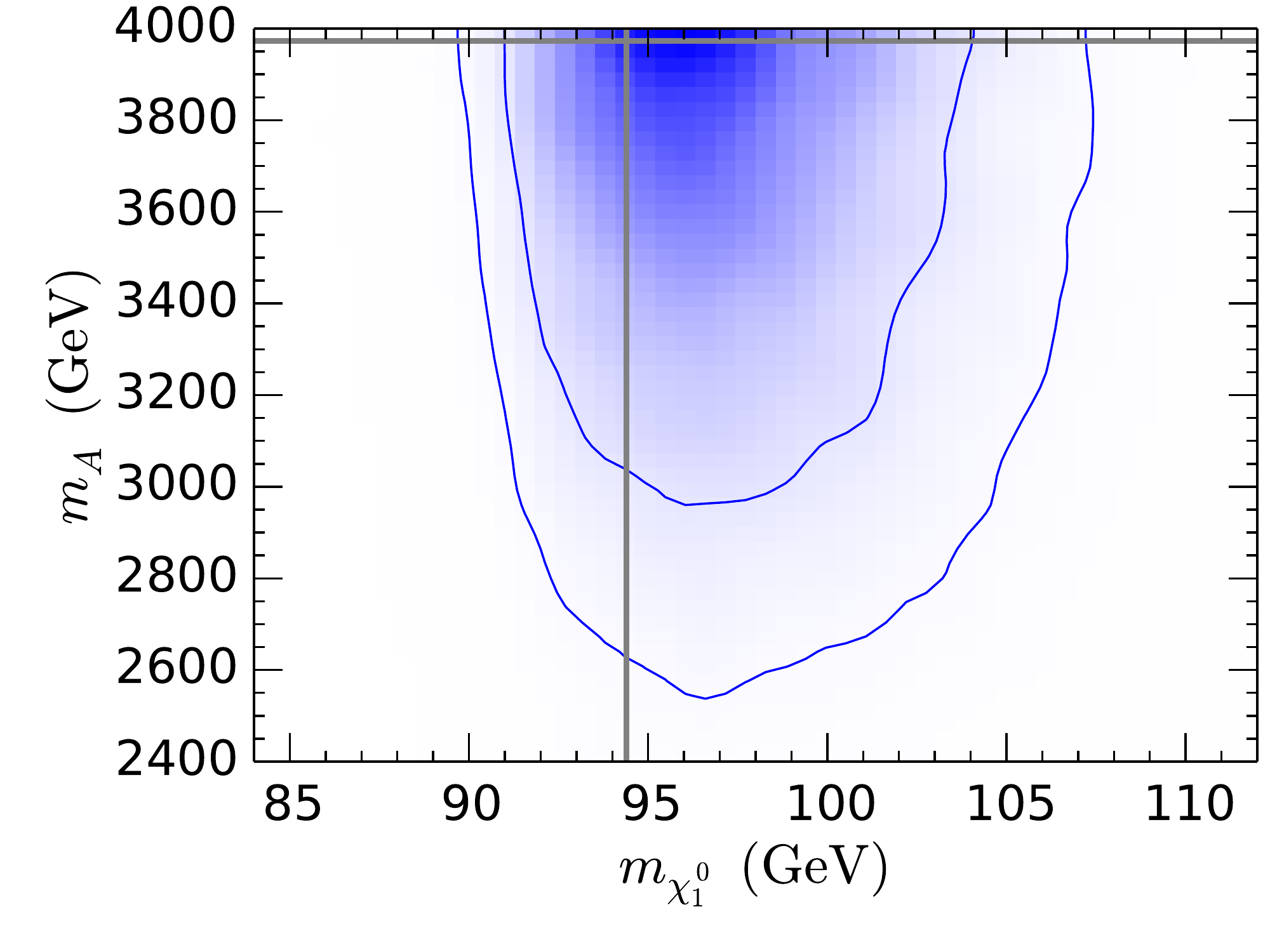}
  \includegraphics[height=0.37\textwidth, angle=0]{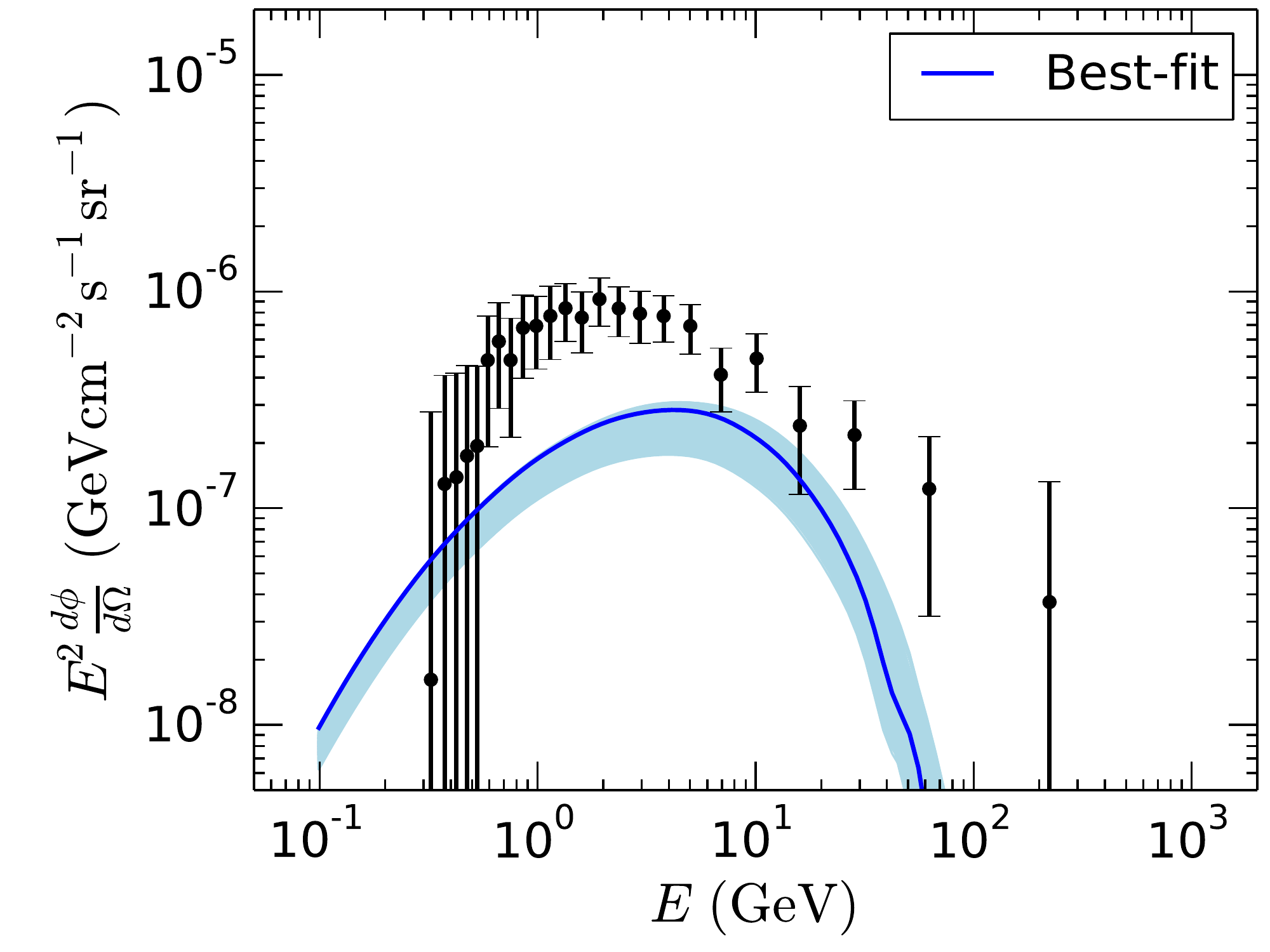}
  \includegraphics[height=0.37\textwidth, angle=0]{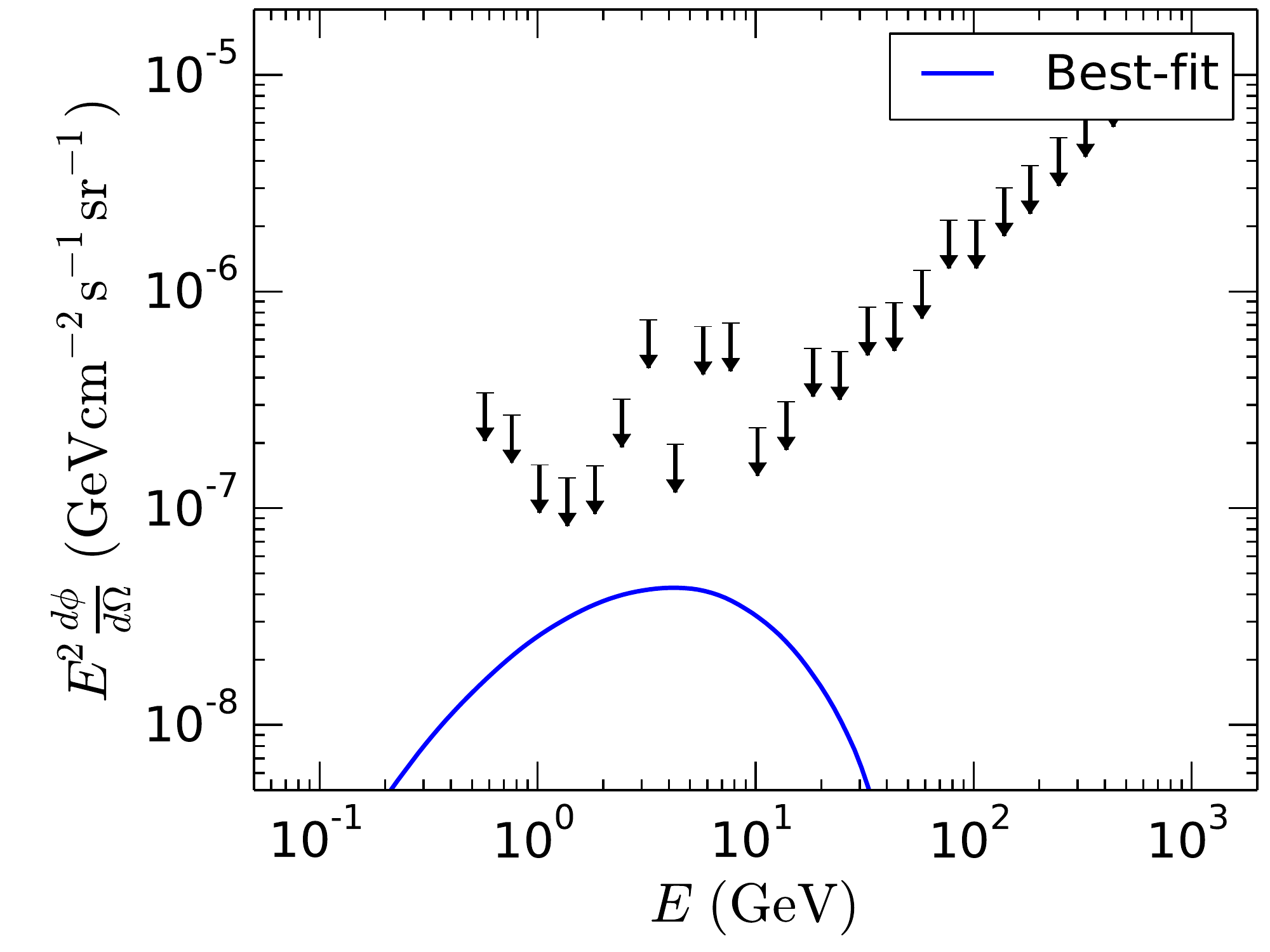}
  
  \caption{1$\sigma$ and 2$\sigma$ contours plots in the plane of $M_1
    - M_2$ (top left), $M_1 - \mu$ (top right), $M_1 - m_A$ (middle
    left), $m_{\chi^0_1} - m_A$ (middle right) for Case 1b. Solid gray lines indicate the
    best fit values. Bottom left: 2$\sigma$ bands of GC excess
    spectrum (light blue region) correspond to this case along with
    Fermi-LAT GC excess data and error bars (diagonal part of the
    covariance matrix). Deep blue line is the spectrum for best-fit
    points. Bottom right: Reticulum II $\gamma$-ray spectrum for the best-fit point (blue
    curve) along with the upper-limit on flux from Pass 8 analysis.}
\label{1b}
\end{figure*}

\underline{Case 1b}: The fits for this case yield $\chi^{2}_{min} = 65.1$, with the same DOF as in the previous case. The best-fit point
corresponds to $M_{1}= -97.56, M_{2}= -106.45, \mu = -341.44, m_A =
3973.12$.  Once more, the dominant channel of annihilation in this
case is $\chi^0_1 \chi^0_1 \rightarrow W^{+} W^{-}$. The $2\sigma$
contour plots corresponding to the fitting of GC flux excess and
Reticulum II upper limits is shown in Fig. \ref{1b}. The range of
neutralino mass in the $2\sigma$ contours for this case comes out to
be in the range $\simeq$ 90 - 108 GeV. The relatively low best-fit
value of $M_2$ and a somewhat higher $\mu$ leads to a $\chi^0_1$ that
is mostly Bino, with a smaller admixture of Wino. For $tan\beta = 50$,
as used here, $m_A$ comes out to be in the range $\simeq$ 2500 - 4000
GeV. This, together with the negligible Higgsino component in
$\chi^0_1$, seals the fate of the resonant annihilation channel and
thus prevents the relic density from becoming too low. On the other
hand, the small mass difference between $\chi_1^{\pm}$ and
$\chi_1^{0}$ once more drives the $W^{+} W^{-}$ channel. Fig. \ref{1b}
shows the 2$\sigma$ band of the GC excess spectrum drawn as earlier. 
As already discussed for the
previous case, the lower limit on the relic density in conjunction
with the position of the peak of the GC spectrum restricts the quality
of the fit for MSSM. The conclusion is similar
for the other benchmark scenarios discussed below, and we do not
repeat this statement for these cases.

\begin{figure*}[h]
\centering
  \includegraphics[height=0.37\textwidth, angle=0]{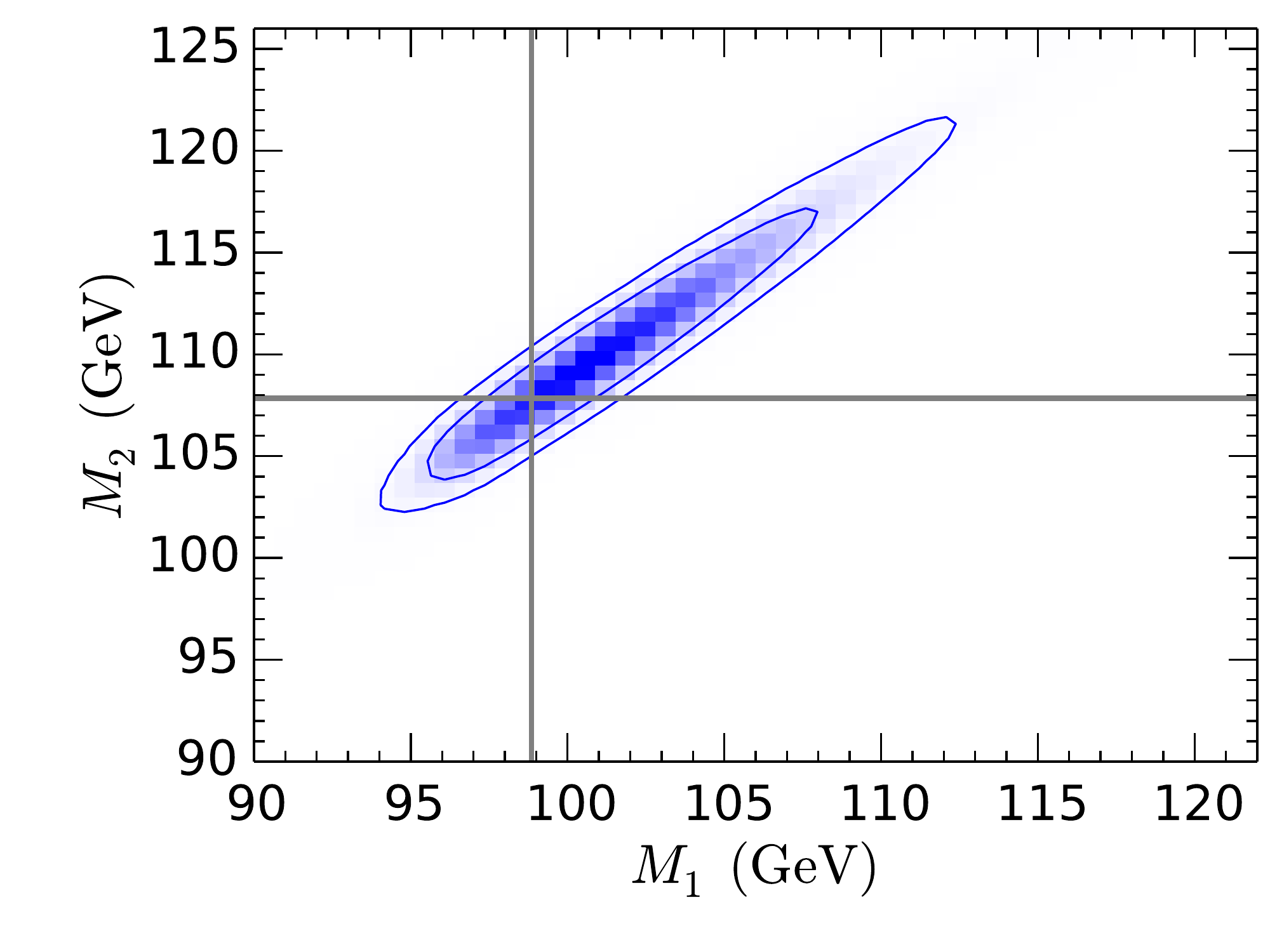}
  \includegraphics[height=0.37\textwidth, angle=0]{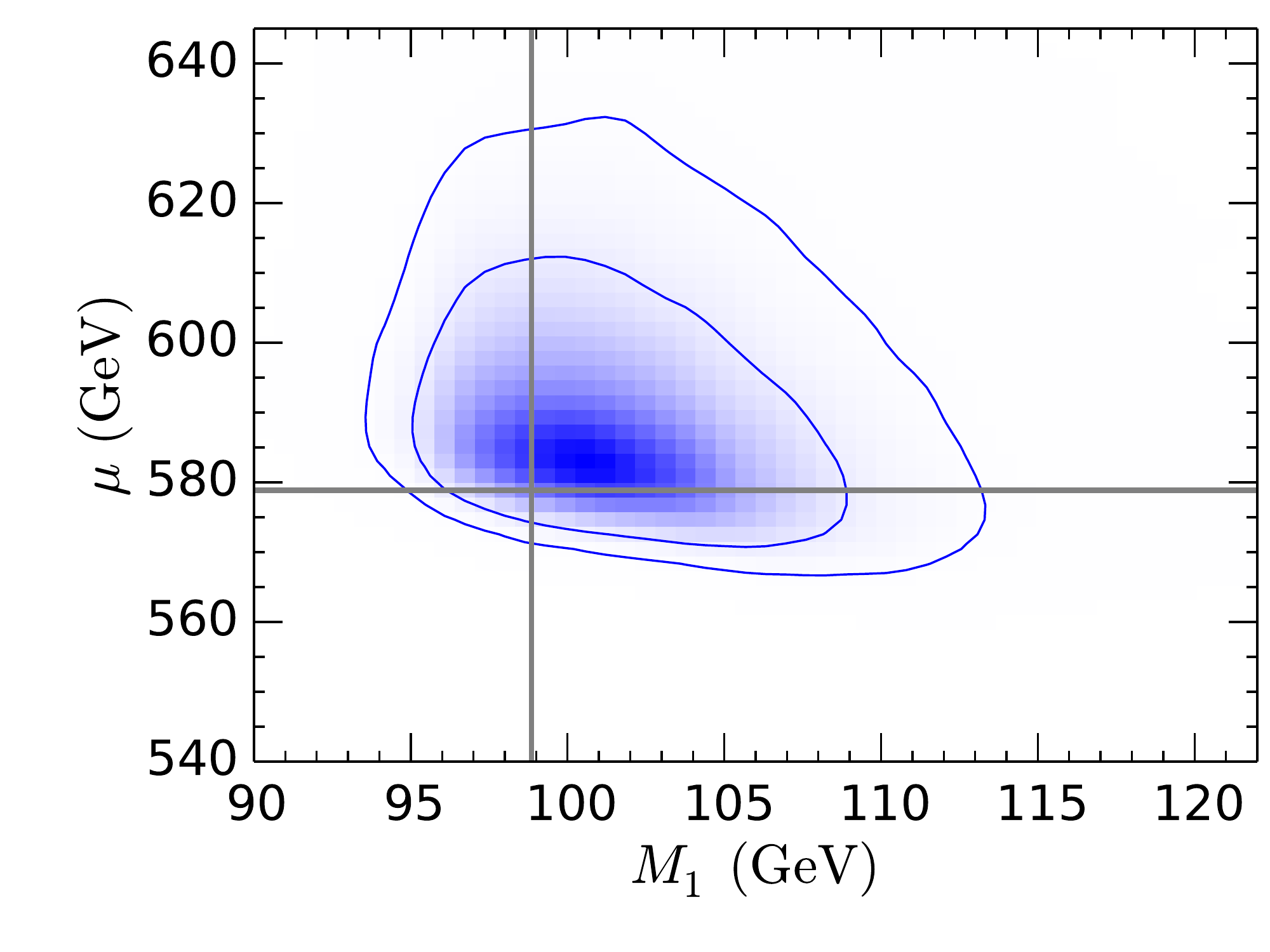}
  \includegraphics[height=0.37\textwidth, angle=0]{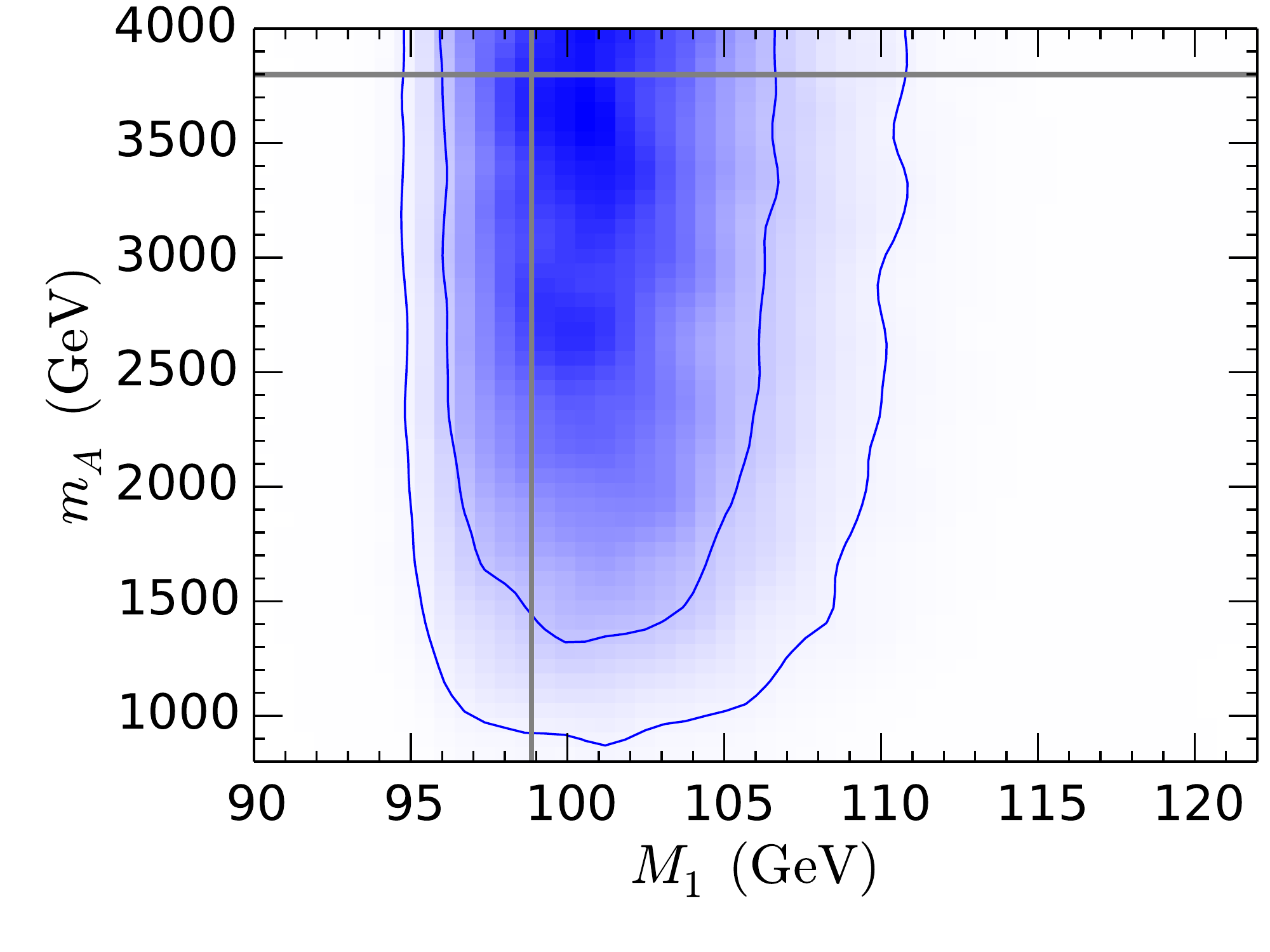}
  \includegraphics[height=0.37\textwidth, angle=0]{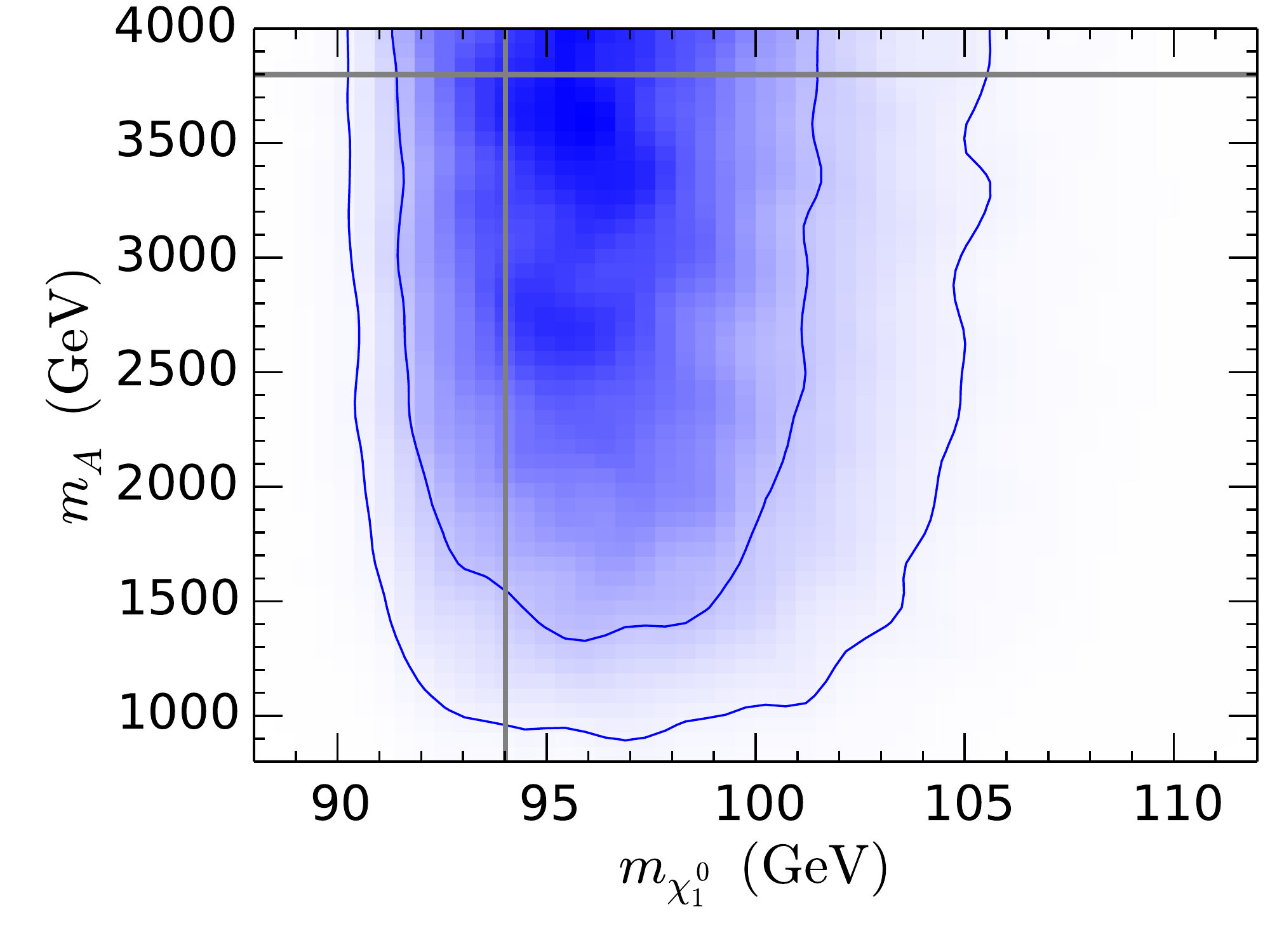}
  \includegraphics[height=0.37\textwidth, angle=0]{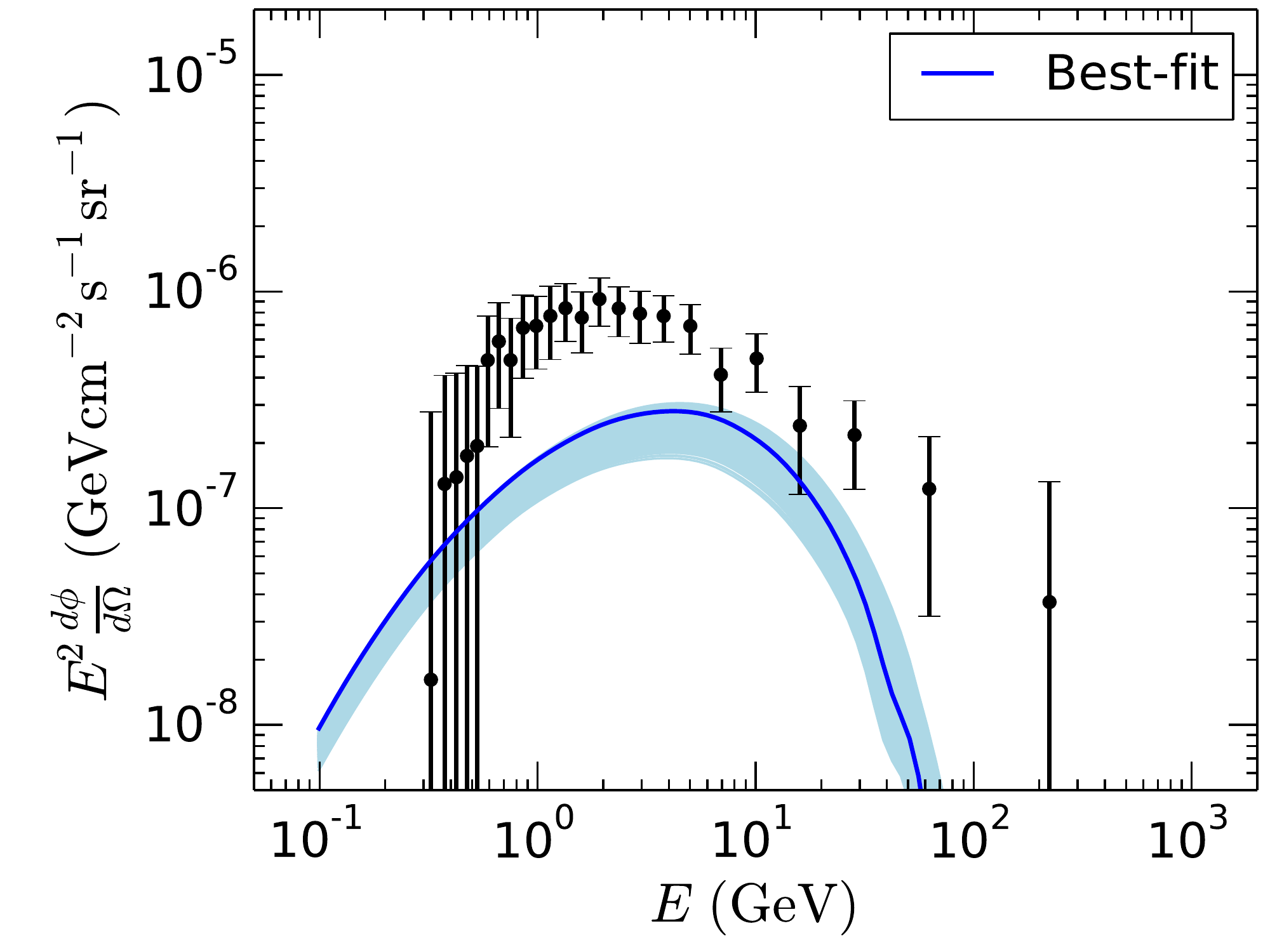}
  \includegraphics[height=0.37\textwidth, angle=0]{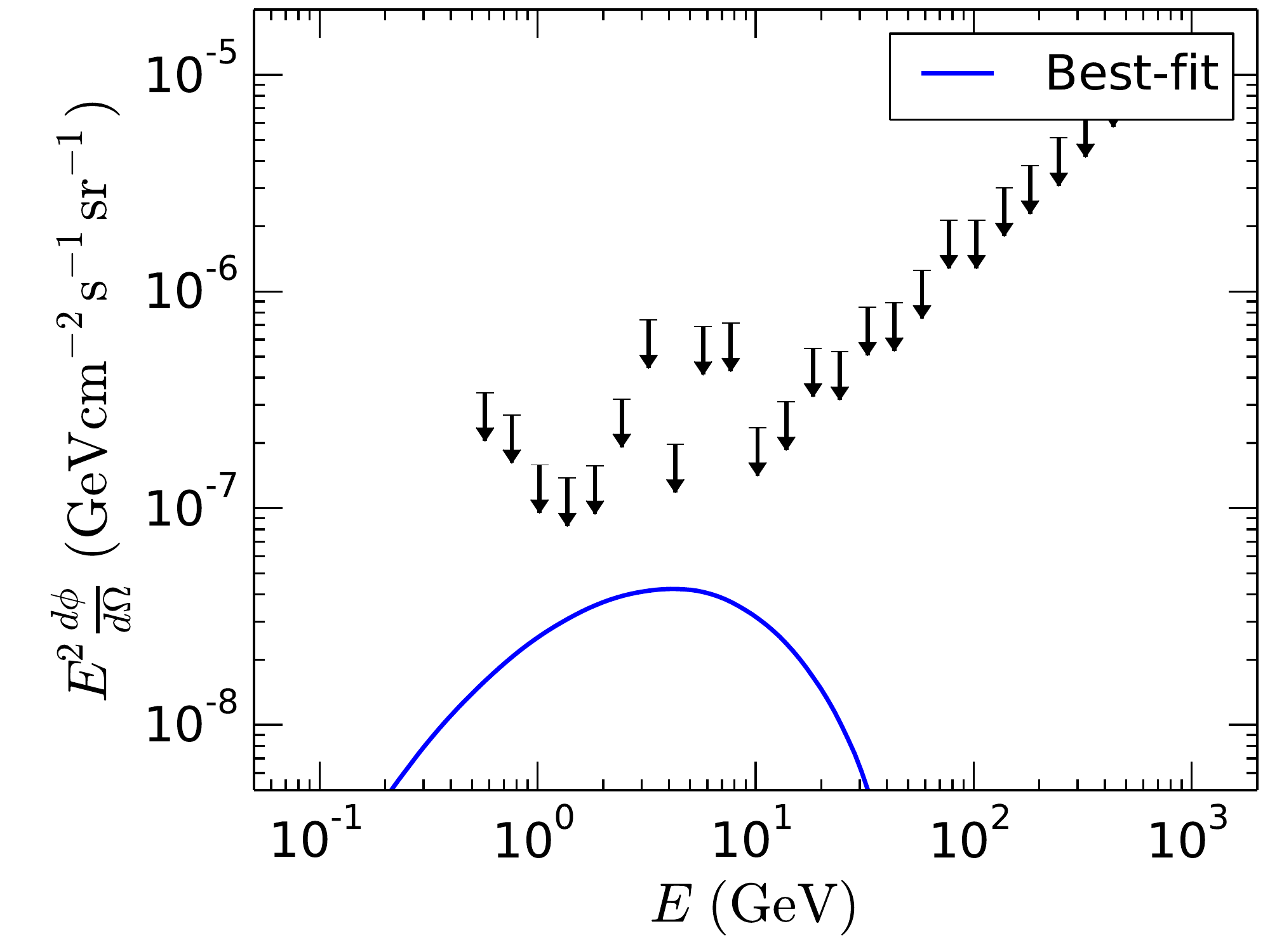}
  
  \caption{1$\sigma$ and 2$\sigma$ contours plots
    in the plane of $M_1 - M_2$ (top left), $M_1 - \mu$ (top right), $M_1 - m_A$ (middle left), $m_{\chi^0_1} - m_A$ (middle right) for Case 1c. Solid gray lines indicate the best fit values. Bottom left:
    2$\sigma$ bands of GC excess spectrum (light blue region)
    correspond to this case along with Fermi-LAT GC excess data and
    error bars (diagonal part of the covariance matrix). Deep blue
    line is the spectrum for best-fit points. Bottom right:
    Reticulum II $\gamma$-ray spectrum for the best-fit point (blue
    curve) along with the upper-limit on flux from Pass 8 analysis.}
\label{1c}
\end{figure*}

\underline{Case 1c}: One obtains $\chi^{2}_{min}$ = 65.2 for this
case, reflecting the fact that both this case and the previous one
yield worse fits compared to Case 1a.  The combination of parameters
corresponding to the best fit are $M_{1}= 98.85, M_{2}= 107.87, \mu =
578.89, m_A = 3798.42$, and the major annihilation channel once more
is $W^+ W^-$. Fig.  \ref{1c} contains the 2$\sigma$ contour plots for
this case. At 95.6$\%$ C.L., the neutralino mass lies in the range
$\simeq$ 90 - 105 GeV and is composed of Bino and Wino as above. Here
one ends up with a high-mass $m_A$ which is in the range 800 - 4000
GeV.  Once more, the 2$\sigma$ band of the GC excess spectrum and the best
fit Reticulum II  spectrum predicion for the best-fit model are shown in
Fig. \ref{1c}. In this case,
too, the lower limit on the relic density restricts large
$\left\langle \sigma v \right\rangle$ and low $m_{\chi}$, so that the
GC $\gamma$-ray data cannot fitted in a completely satisfactory
manner.

\subsection{Case 2: LST}

In this scenario we have set the right-chiral stop mass parameter
$m_{\tilde{t_R}}$ at 300 GeV and retained other squarks and sleptons
above 2 TeV. The latter choice, as has been mentioned already, is not
seen to affect the fits appreciably. The bigger mass parameters
(including the A-parameters corresponding to the trilinear soft SUSY
breaking terms) are varied in such a way as to reproduce the lighter
neutral scalar mass in the appropriate band. This keeps the
lighter stop around $\sim$ 260 - 300 GeV for all benchmark
points. For higher values of the lighter stop mass eigenstate than 300
GeV, the existing collider constraints would force us to a heavier
$\chi^0_1$ \cite{Sirunyan:2017cwe, Aaboud:2017ayj, Aaboud:2017nfd} and
as a result the fitting of $\gamma$-ray data would be worse. Values
lower than 260 GeV for $m_{\tilde{t_R}}$, on the other hand, can create
problems with the Higgs mass when one tries to maintain consistency
with other  constraints on MSSM spectra.

In principle, it may be curious to check the implications of 
scenarios with other third family sfermions like the sbottom
or the stau as well. For a light sbottom, however, the lower 
limit on the lightest neutralino is higher than in the case of
a light stop \cite{Sirunyan:2017kiw,
  Sirunyan:2017cwe}, thus worsening the GC $\gamma$-ray fits. The case 
of a light stau can make a noticeable difference only when it is
close enough to the $\chi^0_1$ to co-annihilate, a case that has 
been reported separately below.

As in the HSS scenario, here, too, we shall consider the results
with  three snapshot values of $\tan\beta$. However, we begin by 
pointing out a few salient features that are common to all three cases.

\begin{itemize}
\item The candidature of a light stop in our context thrives 
on the $t\bar{t}$ channel  of annihilation. The subprocesses
include the pseudoscalar-mediated s-channel diagram as well 
as the stop-mediated t-and u-channel ones.

\item The $t\bar{t}A$ couping is proportional to
$\cot\beta$, thus making the s-channel dominate in the
annihilation process for $\tan\beta$ = 5. 
As for the top-stop-$\chi^0_1$ interaction,
$\tan\beta$ enters in two ways: via the coupling to the
Higgsino component, and through the neutralino mass matrix itself.
The final results are related to all of these factors, and also to 
the fact that the t-and u-channels interfere destructively.

\item In order to annihilate into a top-antitop pair, one requires
  $m_{\chi^0_1} \ge$ 175 GeV approximately, a requirement that comes
  from the LHC constraints on the $m_{\chi^0_1} -
  m_{\tilde{t}_1}$ plane \cite{Sirunyan:2017cwe, Aaboud:2017ayj,
    Aaboud:2017nfd}. However, this inevitably tends to shift the peak
  of the GC $\gamma$-ray spectrum to the region of higher frequency
  than where the observed peak lies.  This, together with the limit on
  upward scaling of the profile coming from the requirement of a
  minimum relic density, poses a challenge to good fits of all data in
  the light stop scenario.
\end{itemize}

\begin{figure*}[h]
\centering
  \includegraphics[height=0.37\textwidth, angle=0]{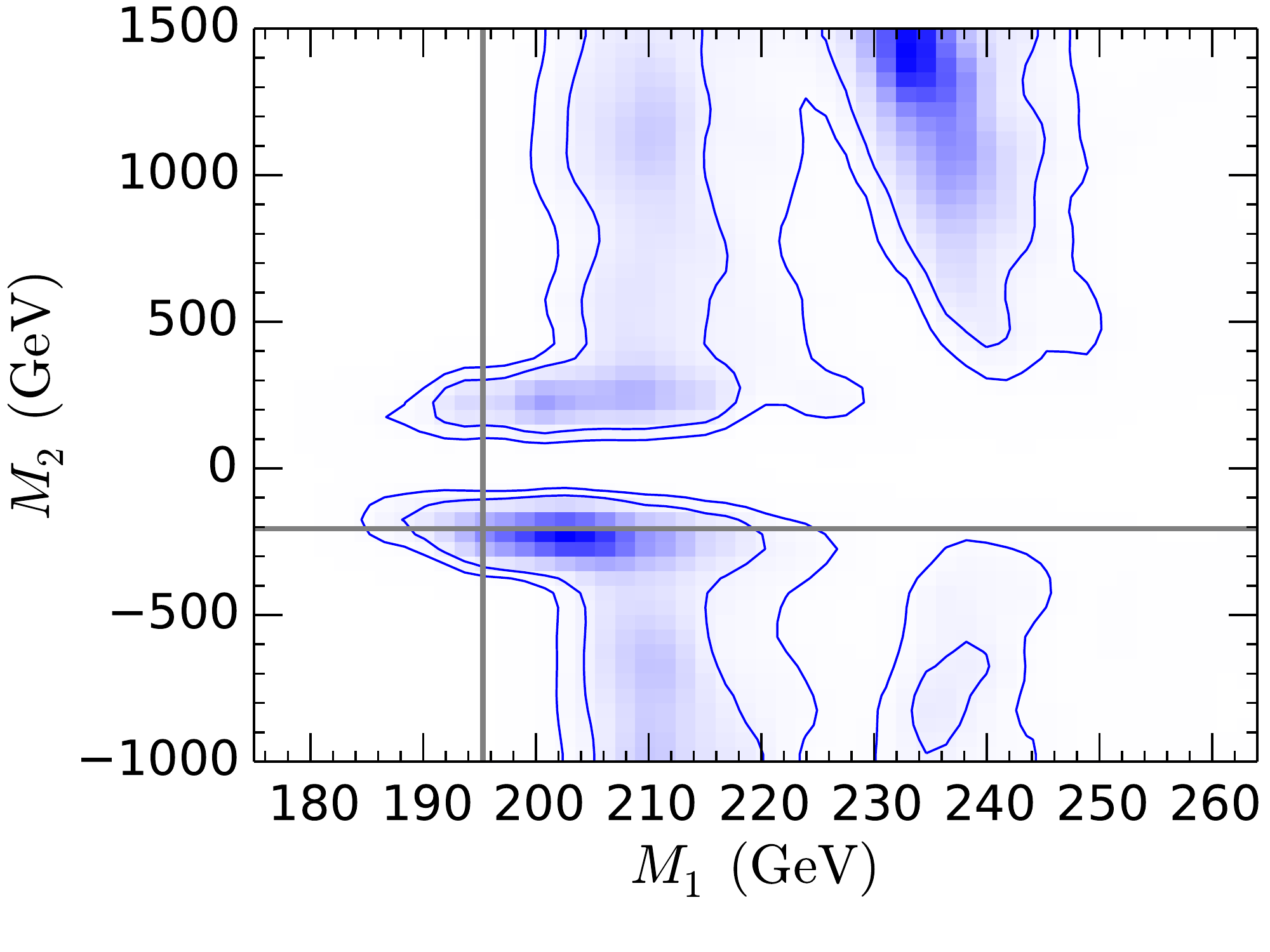}
  \includegraphics[height=0.37\textwidth, angle=0]{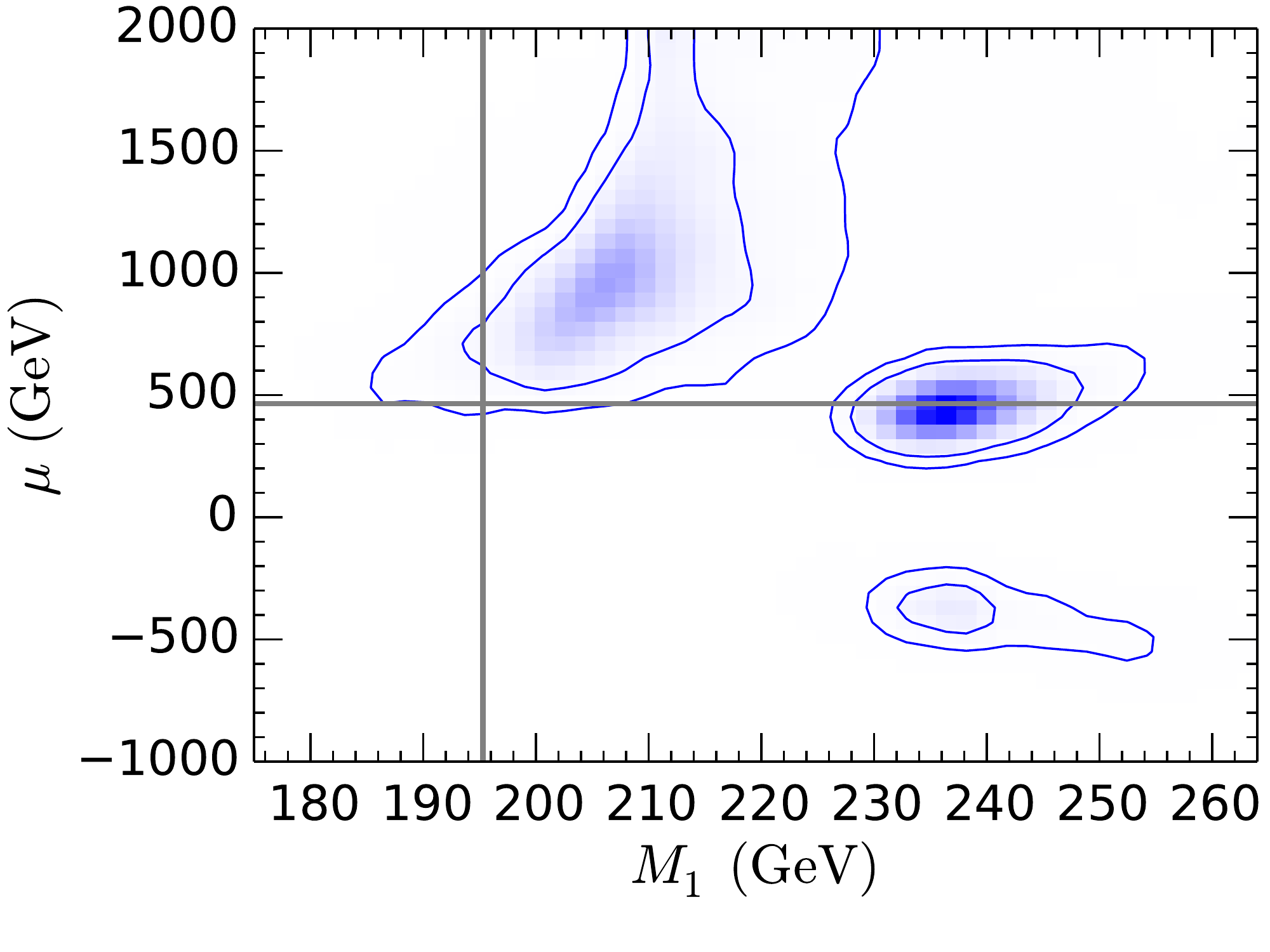}
  \includegraphics[height=0.37\textwidth, angle=0]{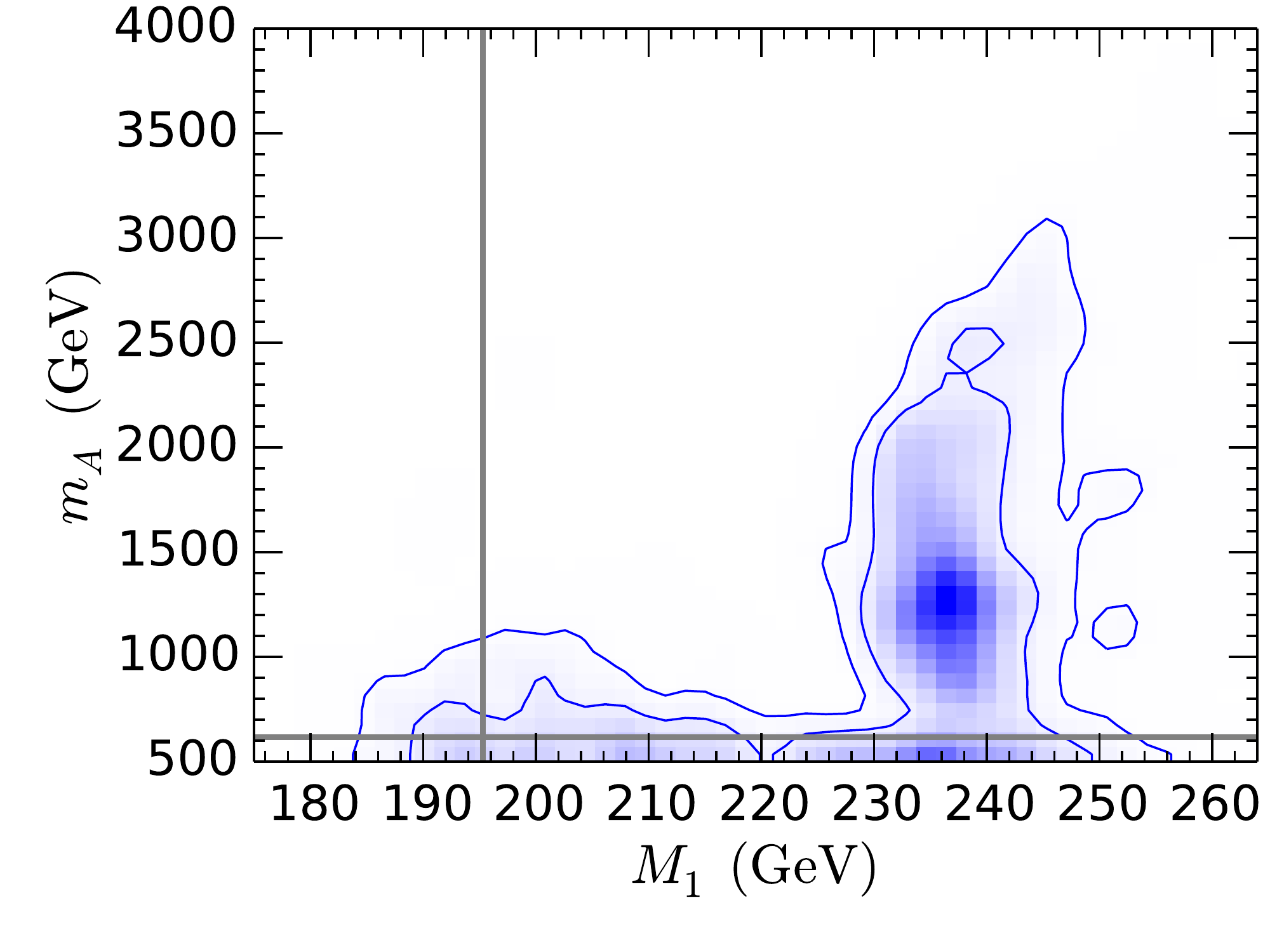}
  \includegraphics[height=0.37\textwidth, angle=0]{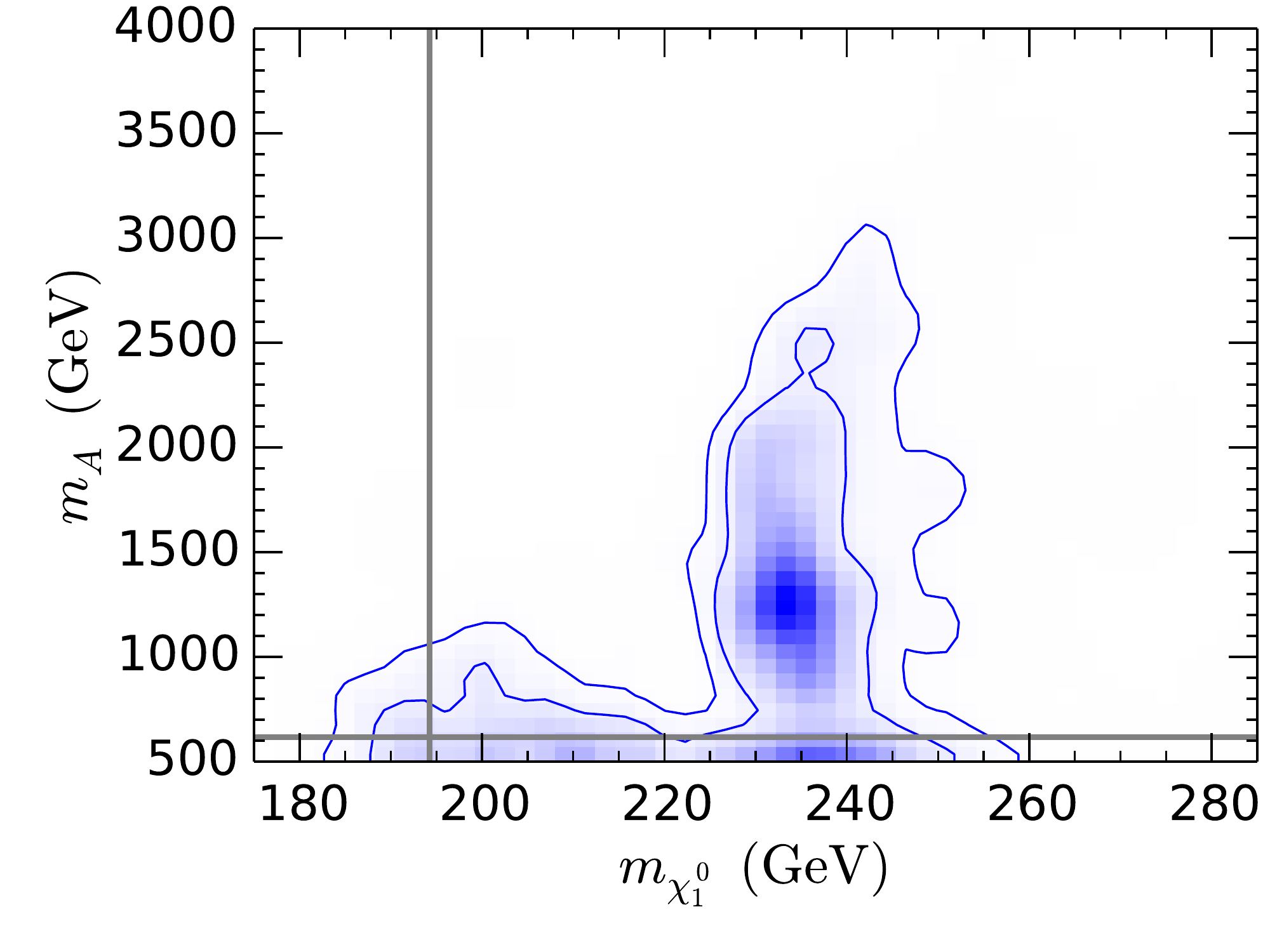}
  \includegraphics[height=0.37\textwidth, angle=0]{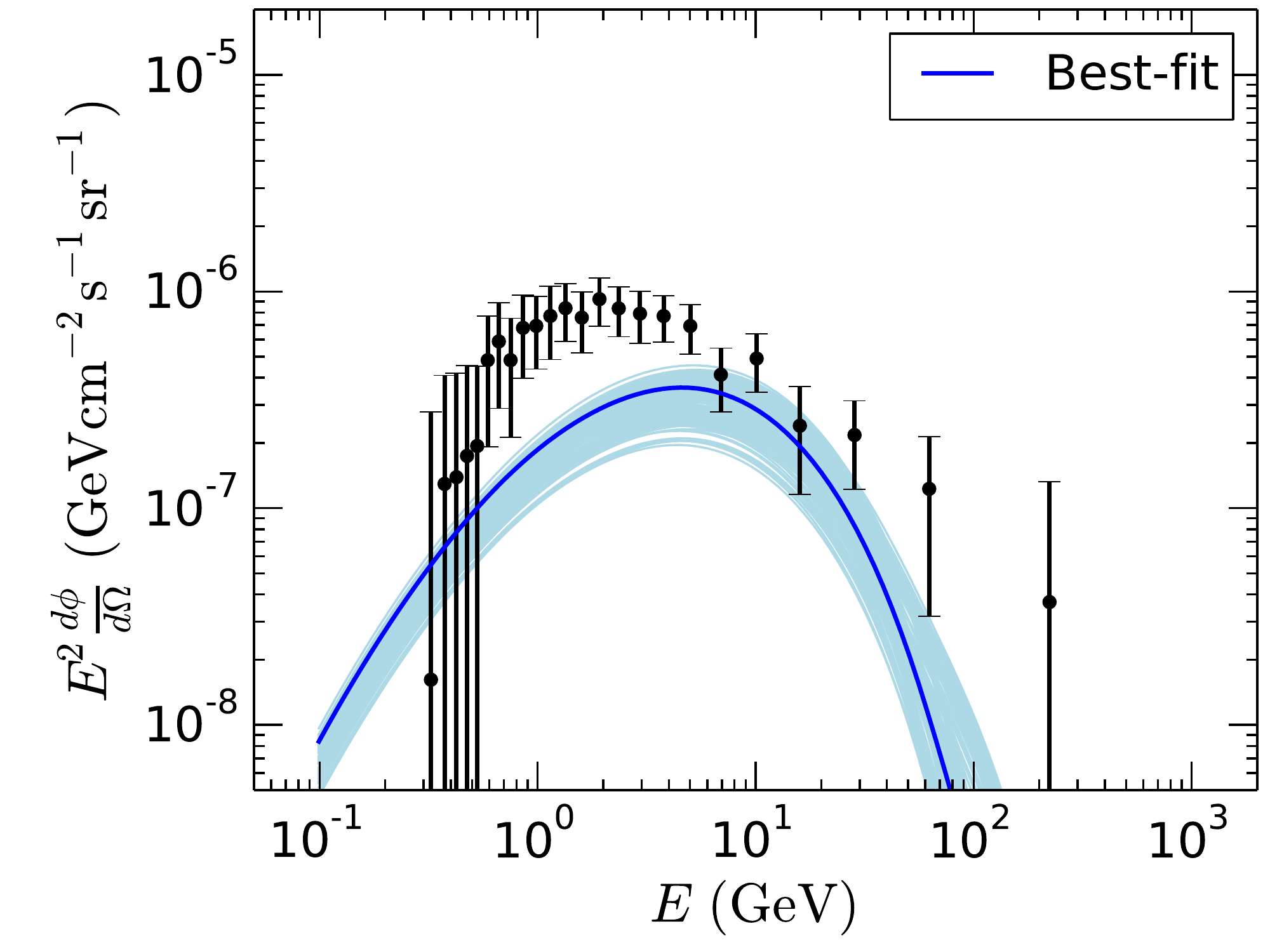}
  \includegraphics[height=0.37\textwidth, angle=0]{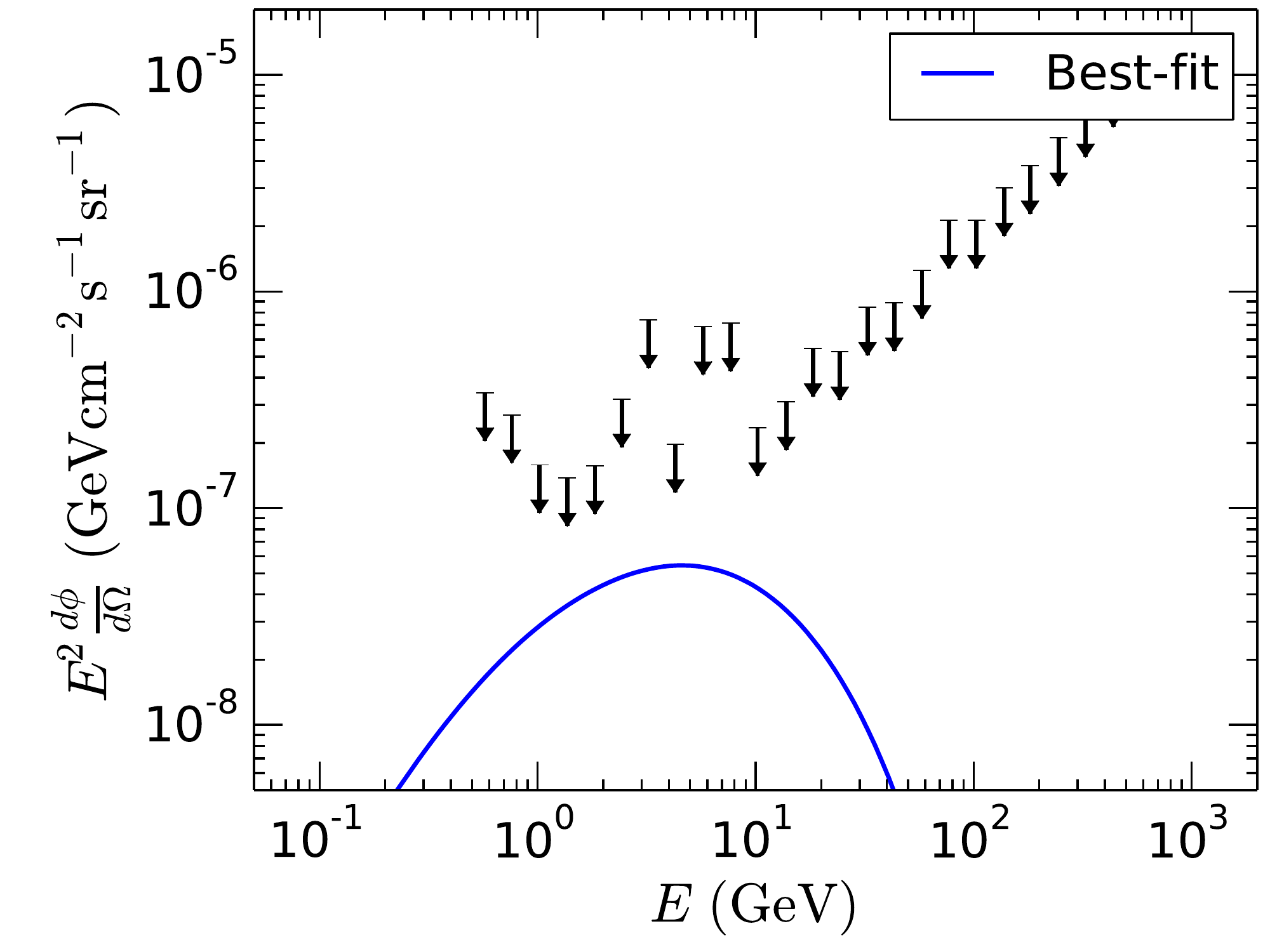}
  
  \caption{1$\sigma$ and 2$\sigma$ contours plots
    in the plane of $M_1 - M_2$ (top left), $M_1 - \mu$ (top right), $M_1 - m_A$ (middle left), $m_{\chi^0_1} - m_A$ (middle right) for Case 2a. Solid gray lines indicate the best fit values. Bottom left:
    2$\sigma$ bands of GC excess spectrum (light blue region)
    correspond to this case along with Fermi-LAT GC excess data and
    error bars (diagonal part of the covariance matrix). Deep blue
    line is the spectrum for best-fit points. Bottom right:
    Reticulum II $\gamma$-ray spectrum for the best-fit point (blue
    curve) along with the upper-limit on flux from Pass 8 analysis.}
\label{2a}
\end{figure*}

\begin{figure*}[h]
\centering
  \includegraphics[height=0.37\textwidth, angle=0]{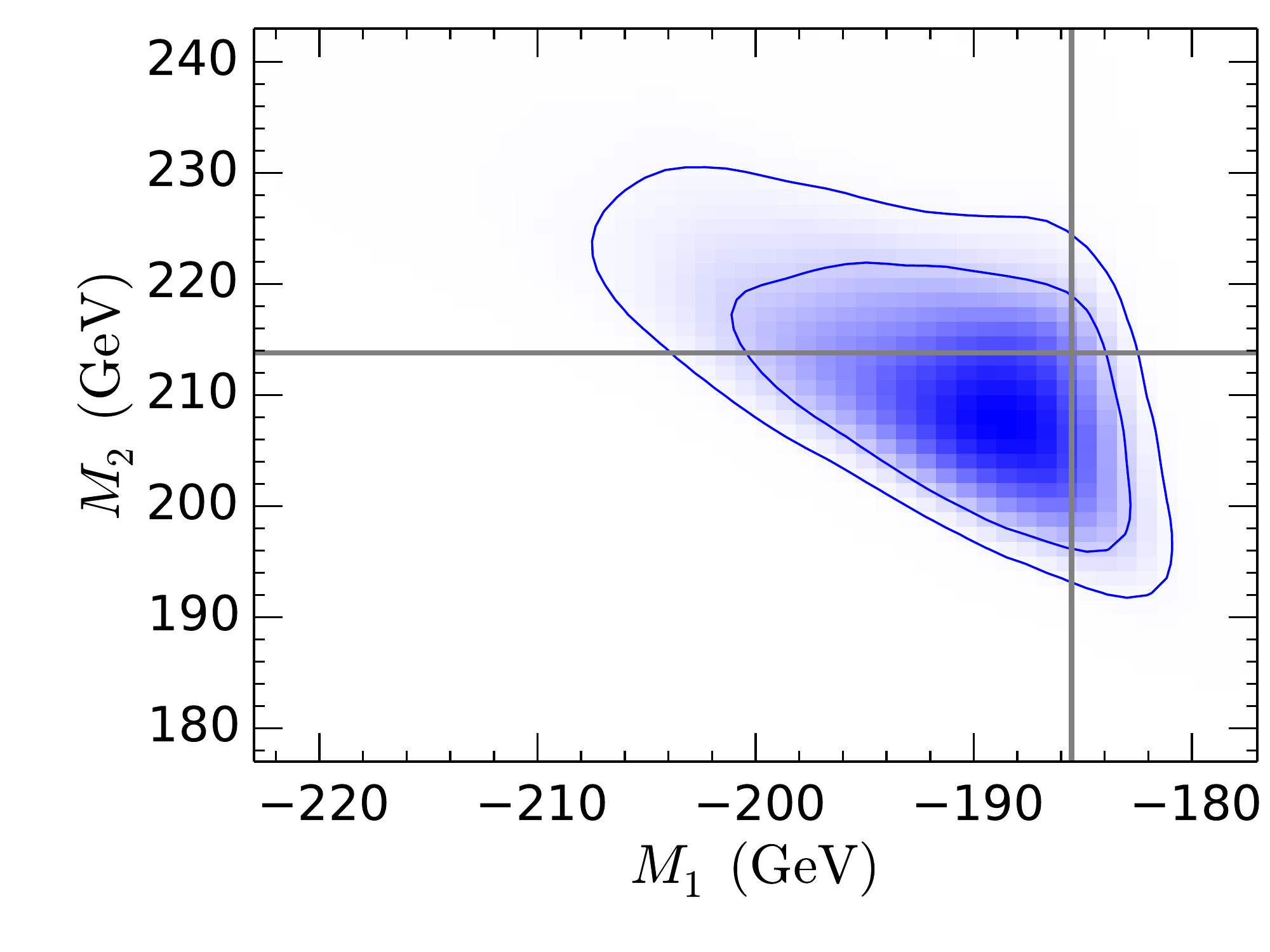}
  \includegraphics[height=0.37\textwidth, angle=0]{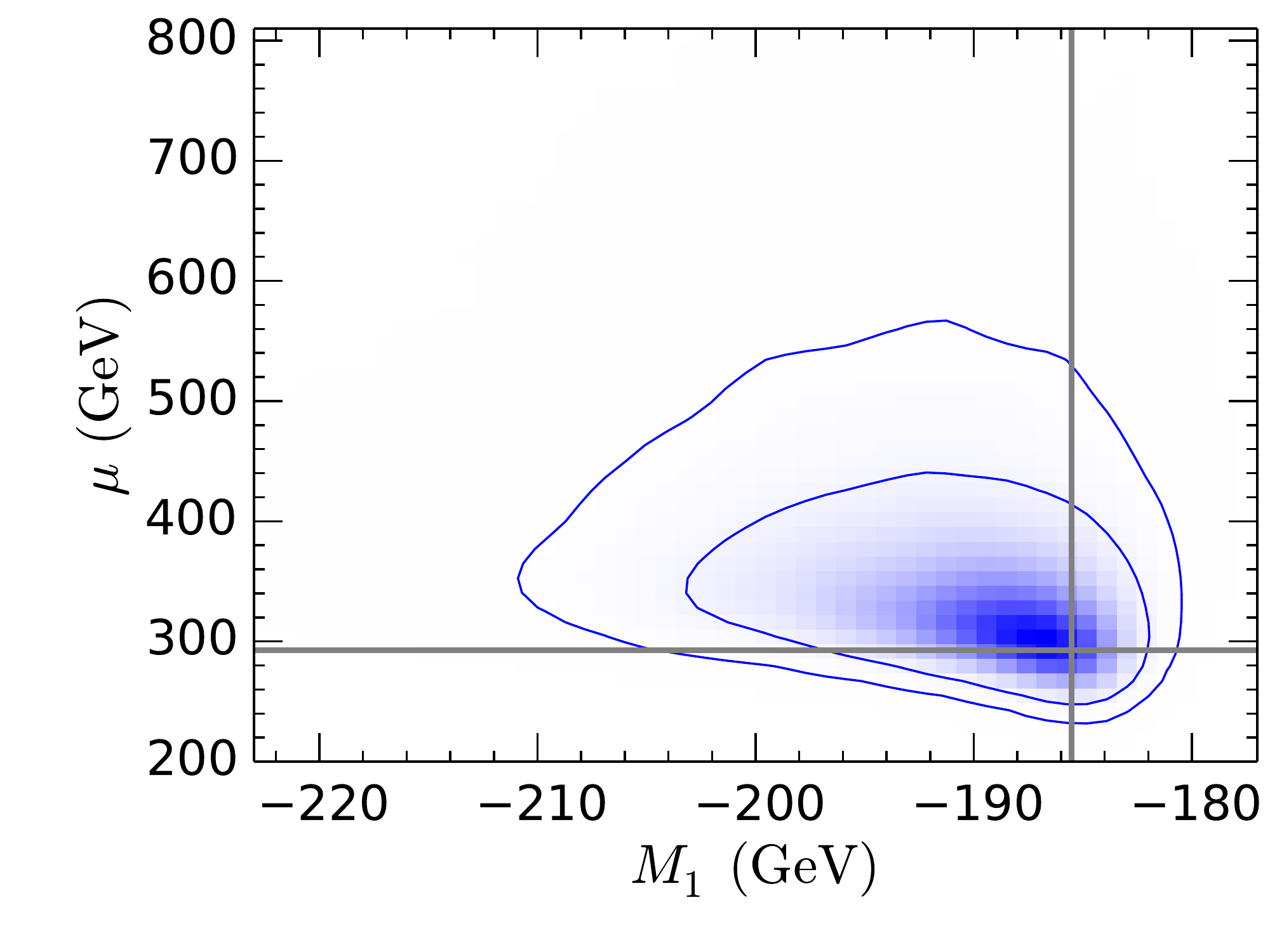}
  \includegraphics[height=0.37\textwidth, angle=0]{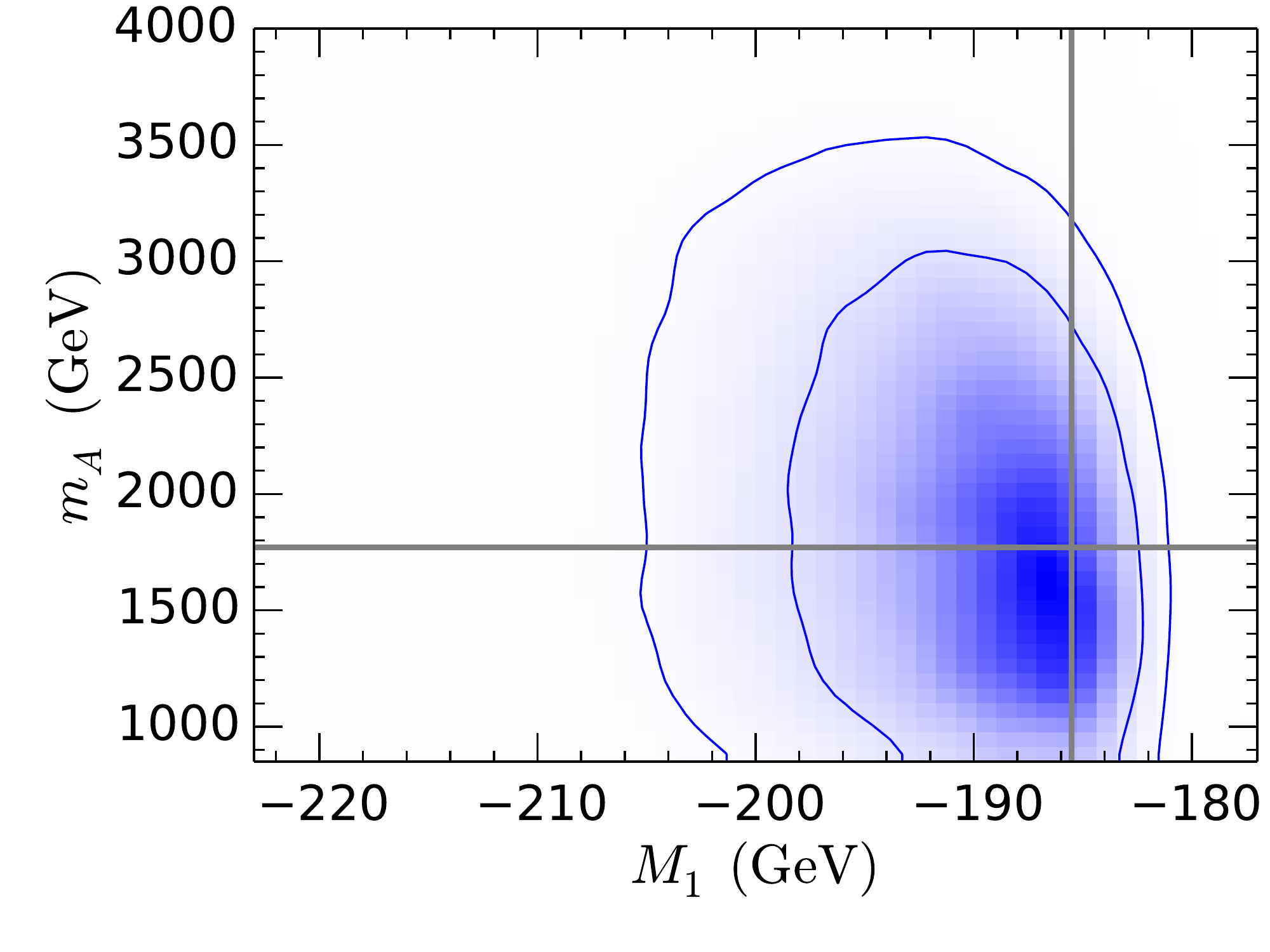}
  \includegraphics[height=0.37\textwidth, angle=0]{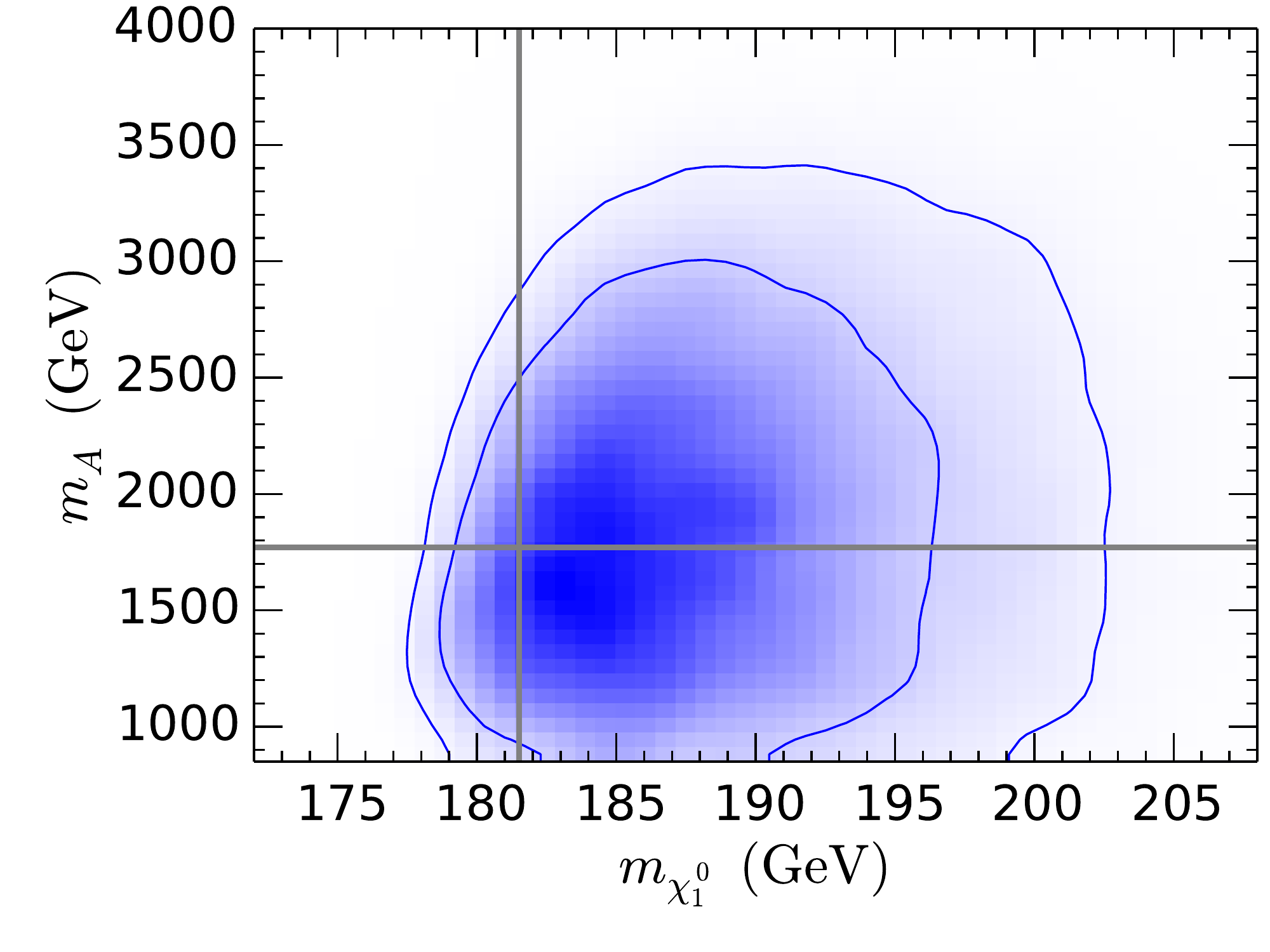}
  \includegraphics[height=0.37\textwidth, angle=0]{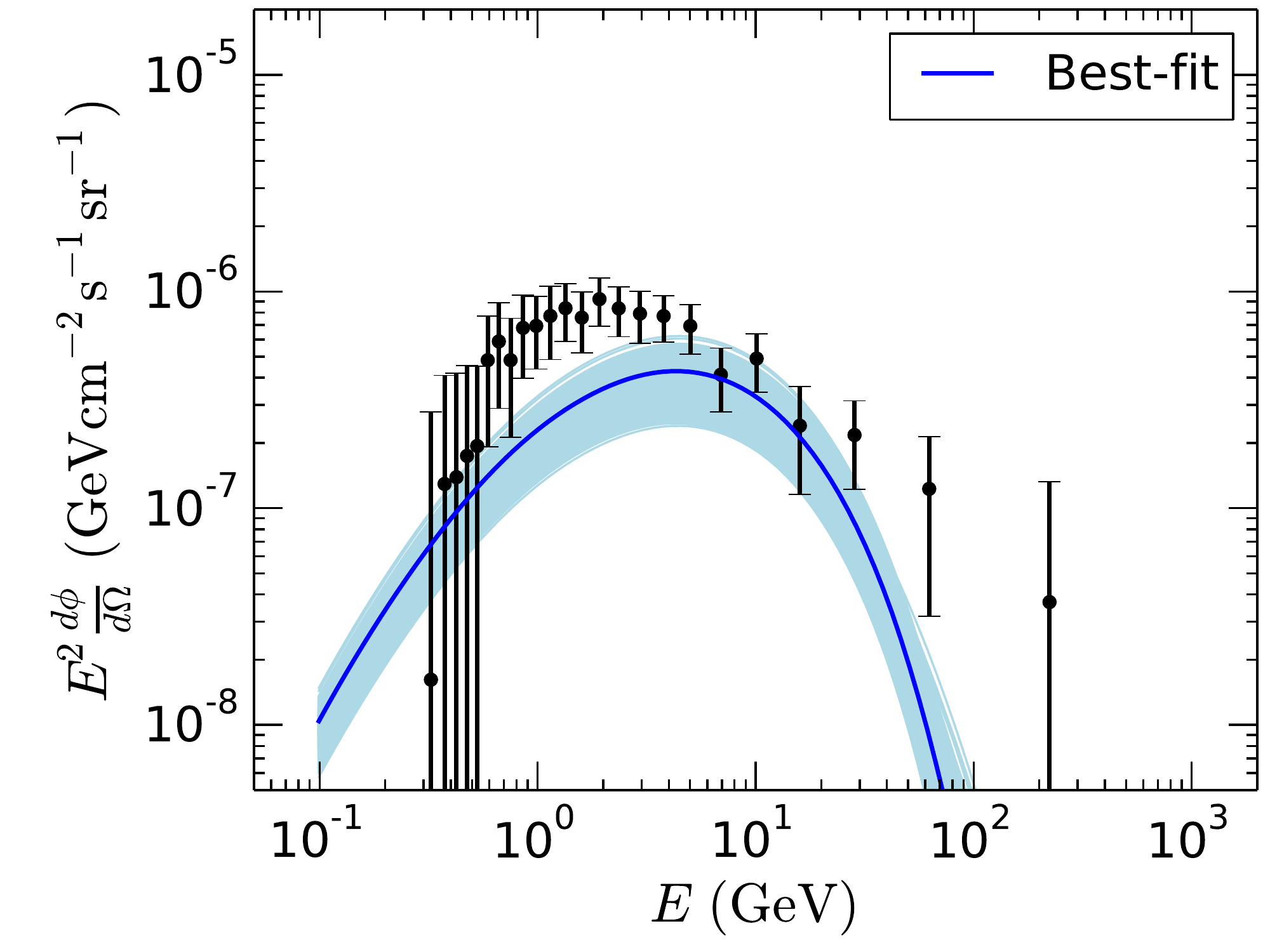}
  \includegraphics[height=0.37\textwidth, angle=0]{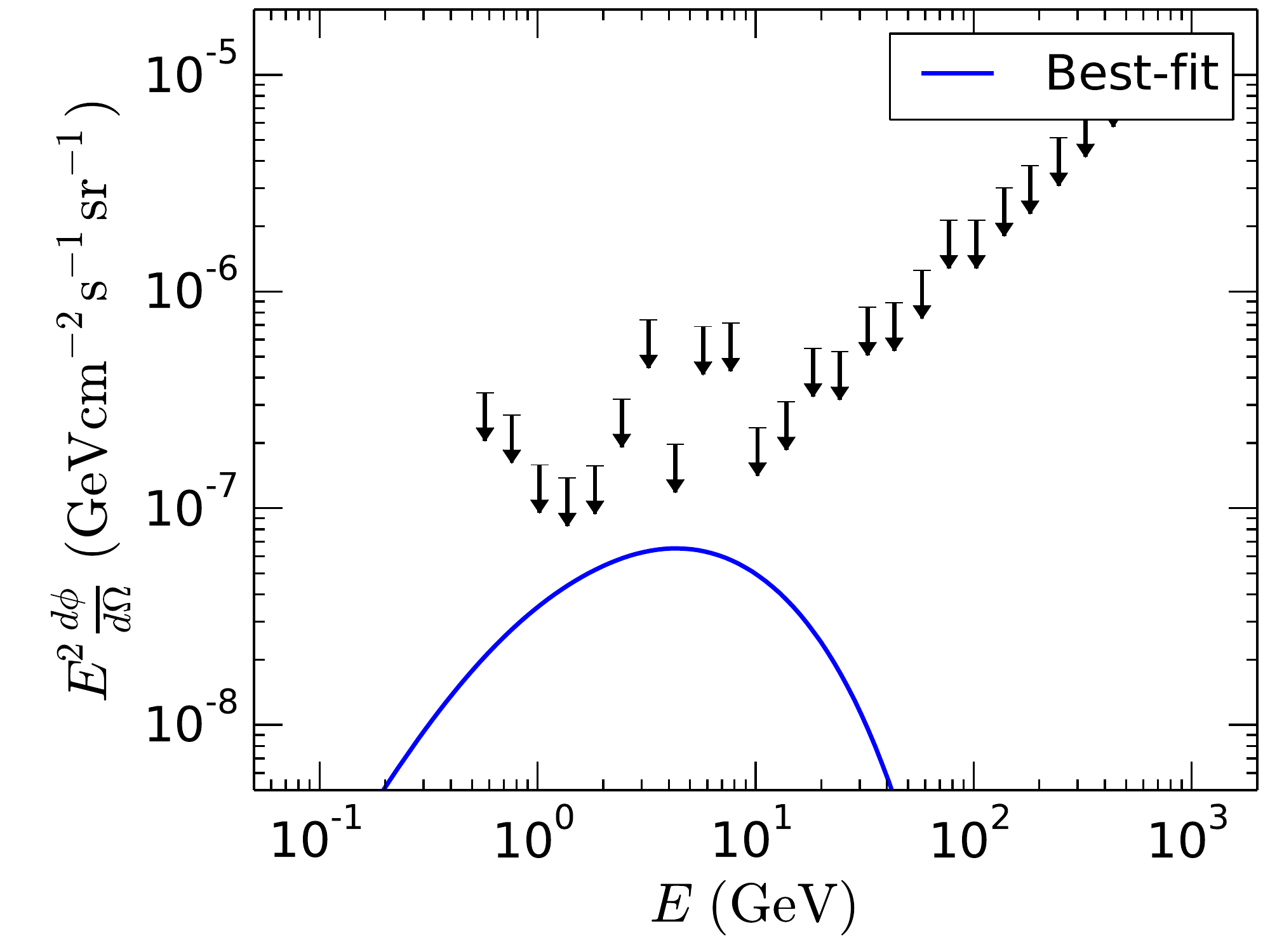}
  
  \caption{1$\sigma$ and 2$\sigma$ contours plots
    in the plane of $M_1 - M_2$ (top left), $M_1 - \mu$ (top right), $M_1 - m_A$ (middle left), $m_{\chi^0_1} - m_A$ (middle right) for Case 2b. Solid gray lines indicate the best fit values. Bottom left:
    2$\sigma$ bands of GC excess spectrum (light blue region)
    correspond to this case along with Fermi-LAT GC excess data and
    error bars (diagonal part of the covariance matrix). Deep blue
    line is the spectrum for best-fit points. Bottom right:
    Reticulum II $\gamma$-ray spectrum for the best-fit point (blue
    curve) along with the upper-limit on flux from Pass 8 analysis.}
\label{2b}
\end{figure*}

\begin{figure*}[h]
\centering
  \includegraphics[height=0.37\textwidth, angle=0]{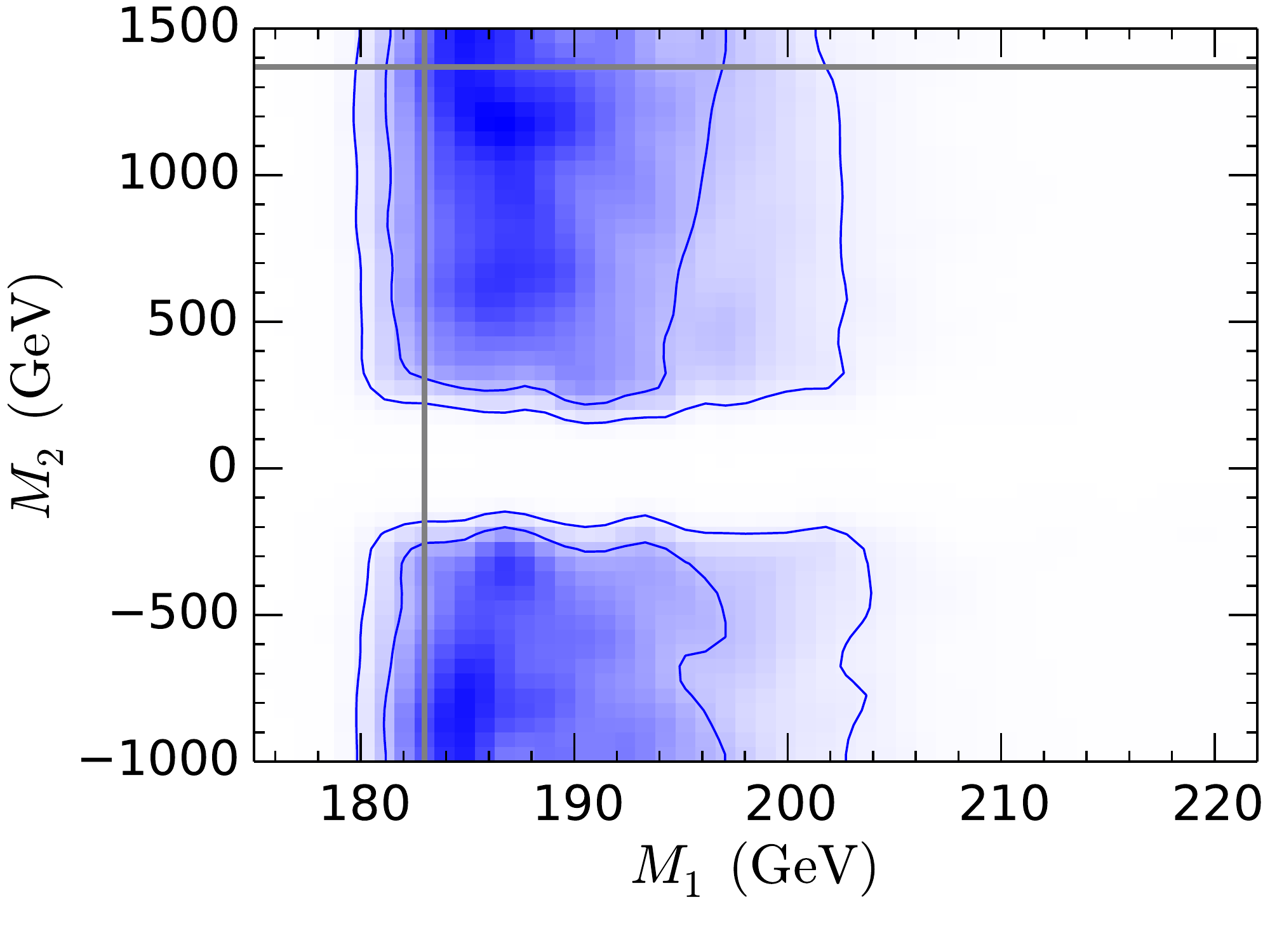}
  \includegraphics[height=0.37\textwidth, angle=0]{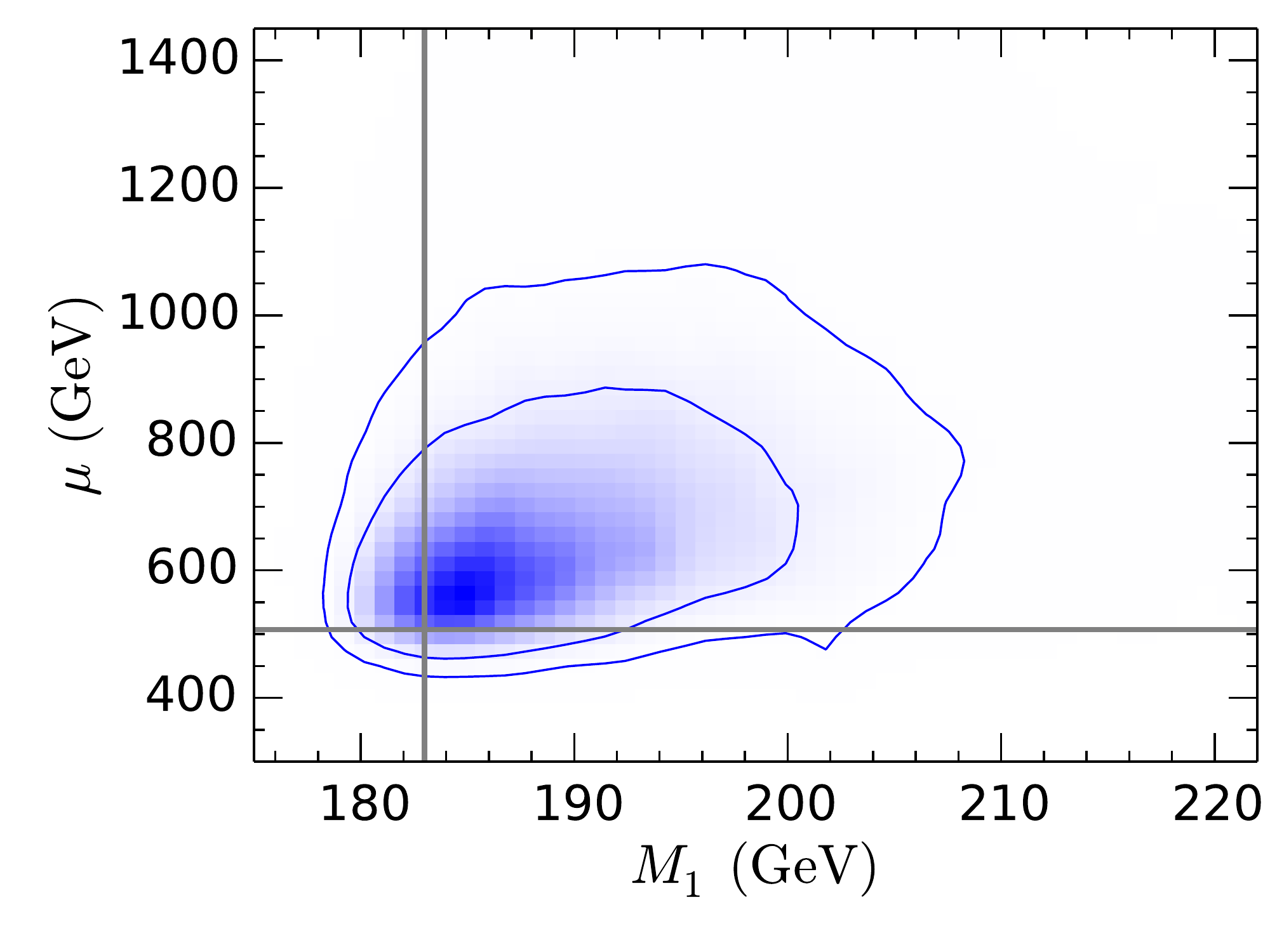}
  \includegraphics[height=0.37\textwidth, angle=0]{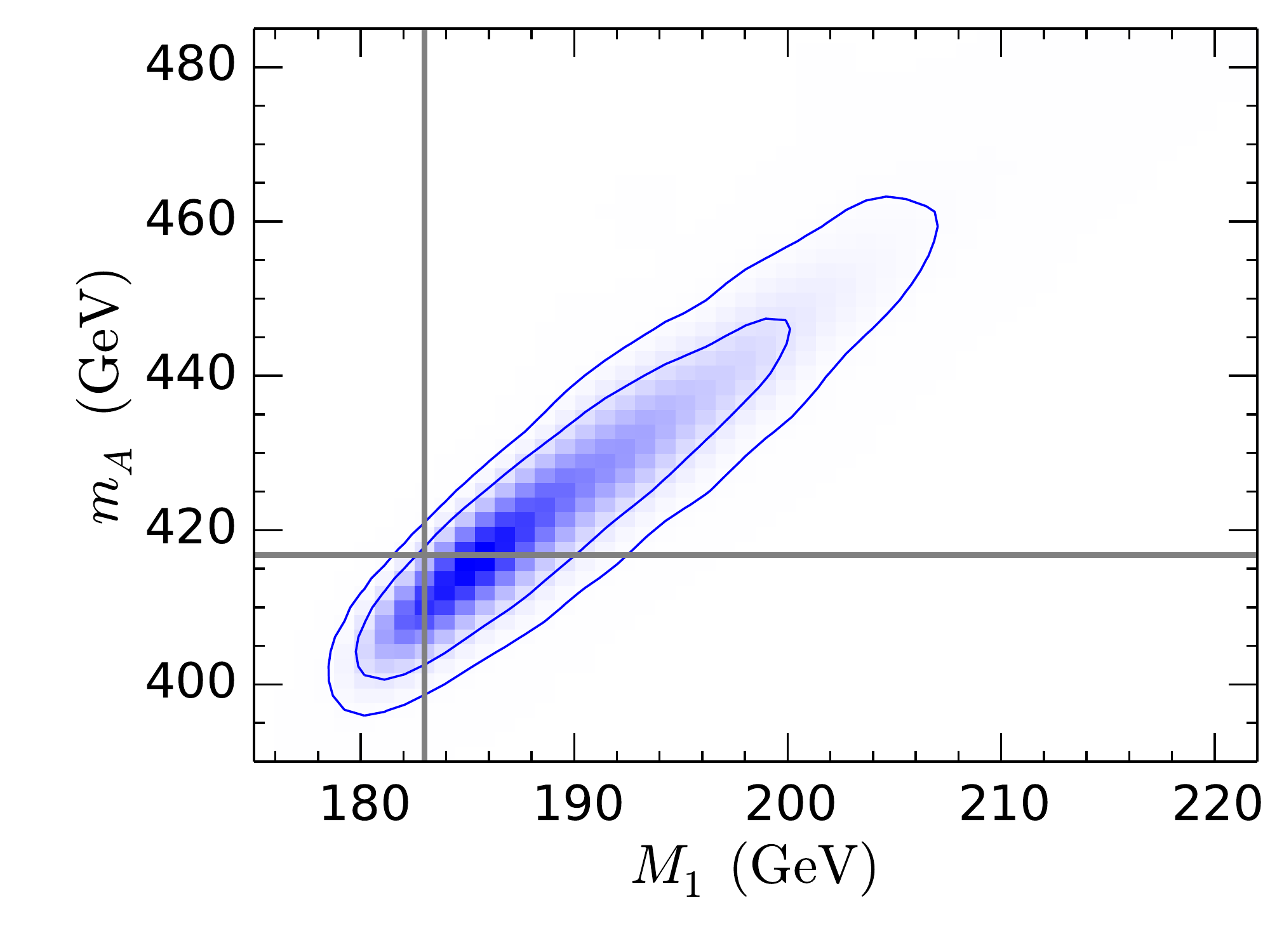}
  \includegraphics[height=0.37\textwidth, angle=0]{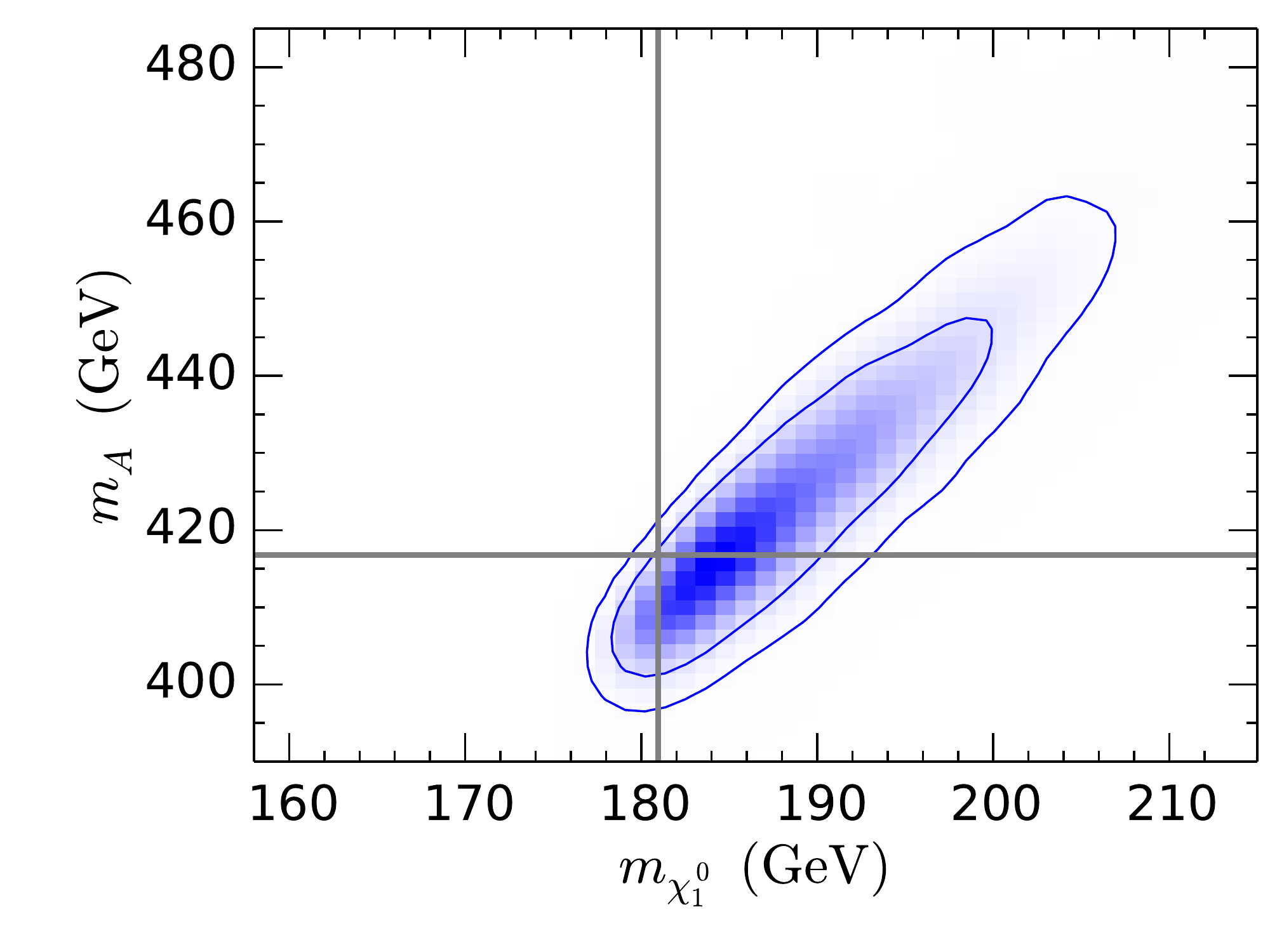}
  \includegraphics[height=0.37\textwidth, angle=0]{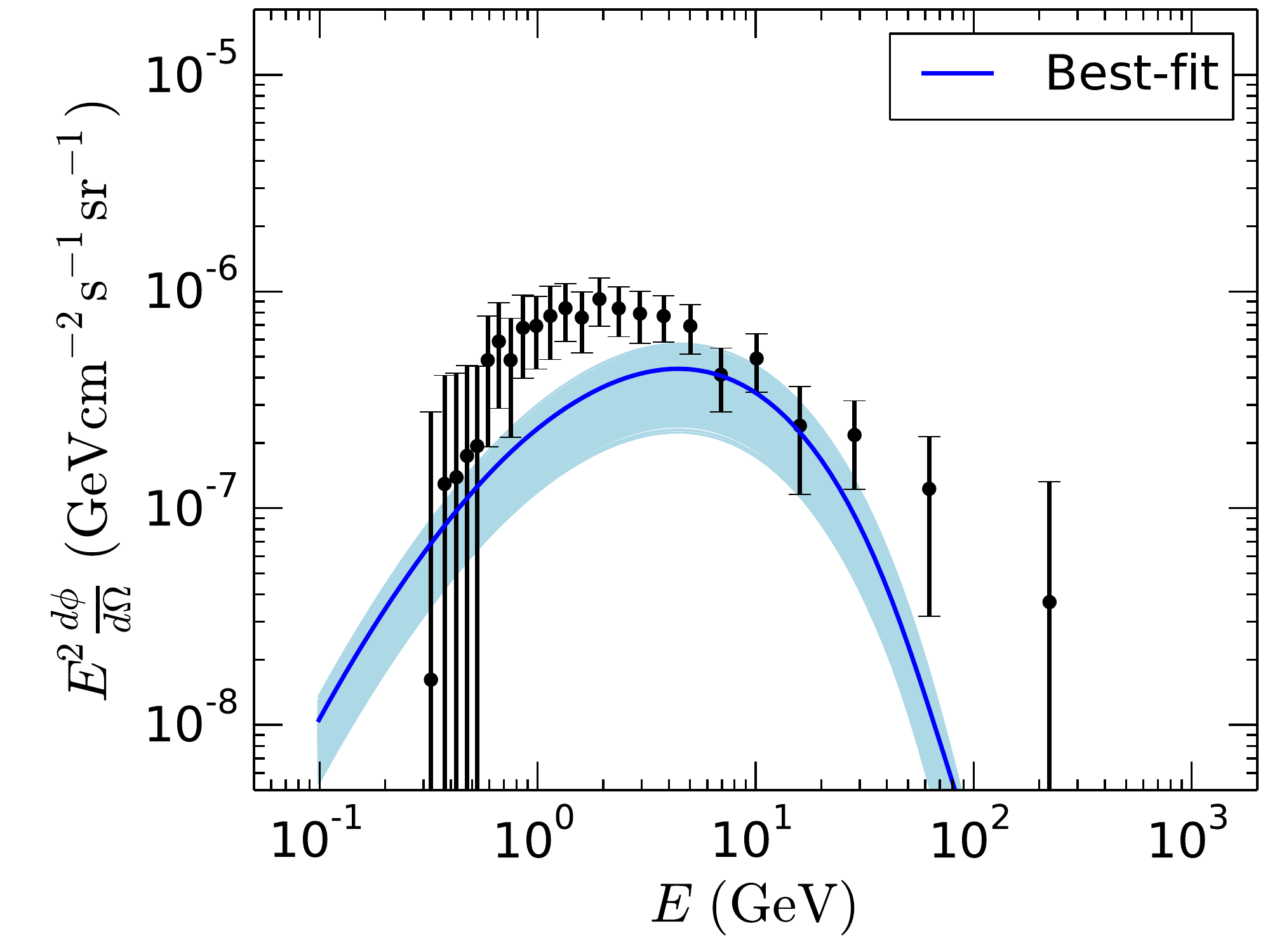}
  \includegraphics[height=0.37\textwidth, angle=0]{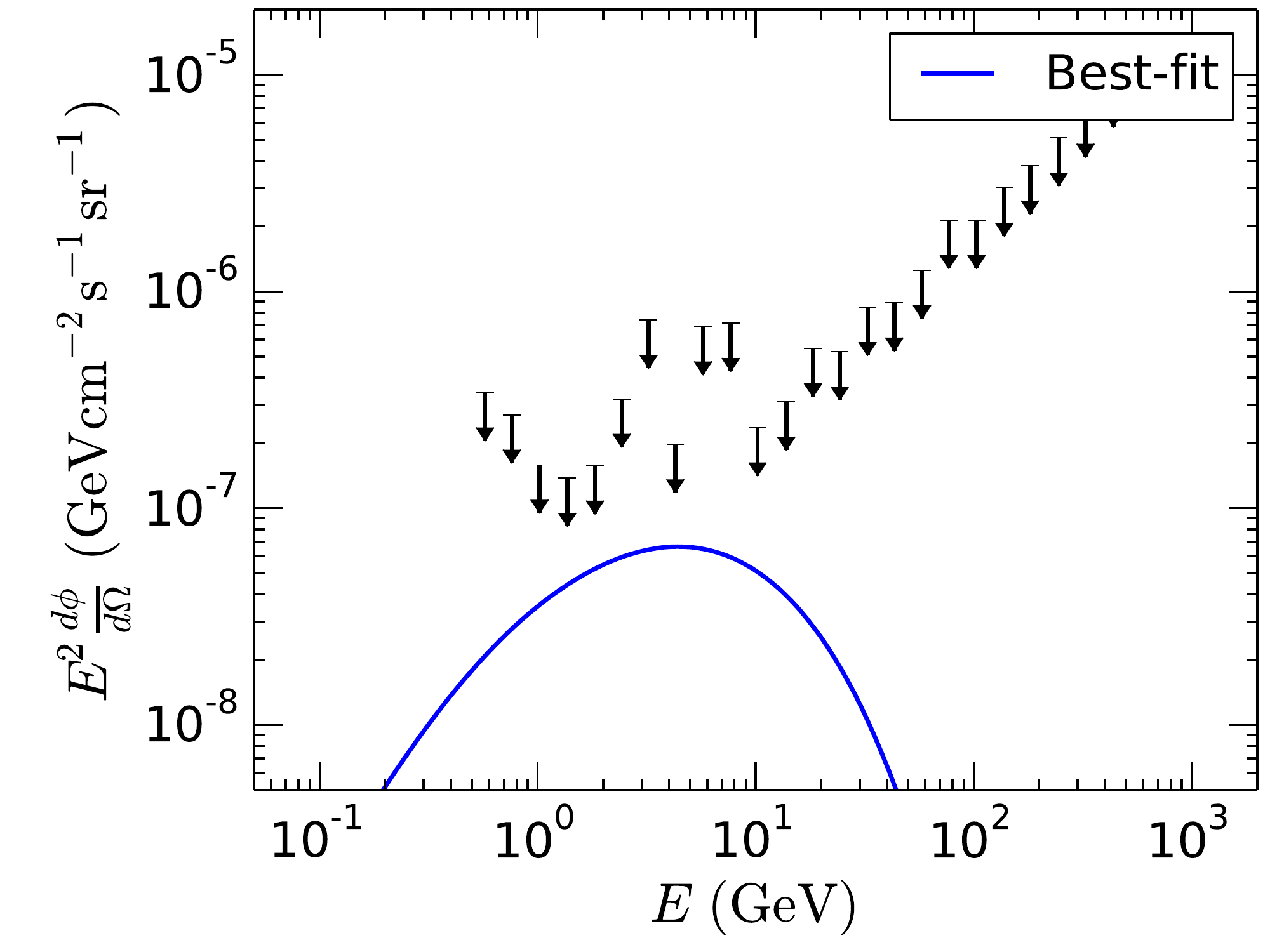}
  
  \caption{1$\sigma$ and 2$\sigma$ contours plots
    in the plane of $M_1 - M_2$ (top left), $M_1 - \mu$ (top right), $M_1 - m_A$ (middle left), $m_{\chi^0_1} - m_A$ (middle right) for Case 2c. Solid gray lines indicate the best fit values. Bottom left:
    2$\sigma$ bands of GC excess spectrum (light blue region)
    correspond to this case along with Fermi-LAT GC excess data and
    error bars (diagonal part of the covariance matrix). Deep blue
    line is the spectrum for best-fit points. Bottom right:
    Reticulum II $\gamma$-ray spectrum for the best-fit point (blue
    curve) along with the upper-limit on flux from Pass 8 analysis.}
\label{2c}
\end{figure*}

\underline{Case 2a}: For this case, the best fit point corresponds 
to  $M_{1}= 195.32, M_{2}= -205.55, \mu = 464.85, m_A = 617.18$, 
with $\chi^{2}_{min}$ = 65.7 for DOF = 24, and  the DM mass
in the range $\simeq$ 180 - 260 GeV at 95.6$\%$ C.L.  
The aforementioned trend of upward movement of the GC $\gamma$-ray
peak in the LST scenario causes more mismatch with data
if  $\left\langle \sigma v \right\rangle$ goes to the
higher side. This favours values of $m_A$ which are 
always above 600 GeV.  

As before, Fig. \ref{2a}  
demonstrates the viability of the preferred regions with respect 
to the GC excess spectrum and the observations on Reticulum II.

\underline{Case 2b}: This is the second best case
(after Case 1a) for fitting the GC excess, with $\chi^{2}_{min}$ = 61.2. 
The best fit point corresponds to $M_{1}= -185.52, M_{2}= 213.8, \mu = 292.79, 
m_A = 1769.64$, the lightest neutralino mass being in the range $\simeq$
178 - 205 GeV at 95.6$\%$ C.L. Fig. \ref{2b} includes the 2$\sigma$ contour
plots. This is a situation where the s-channel annihilation diagram
is least significant.  This is also because the favoured values
of $m_A$ are high for $\tan\beta$ = 50.

The contribution to annihilation comes from
stop mediated t-channel annihilation, yielding also a relic density
in the right range. Our earlier observations related to the
GC spectrum hold here as well, and the comparison with data
for GC as well as Reticulum II are found in Fig. \ref{2b}.


\underline{Case 2c}: With $\tan\beta$ = 5, this is the case where the
pseudoscalar-mediated s-channel annihilation dominates.  The best fit
point corresponds to the parameter values $M_{1}= 182.99, M_{2}=
1369.44, \mu = 507.05, m_A = 416.75$, with $\chi^{2}_{min}$ = 62.2.  The
lightest neutralino mass at 95.6$\%$ C.L. comes out to be in the range
$\sim$ 175 - 210 GeV. Fig. \ref{2c} represents the 2$\sigma$ contour
plots for this case. For this value of $\tan\beta$, $m_A$ can be as
low as 350 GeV. However, the fact that $m_{\chi_1^0}$ must be at least
equal to $m_t$ pushes the lowest value of $m_A$ so close to resonance
that the annihilation rate overshoots the upper limit imposed by the
requirement of a minimum relic density. In addition, this also leads
to the previously mentioned problem in fitting the GC $\gamma$-ray
spectrum.  Thus the 2$\sigma$ minimum in the marginalised plot
involving $m_A$ does not go below $m_A = 400$ GeV approximately.
Figs. \ref{2c}, drawn as before, are
self-explanatory.

\subsection{Case 3: STC}

We consider next the stau co-annihilation region, where the lighter
stau and lightest neutralino masses are within 3$\%$ of each
other. All other slepton and squark masses are set at values ($>$ 2
TeV).  The existing lower limits on the stau mass from the LEP and LHC
data \cite{Aad:2015baa} have been respected throughout our analysis.
Obviously, the $\chi_1^0$ mass, too, get constrained in a correlated
fashion. We remind the reader here that such co-annihilation regions
bring in additional angles when it comes to the neutralino freeze-out
process, and thus may cause modification in the relic density
constraints. However, the DM annihilation process in the GC, dwarf
spheroidal galaxies or even galactic clusters is solely of the form
$\chi_1^0 \chi_1^0 \rightarrow X \bar{X}$, as the co-annihilation
partner is not available there.

In principle, one can also think of MSSM spectra with the $\chi^{\pm}_1$ or the
lighter stop/sbottom as the co-annihilation partner of the  $\chi_1^0$.
Of these, the possibility of coannhilation with  $\chi^{\pm}_1$ is 
implicitly included in all of our scans. With stop co-annihilation,
only a limited region in the LST scenario, already included in our
analysis, can be effective. However, as we have already observed, 
these almost always require the neutralino to be of such mass where
the $\gamma$-ray distribution become peaked at somewhat high values
compared to what is indicated by the data. 
This problem is more pronounced
for sbottom co-annihilation, since the sbottom mass has still higher
bounds, as has been mentioned above.

A feature of the STC scenario is that the $\chi_1^0$  pair-annihilation
cross section can have a more substantial branching fraction in the
$\tau^+ \tau^-$ final state, driven by either the $A_0$ in s-channel or
the lighter stau in the t/u-channel . The $\gamma$-ray energy spectrum is 
affected accordingly. 

The fits obtained for the three regions corresponding to STC 
are in fact worse than those for both HSS and
LST. This is because of a two-fold constraint applicable
here. Firstly, co-annihilation tends to lower the relic density below
the stipulated lower limit.  Therefore, if the MSSM has to account for
all dark matter, it is imperative to keep the contribution to
$\left\langle \sigma v \right\rangle$ in the freeze-out process (as
opposed to annihilation within stellar objects) within limits. One is
thus forced to go to higher mass regions (as can be seen from
\cite{gondolo}) for the $\chi_1^0 - \tilde{\tau_1}$ duo, leading again
to the undesirable consequence of the $\gamma$-ray peak shifting
towards high frequencies.

\begin{figure*}[h]
\centering
  \includegraphics[height=0.37\textwidth, angle=0]{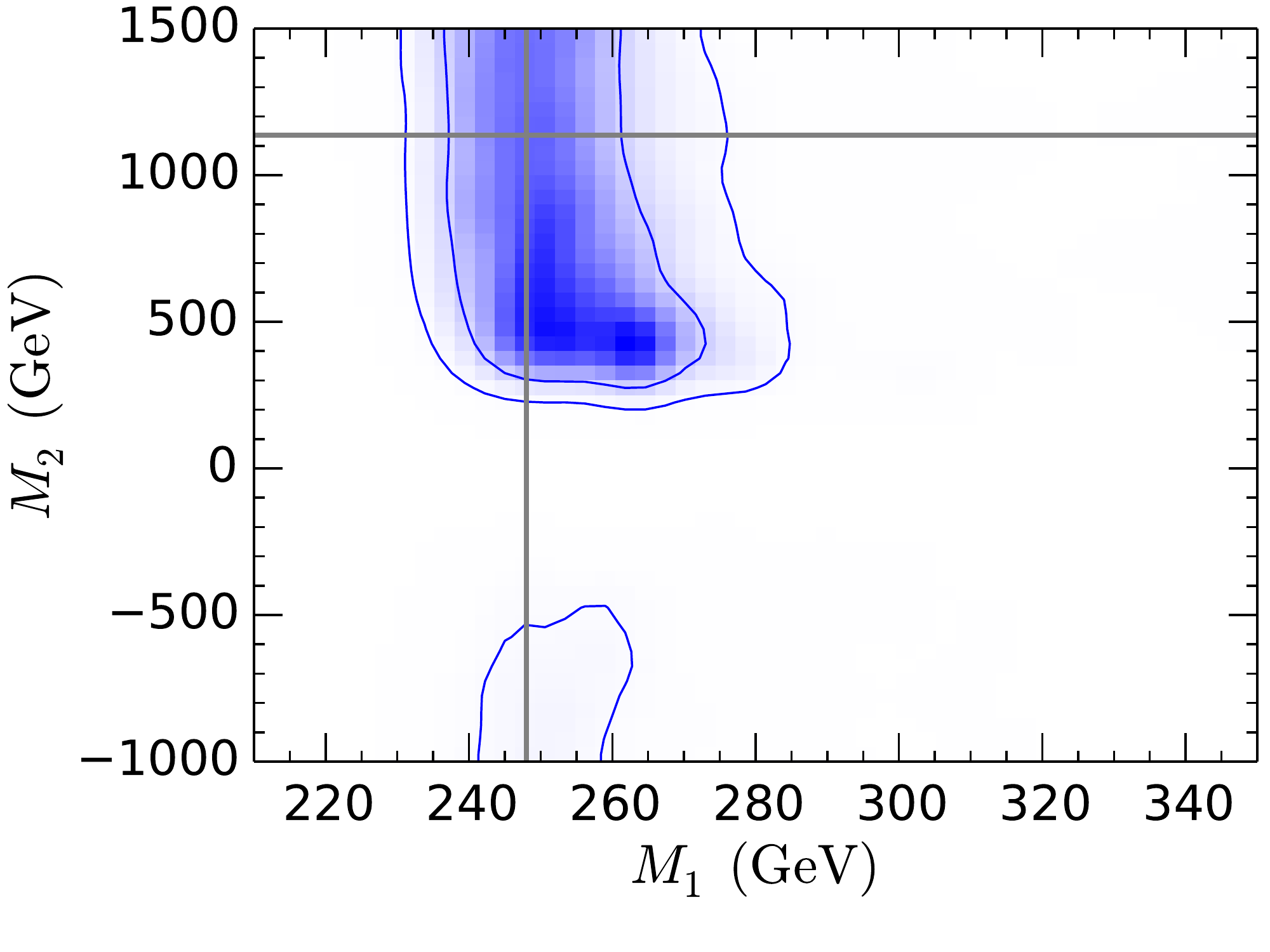}
  \includegraphics[height=0.37\textwidth, angle=0]{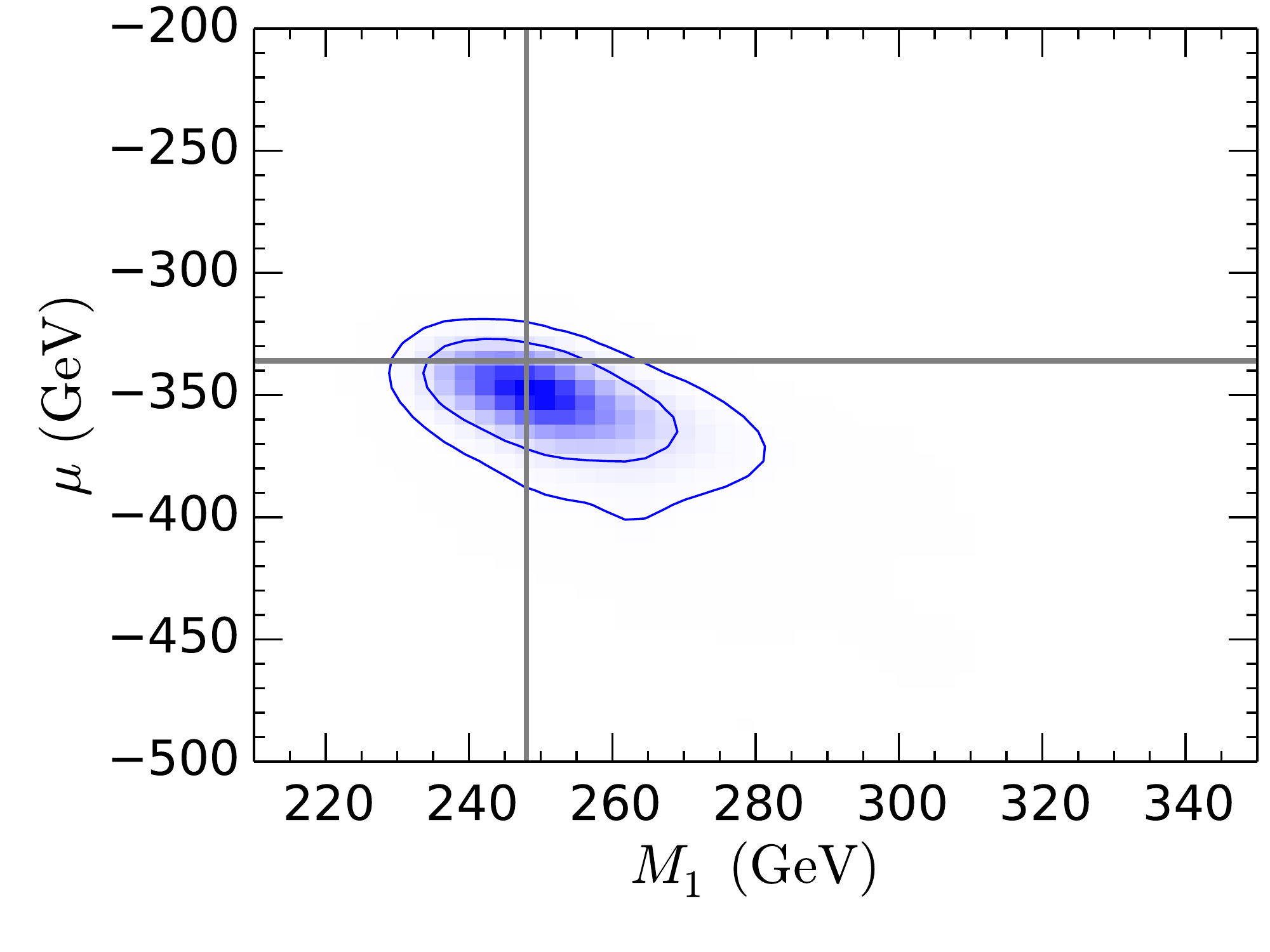}
  \includegraphics[height=0.37\textwidth, angle=0]{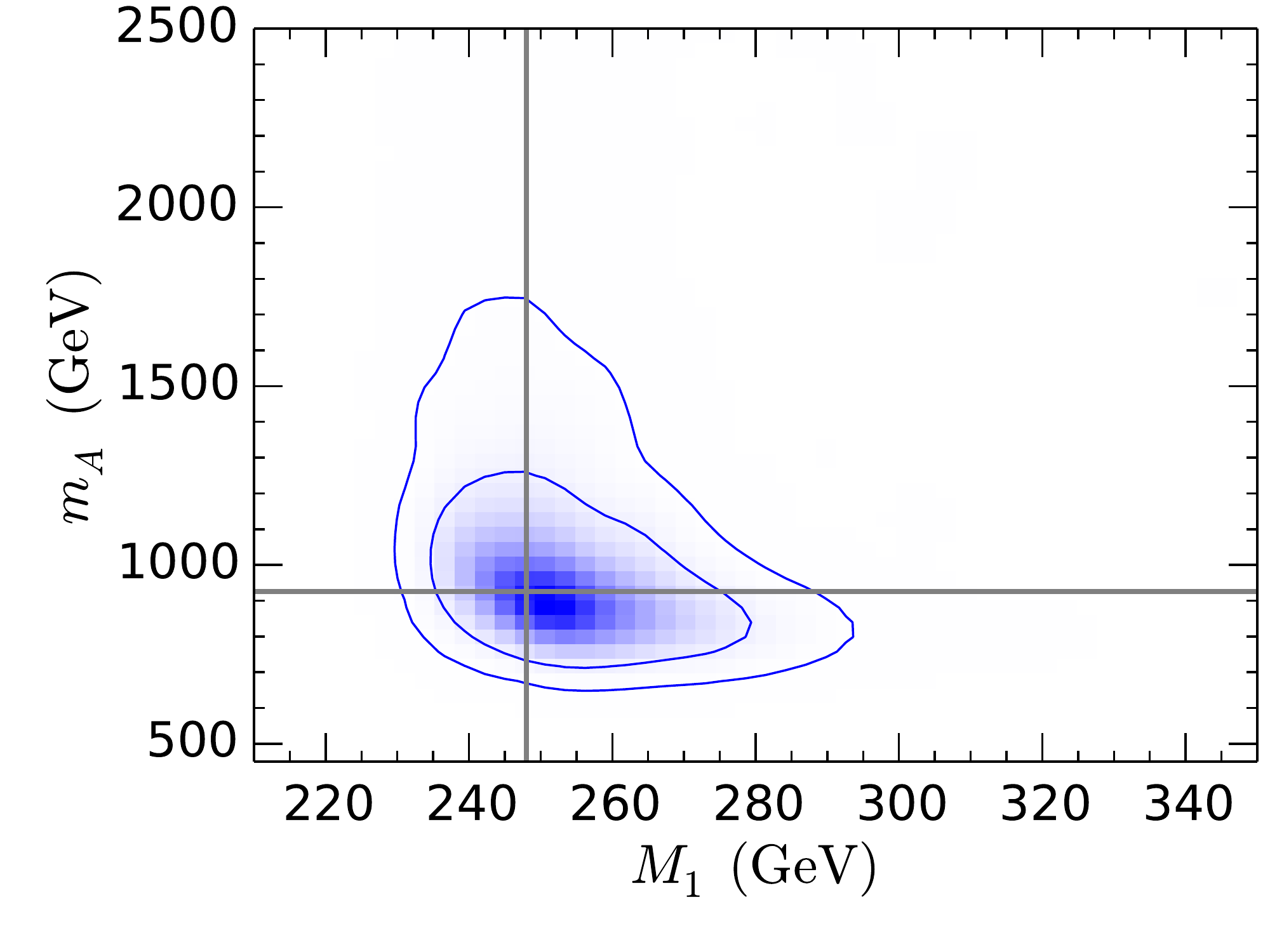}
  \includegraphics[height=0.37\textwidth, angle=0]{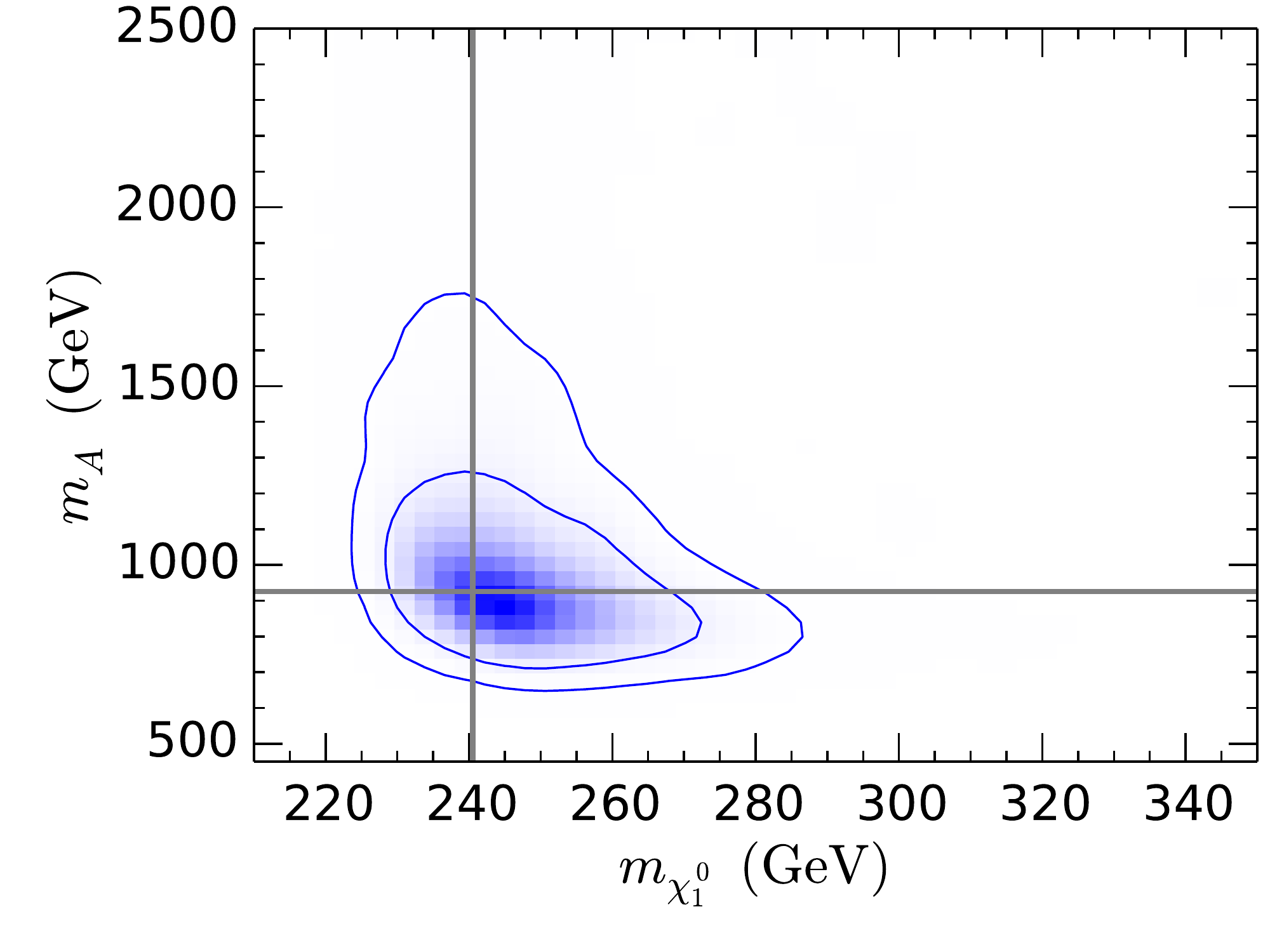}
  \includegraphics[height=0.37\textwidth, angle=0]{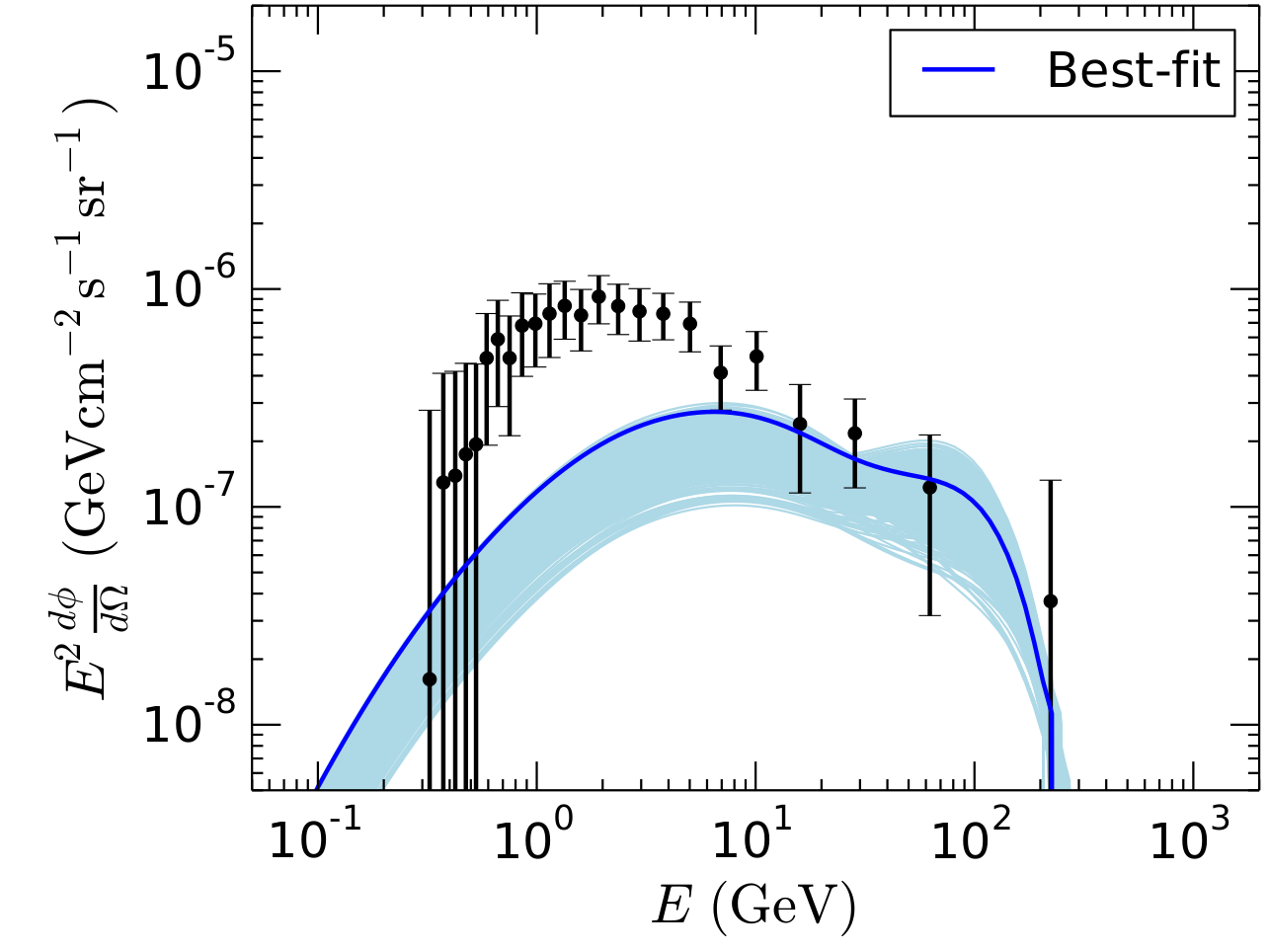}
  \includegraphics[height=0.37\textwidth, angle=0]{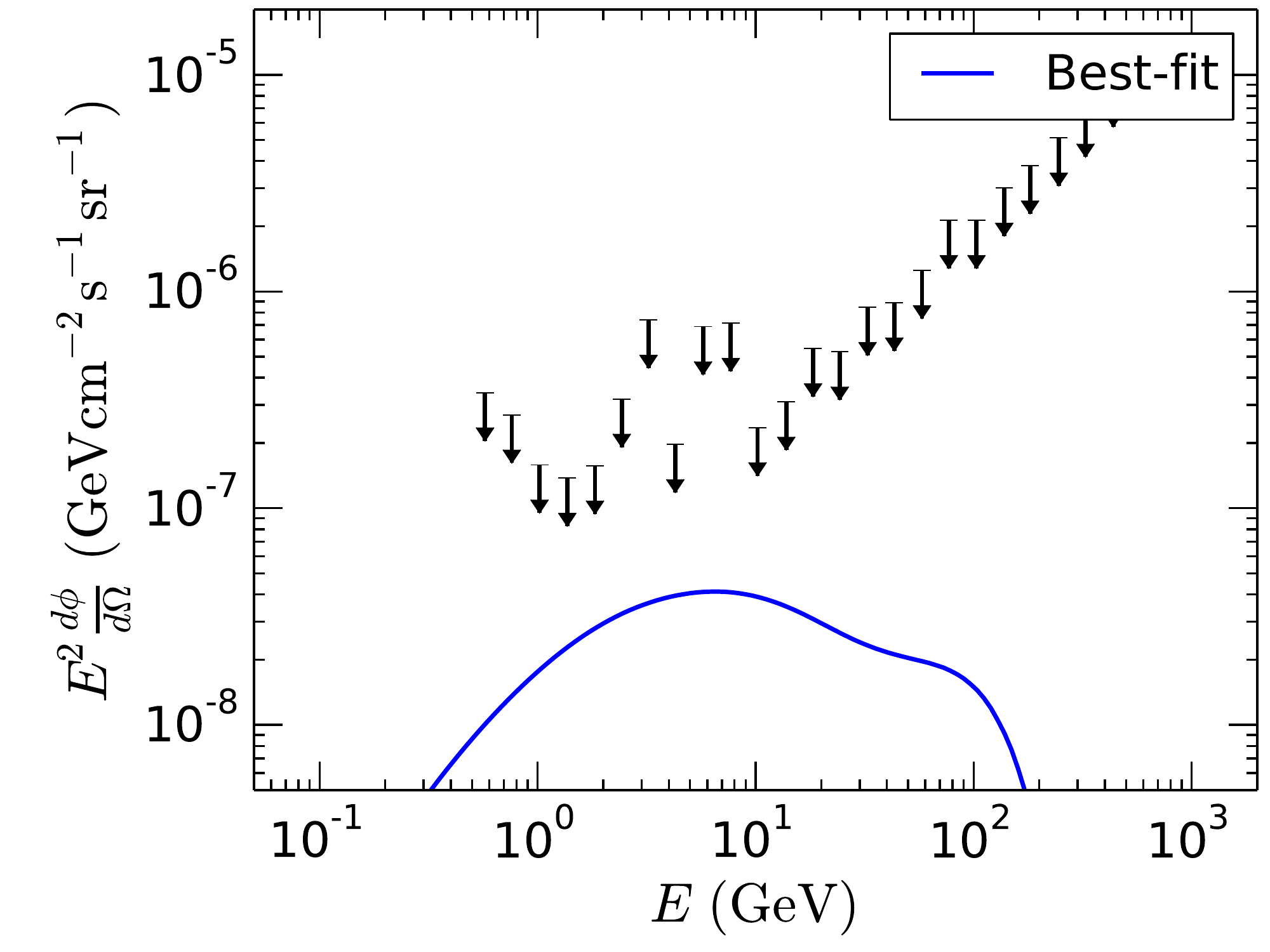}
  
  \caption{1$\sigma$ and 2$\sigma$ contours plots
    in the plane of $M_1 - M_2$ (top left), $M_1 - \mu$ (top right), $M_1 - m_A$ (middle left), $m_{\chi^0_1} - m_A$ (middle right) for Case 3a. Solid gray lines indicate the best fit values. Bottom left:
    2$\sigma$ bands of GC excess spectrum (light blue region)
    correspond to this case along with Fermi-LAT GC excess data and
    error bars (diagonal part of the covariance matrix). Deep blue
    line is the spectrum for best-fit points. Bottom right:
    Reticulum II $\gamma$-ray spectrum for the best-fit point (blue
    curve) along with the upper-limit on flux from Pass 8 analysis.}
\label{3a}
\end{figure*}

\begin{figure*}[h]
\centering
  \includegraphics[height=0.37\textwidth, angle=0]{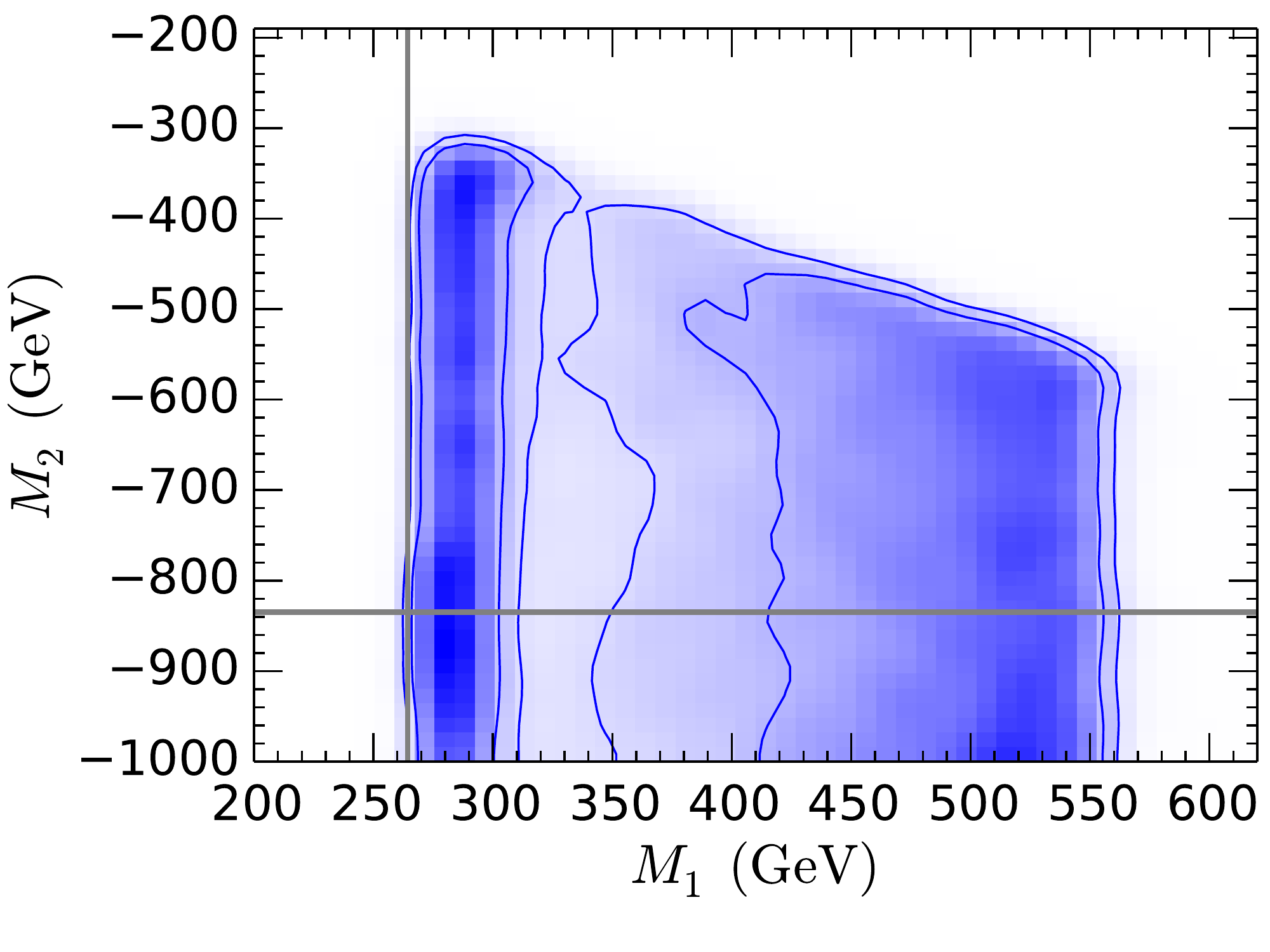}
  \includegraphics[height=0.37\textwidth, angle=0]{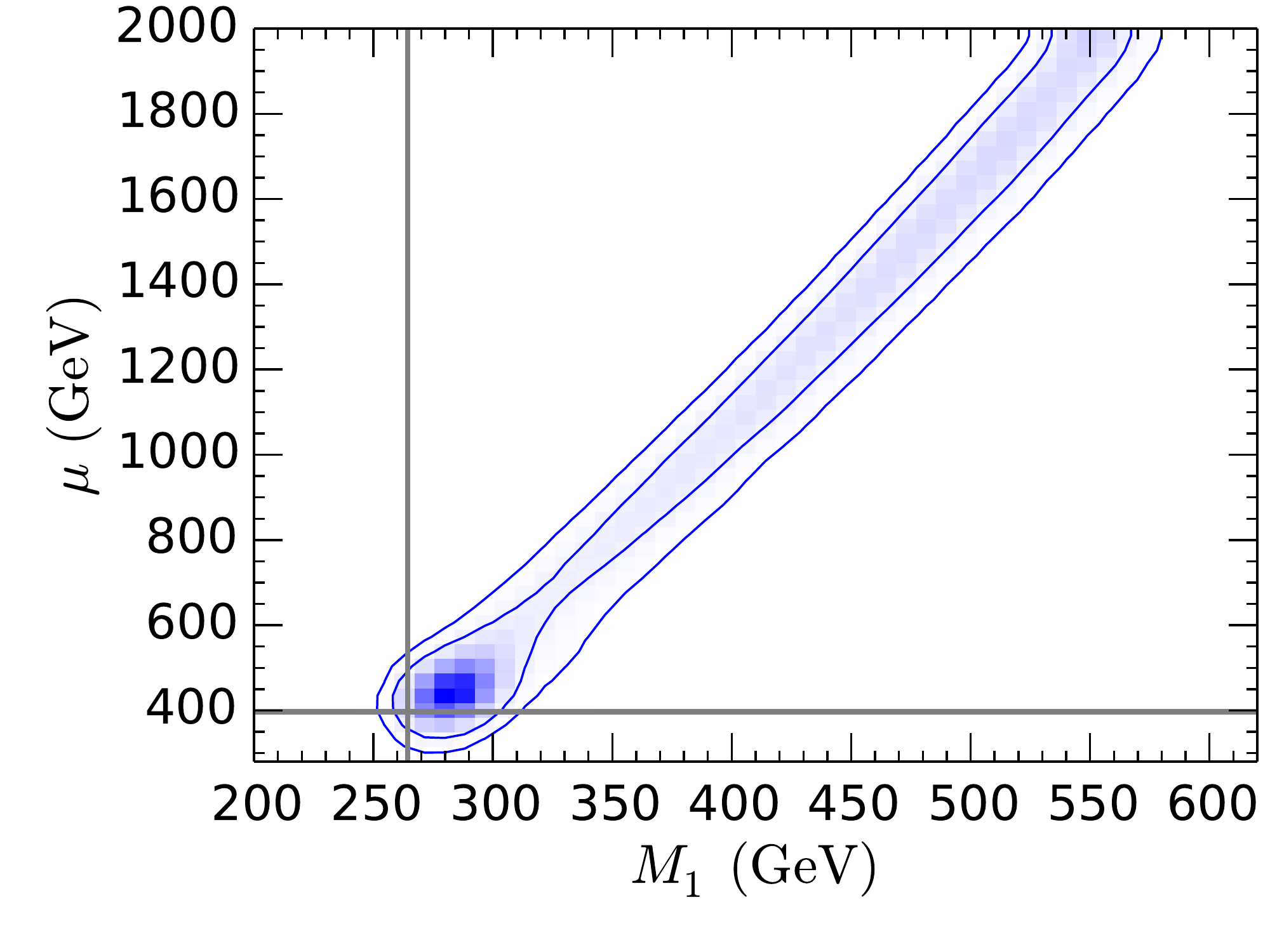}
  \includegraphics[height=0.37\textwidth, angle=0]{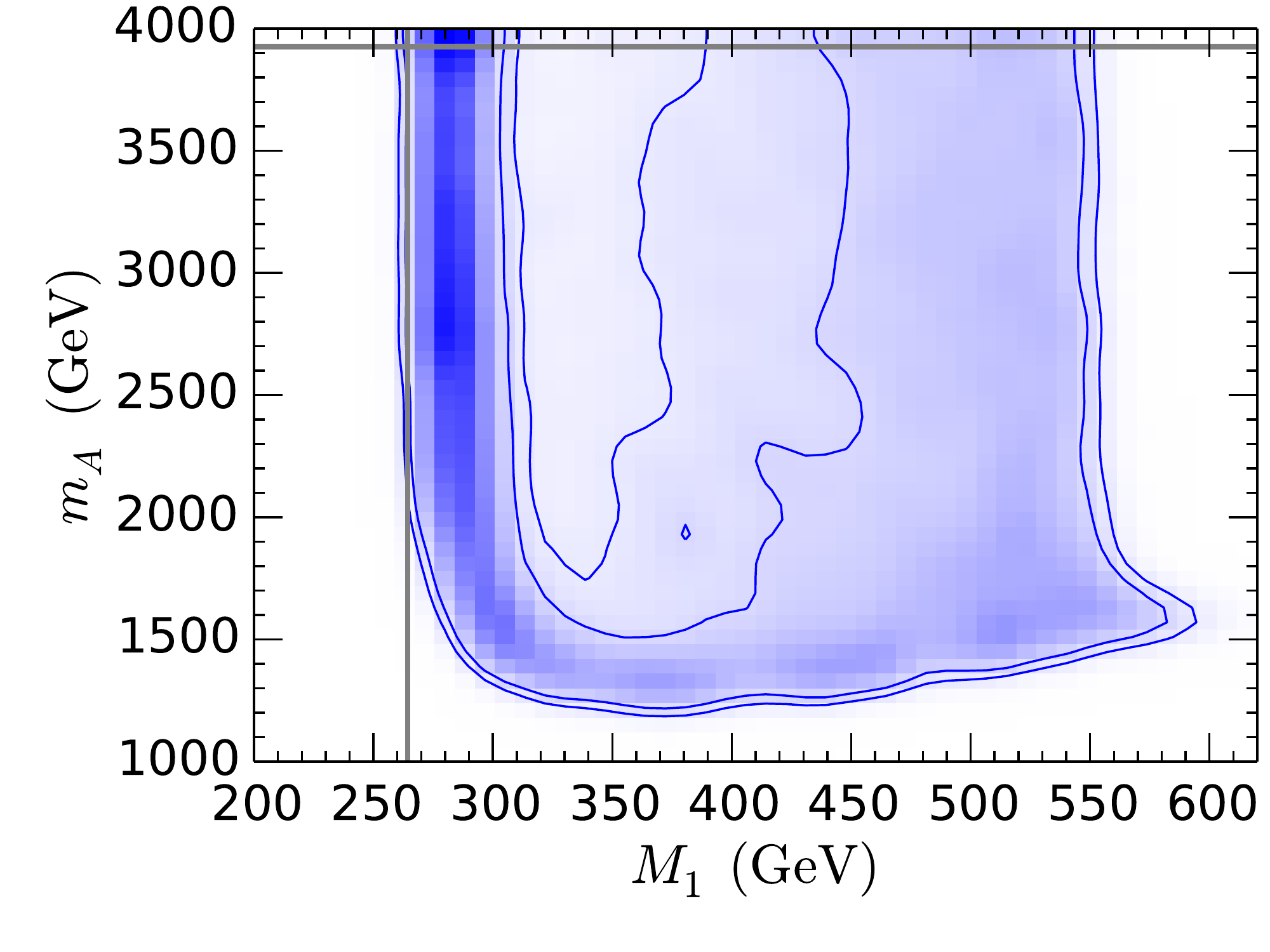}
  \includegraphics[height=0.37\textwidth, angle=0]{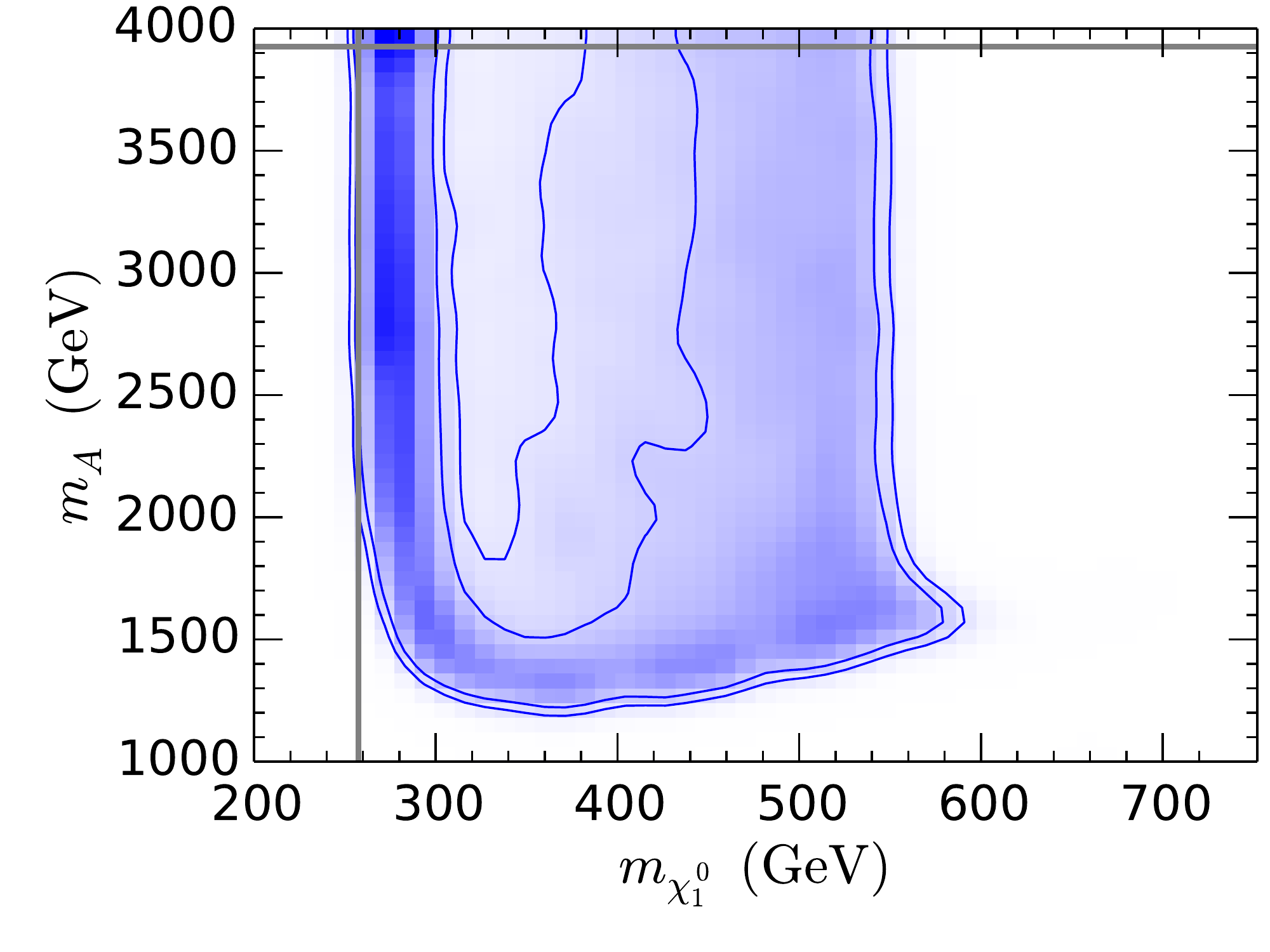}
  \includegraphics[height=0.37\textwidth, angle=0]{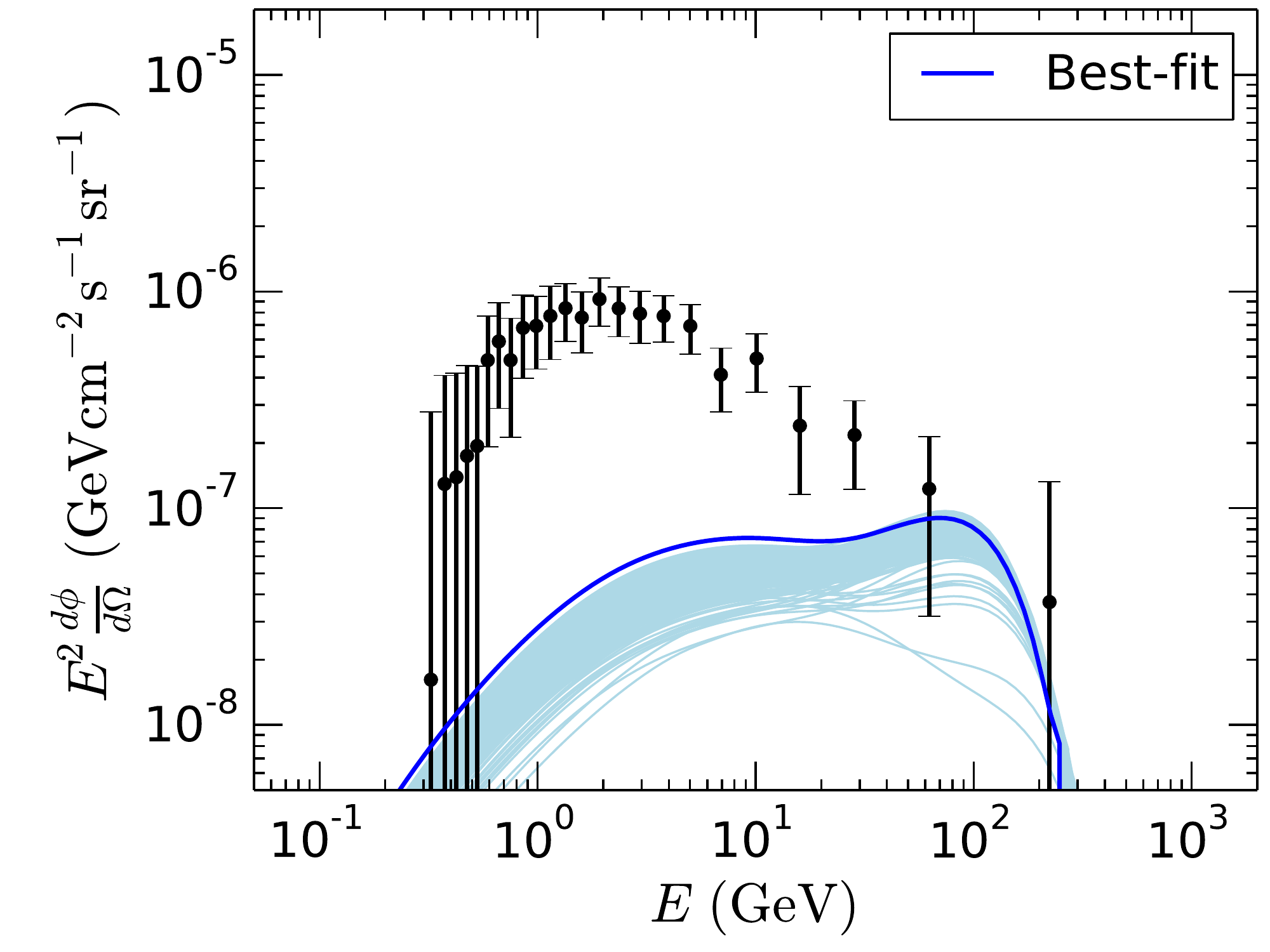}
  \includegraphics[height=0.37\textwidth, angle=0]{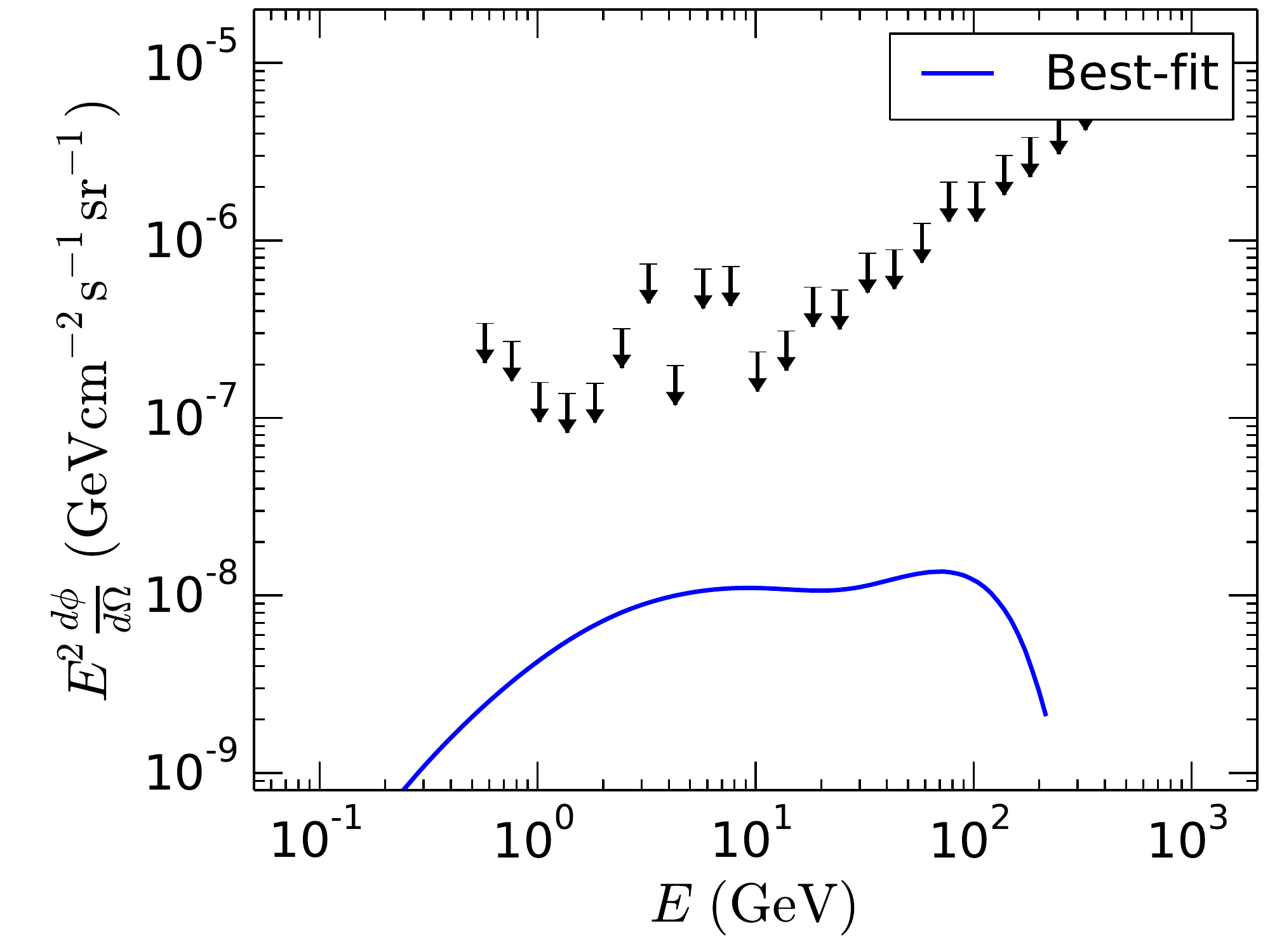}
  
  \caption{1$\sigma$ and 2$\sigma$ contours plots
    in the plane of $M_1 - M_2$ (top left), $M_1 - \mu$ (top right), $M_1 - m_A$ (middle left), $m_{\chi^0_1} - m_A$ (middle right) for Case 3b. Solid gray lines indicate the best fit values. Bottom left:
    2$\sigma$ bands of GC excess spectrum (light blue region)
    correspond to this case along with Fermi-LAT GC excess data and
    error bars (diagonal part of the covariance matrix). Deep blue
    line is the spectrum for best-fit points. Bottom right:
    Reticulum II $\gamma$-ray spectrum for the best-fit point (blue
    curve) along with the upper-limit on flux from Pass 8 analysis.}
\label{3b}
\end{figure*}

\begin{figure*}[h]
\centering
  \includegraphics[height=0.37\textwidth, angle=0]{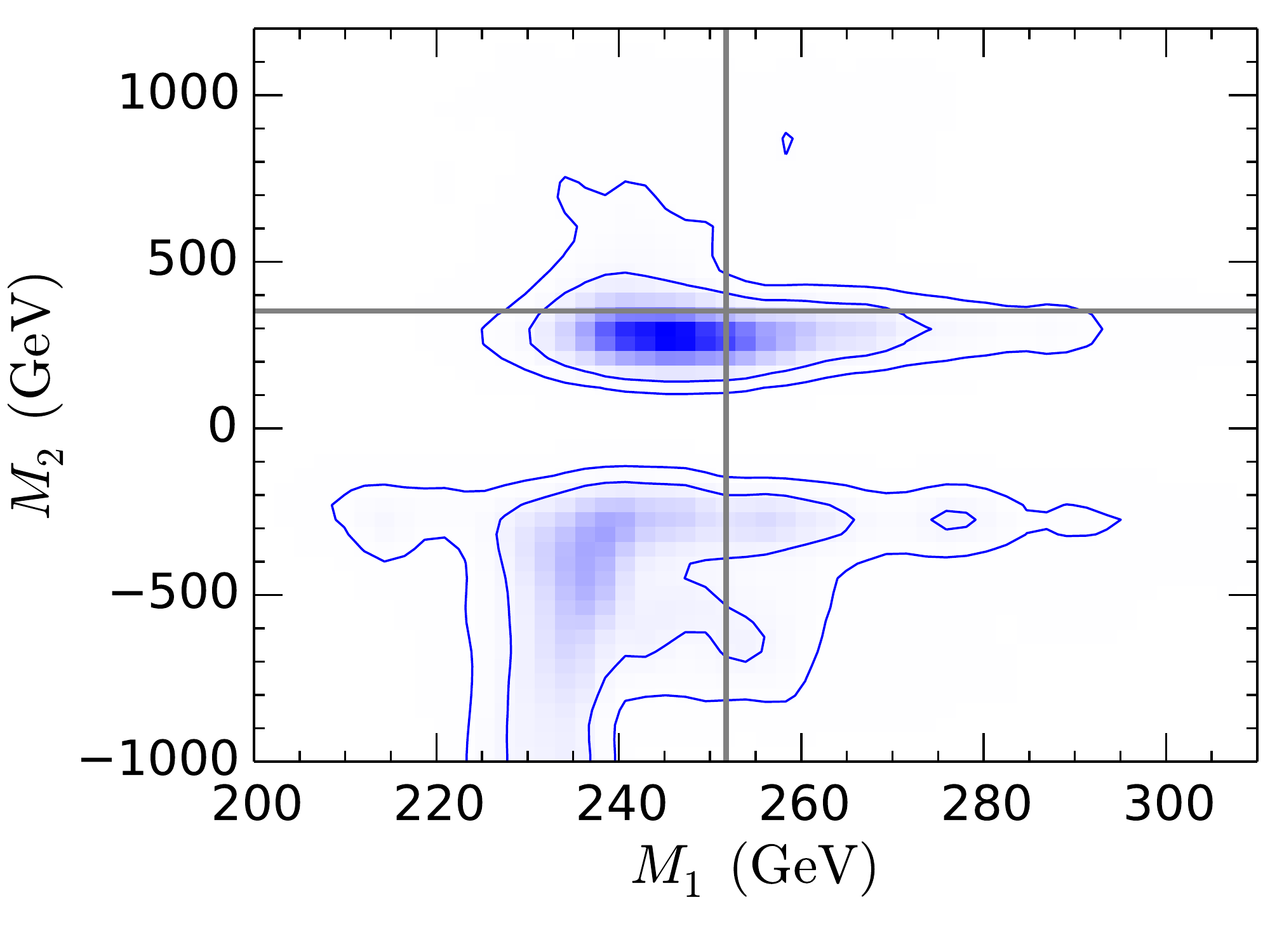}
  \includegraphics[height=0.37\textwidth, angle=0]{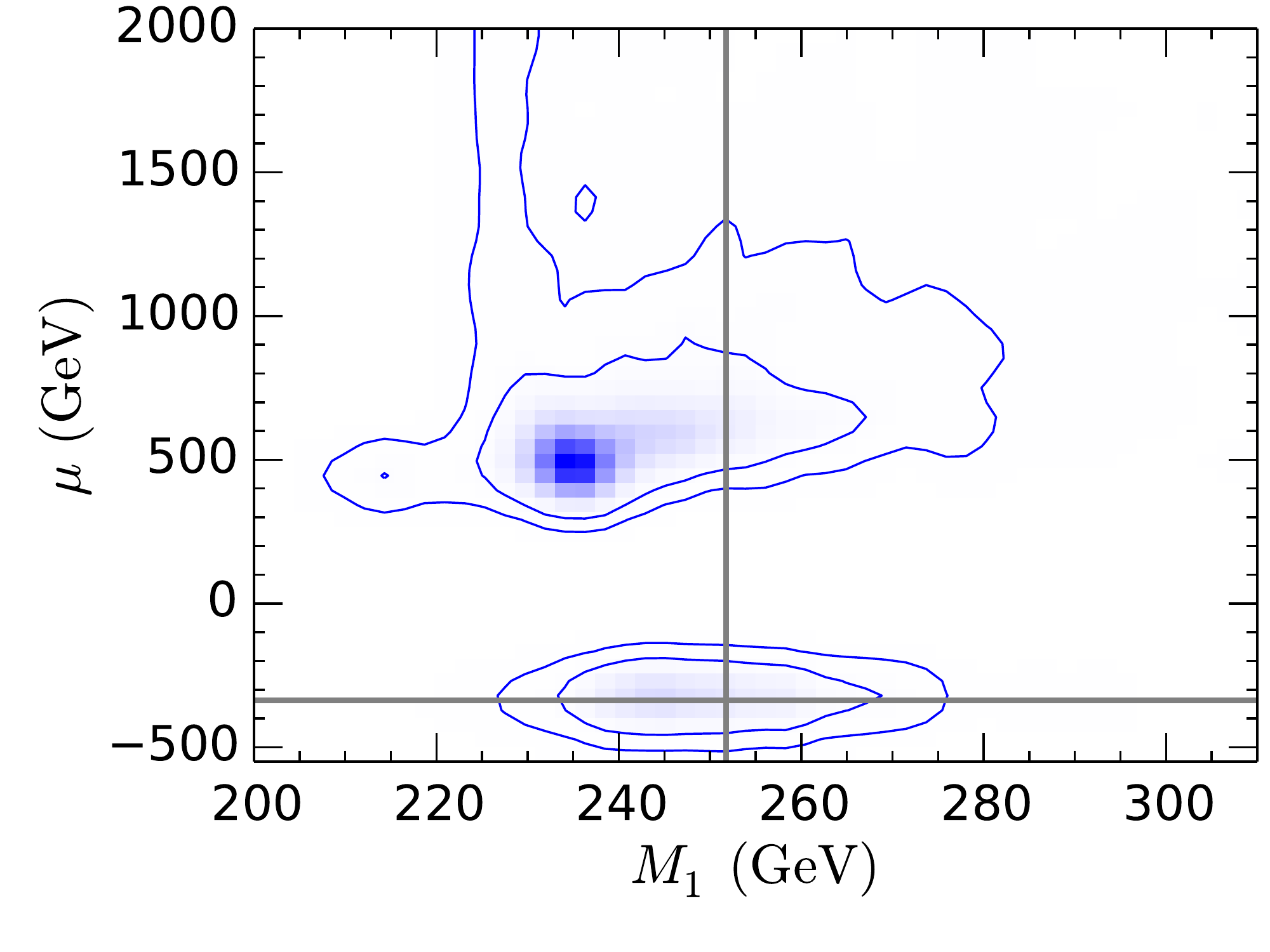}
  \includegraphics[height=0.37\textwidth, angle=0]{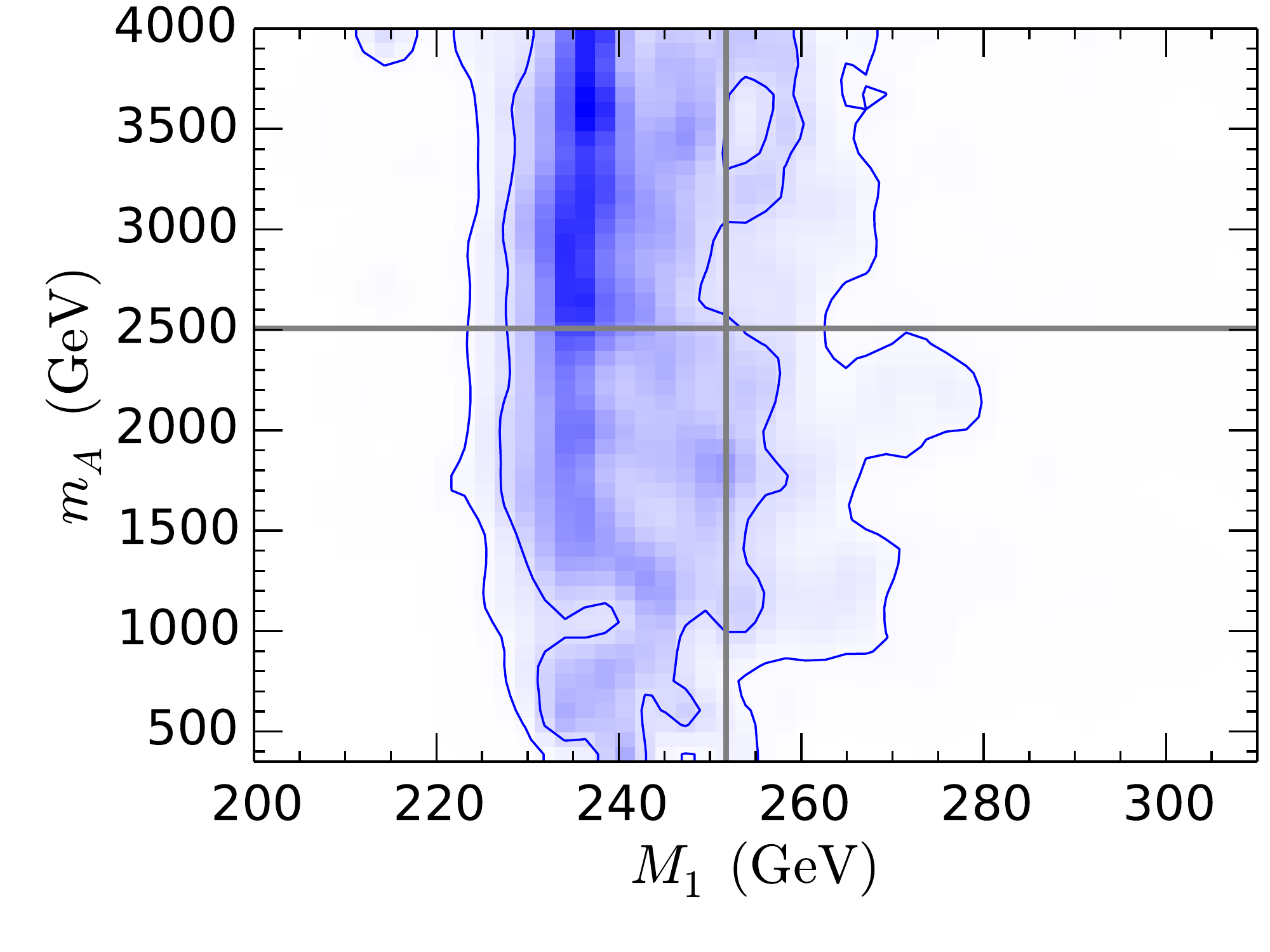}
  \includegraphics[height=0.37\textwidth, angle=0]{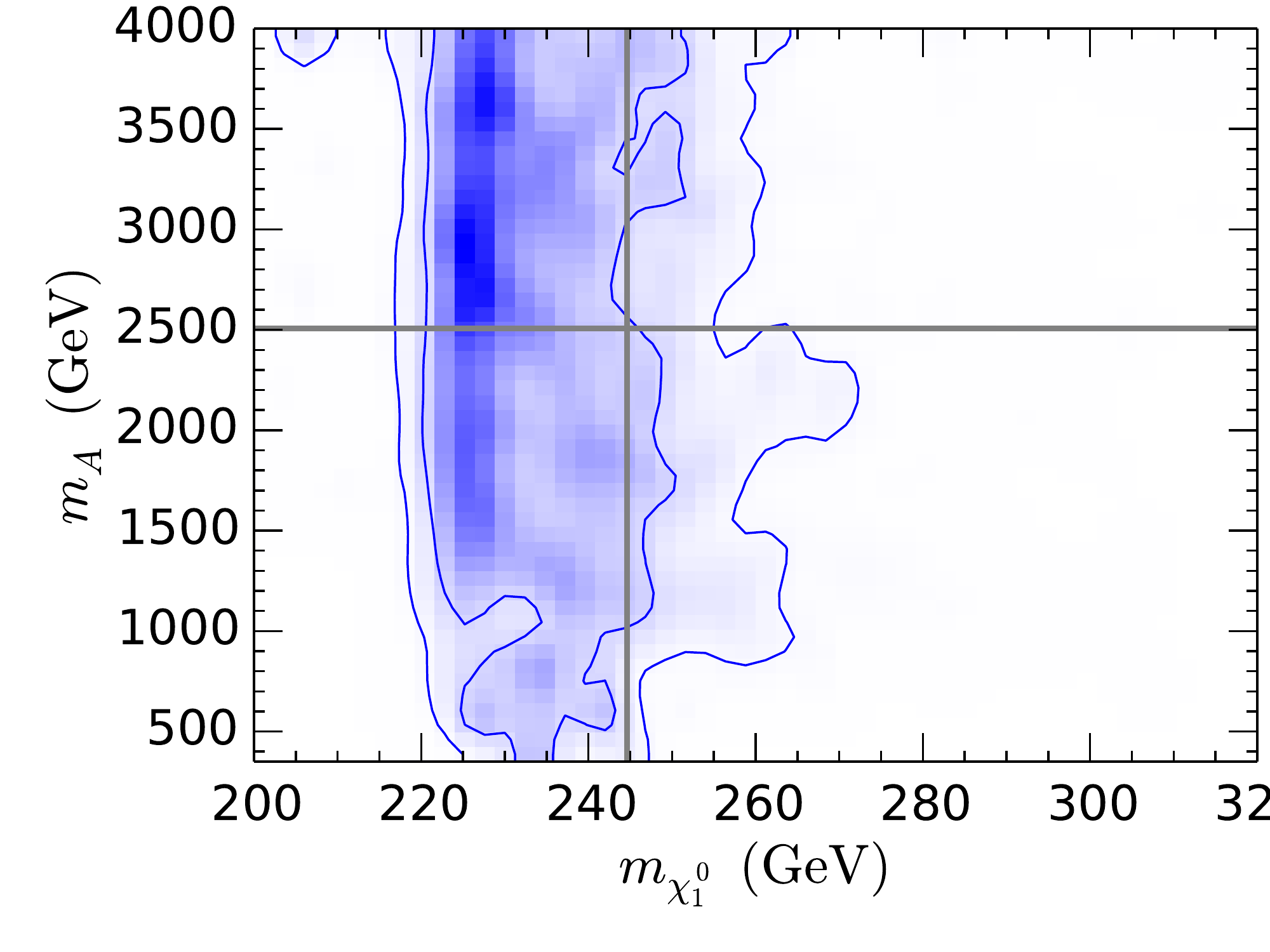}
  \includegraphics[height=0.37\textwidth, angle=0]{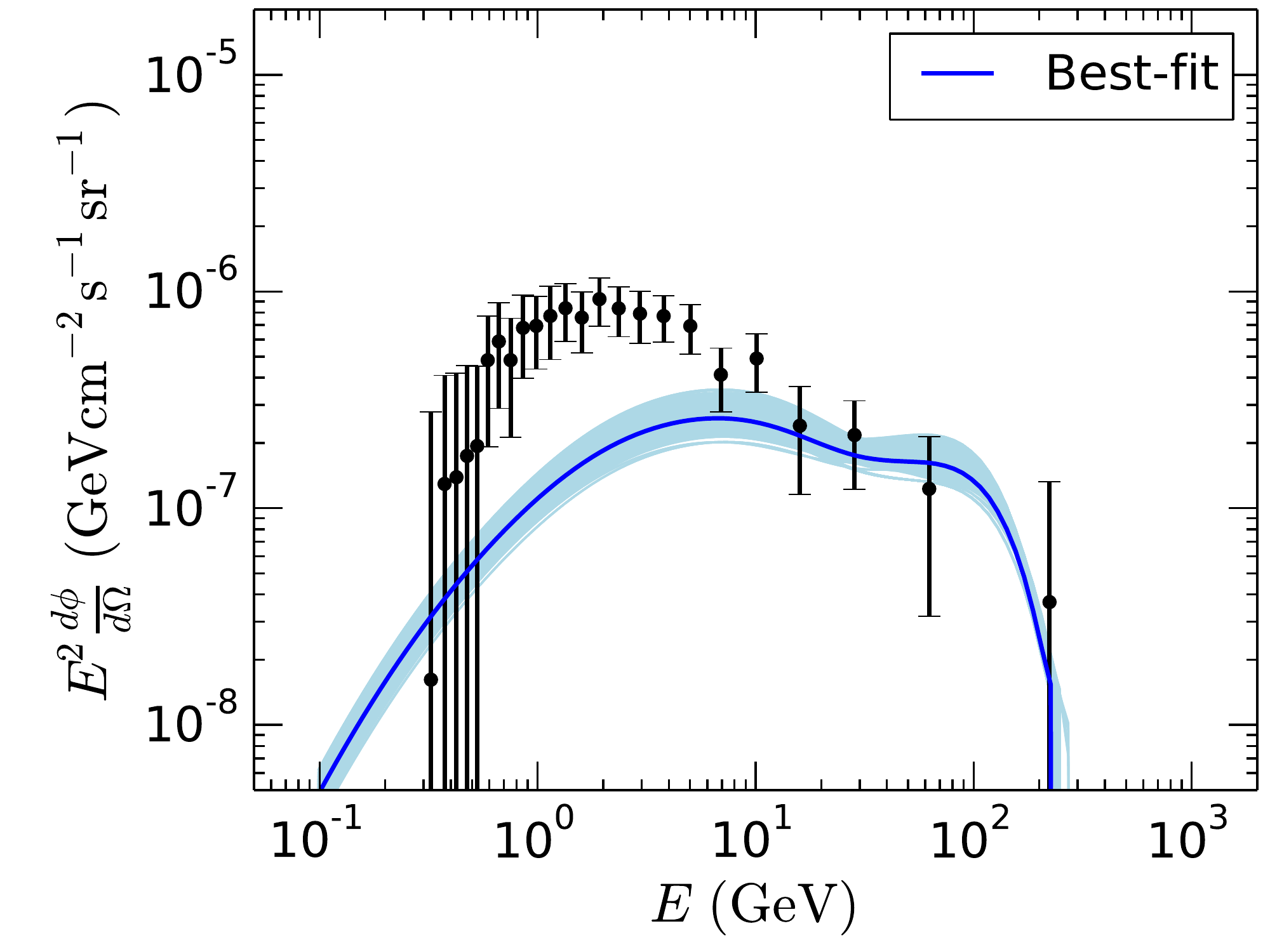}
  \includegraphics[height=0.37\textwidth, angle=0]{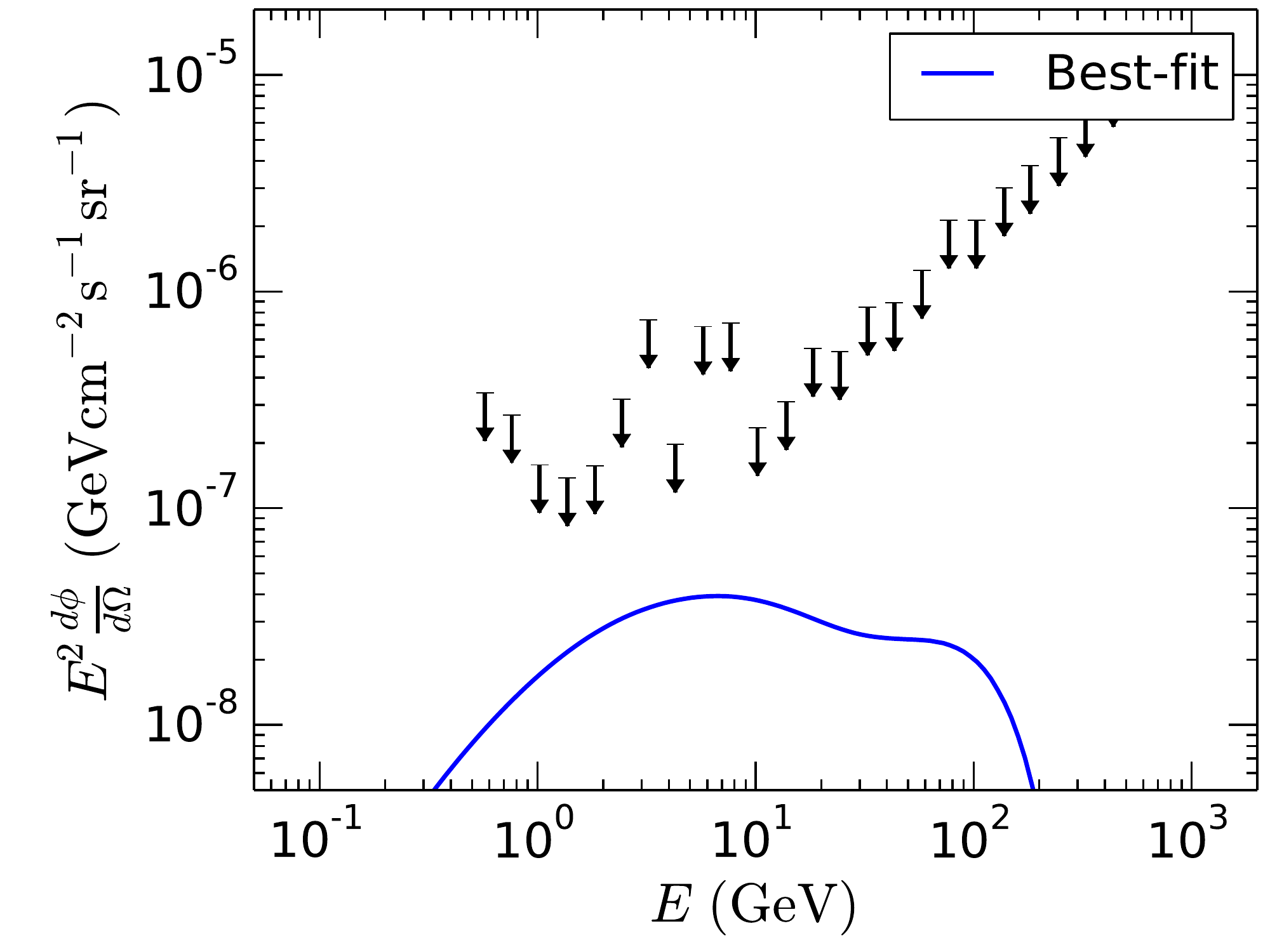}
  
  \caption{1$\sigma$ and 2$\sigma$ contours plots
    in the plane of $M_1 - M_2$ (top left), $M_1 - \mu$ (top right), $M_1 - m_A$ (middle left), $m_{\chi^0_1} - m_A$ (middle right) for Case 3c. Solid gray lines indicate the best fit values. Bottom left:
    2$\sigma$ bands of GC excess spectrum (light blue region)
    correspond to this case along with Fermi-LAT GC excess data and
    error bars (diagonal part of the covariance matrix). Deep blue
    line is the spectrum for best-fit points. Bottom right:
    Reticulum II $\gamma$-ray spectrum for the best-fit point (blue
    curve) along with the upper-limit on flux from Pass 8 analysis.}
\label{3c}
\end{figure*}

\underline{Case 3a}: As mentioned already, this case gives a poor fit
to the GC excess compared to the previous cases, yielding 
$\chi^{2}_{min} = 85.2.$ for DOF = 24. The best fit point corresponds to
$M_{1}= 248.0, M_{2}= 1136.6, \mu = -335.96, m_A = 925.58$. 
$m_{\chi^0_1}$ is in the range $\simeq$ 220 -285 GeV at 95.6$\%$ C.L.
In order to offset the downward pull on the relic density due to
co-annihilation, one requires the lightest neutralino to
be more massive \cite{gondolo} than in any of the previous cases.  
The dominant channel of annihilation of DM pairs in GC/Reticulum II
is $t\overline{t}$ and
$\tau\overline{\tau}$, mediated by the pseudoscalar Higgs. The high mass of 
the $\chi^0_1$ will shift the peak to high-frequency regions, and it 
is only through a sufficiently reduced (co)-annihilation rate that one can
be saved from exceeding the observed limits there. The profile, however,
will display a clear mismatch with what is observed, making 
it a rather poor fit. The various 2$\sigma$
contour plots as well as the band attempting a fit to the GC profile
and the prediction best GC fit on the Reticulum II spectrum are presented in Fig.
 \ref{3a}. As earlier, the GC best fit corresponds to practically
no Reticulum excess and thus comply with the 
Pass 8 upper limits.

Cases 3b and 3c lead to very similar conclusions. For Case 3b, with
$\tan\beta$= 50, the $\tau^+ \tau^-$ annihilation channel rises above
$t\bar{t}$. The various plots can be seen in Figs. \ref{3b} and \ref{3c}.

\subsection{Case 4: LSTSTC}

This scenario has the stau co-annihilation region buttressed with one
light stop. Most observations here are similar to those made for the
STC scenario. However, the requirements brought in by stau
co-annihilation during freeze-out tend to jack up $m_{\chi^0_1}$, with
the additional prospect of making it close to the light stop
mass. This makes the latter an additional co-annihilation partner for
the DM candidate, threatening one with too fast an annihilation rate,
and a consequent violation of the lower limit on the relic density. As
a result, the parameter space consistent with all constraints shrinks
in size, The $\sigma$ contours have to emerge out of whatever is left
after eliminating the regions ruled out by the above consideration.
Together with the problems arising in the GC spectrum fit, this yields
fits at least as bad as those in the previous scenario.

\begin{figure*}[h]
\centering
  \includegraphics[height=0.37\textwidth, angle=0]{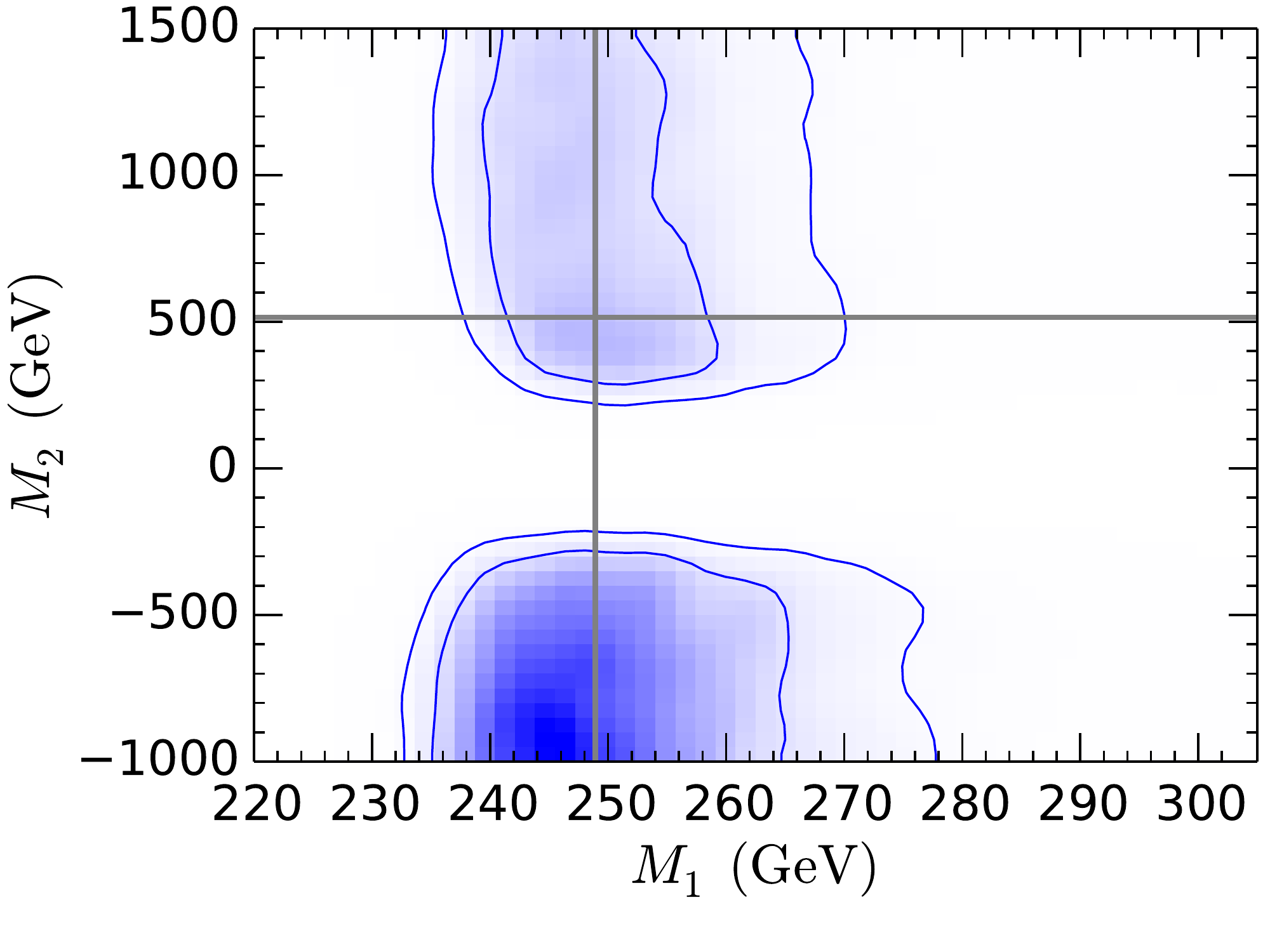}
  \includegraphics[height=0.37\textwidth, angle=0]{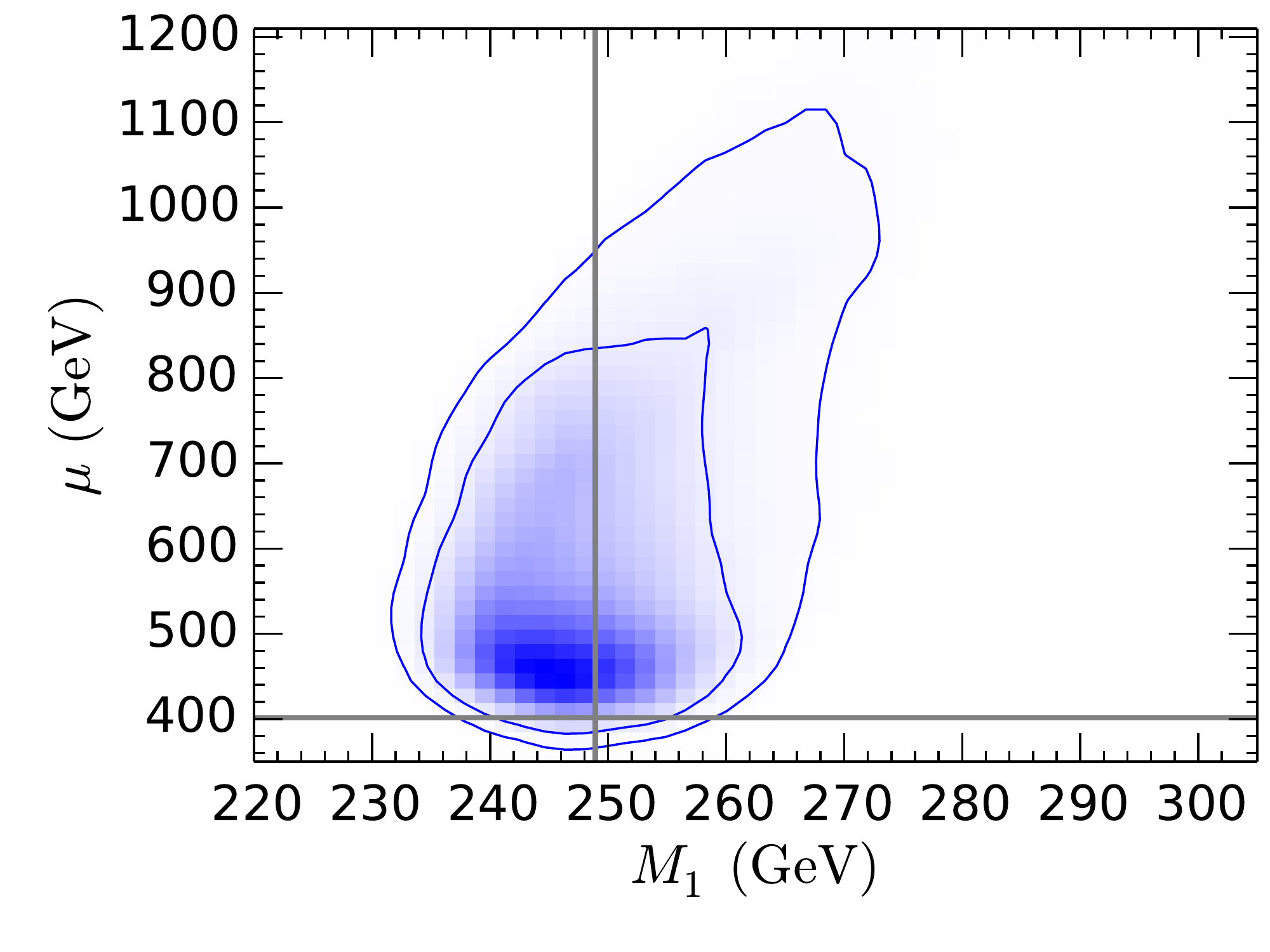}
  \includegraphics[height=0.37\textwidth, angle=0]{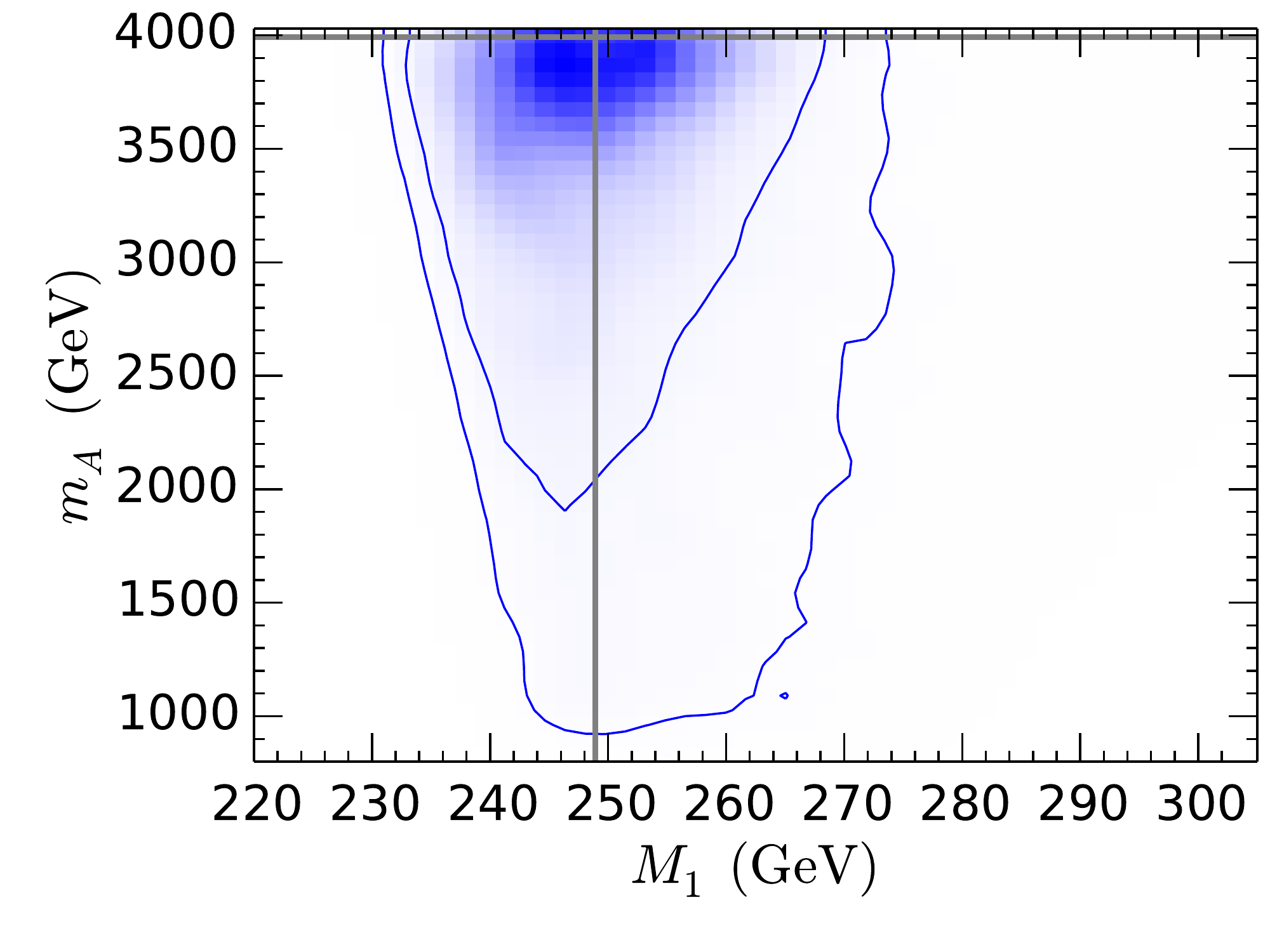}
  \includegraphics[height=0.37\textwidth, angle=0]{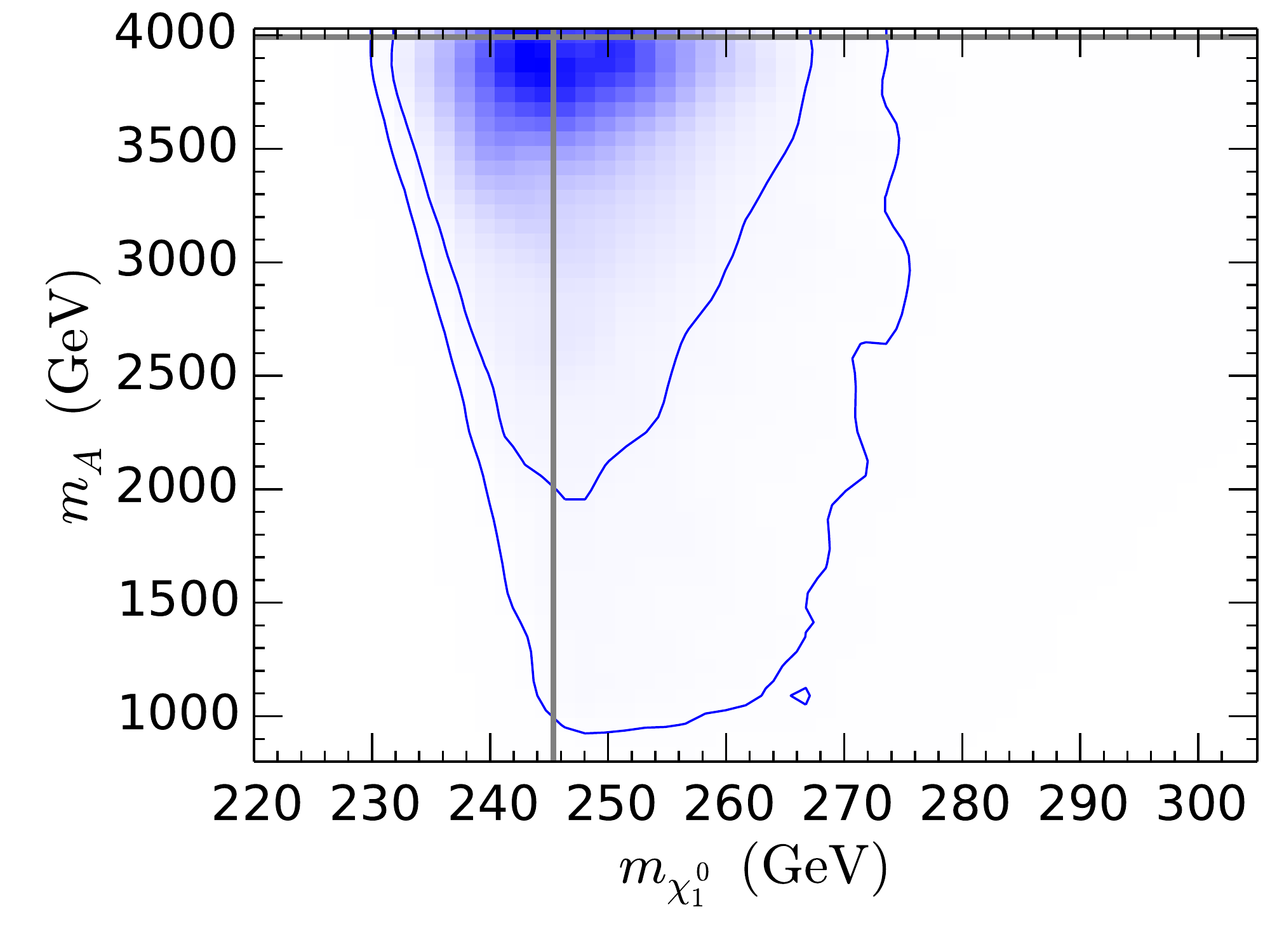}
  \includegraphics[height=0.37\textwidth, angle=0]{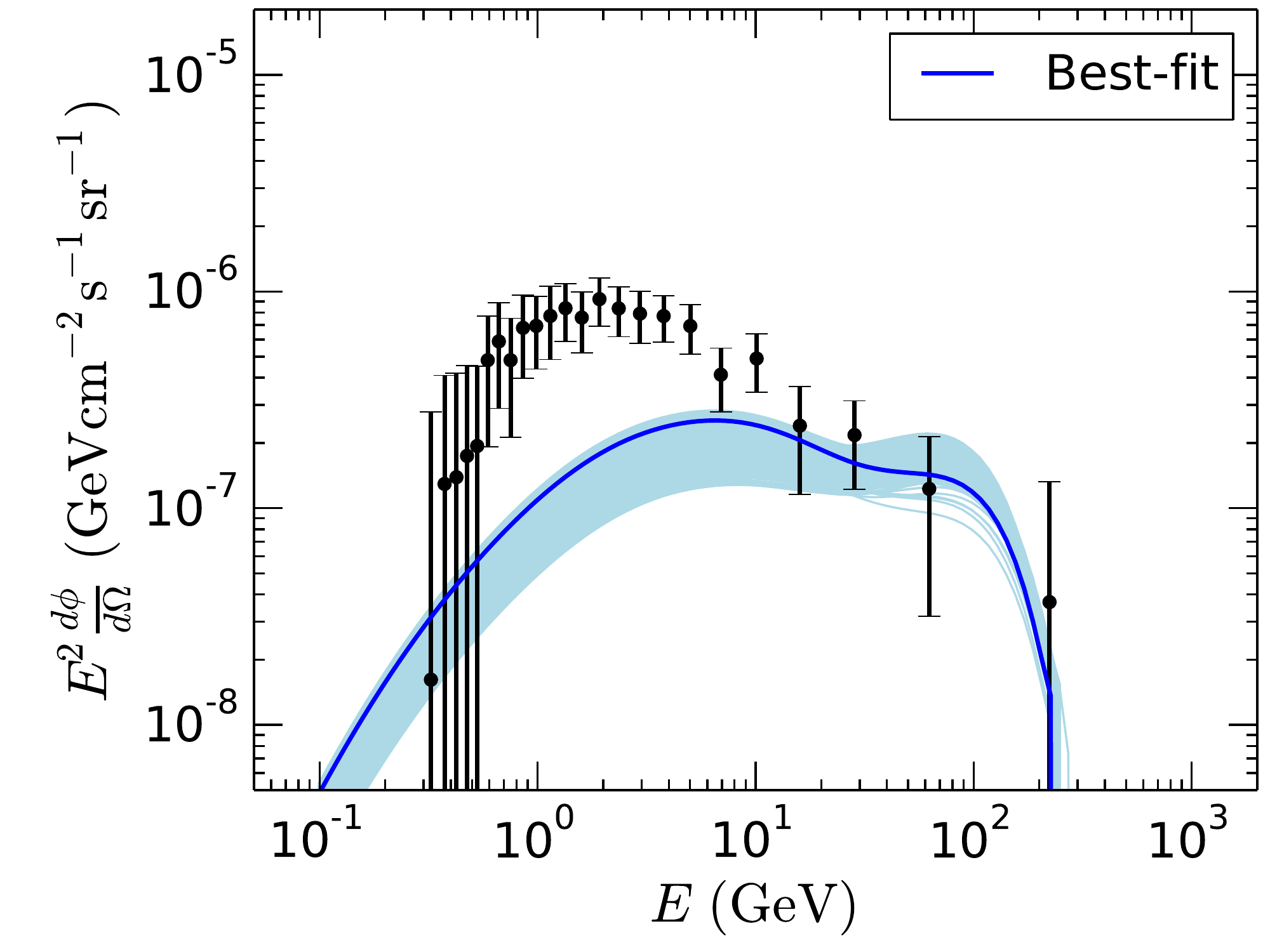}
  \includegraphics[height=0.37\textwidth, angle=0]{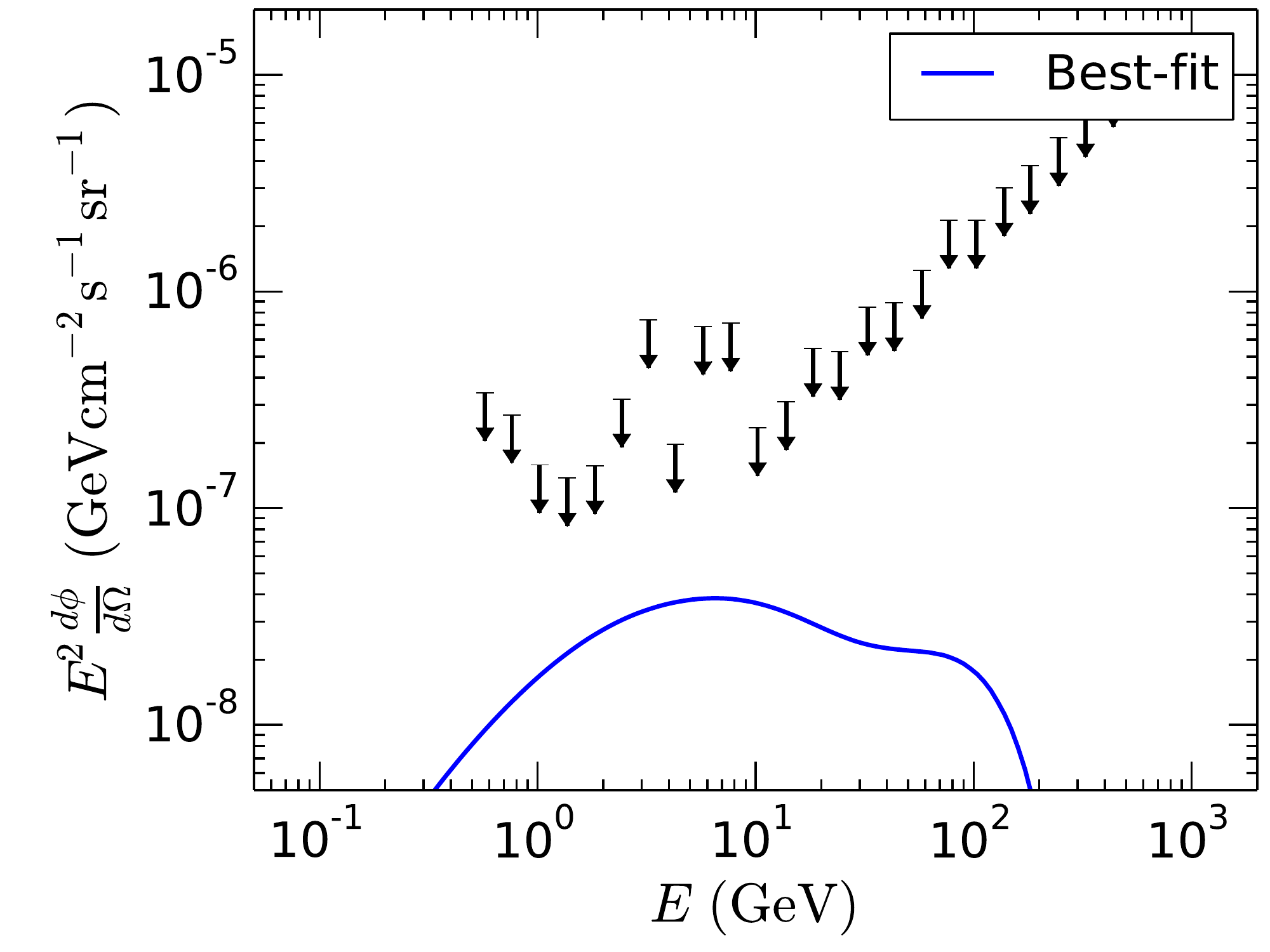}
  
  \caption{1$\sigma$ and 2$\sigma$ contours plots in the plane of $M_1
    - M_2$ (top left), $M_1 - \mu$ (top right), $M_1 - m_A$ (middle
    left), $m_{\chi^0_1} - m_A$ (middle right) for Case 4a. Solid gray lines indicate the
    best fit values. Bottom left: 2$\sigma$ bands of GC excess
    spectrum (light blue region) correspond to this case along with
    Fermi-LAT GC excess data and error bars (diagonal part of the
    covariance matrix). Deep blue line is the spectrum for best-fit
    points. Bottom right:
    Reticulum II $\gamma$-ray spectrum for the best-fit point (blue
    curve) along with the upper-limit on flux from Pass 8 analysis.}
\label{4a}
\end{figure*}

\begin{figure*}[h]
\centering
  \includegraphics[height=0.37\textwidth, angle=0]{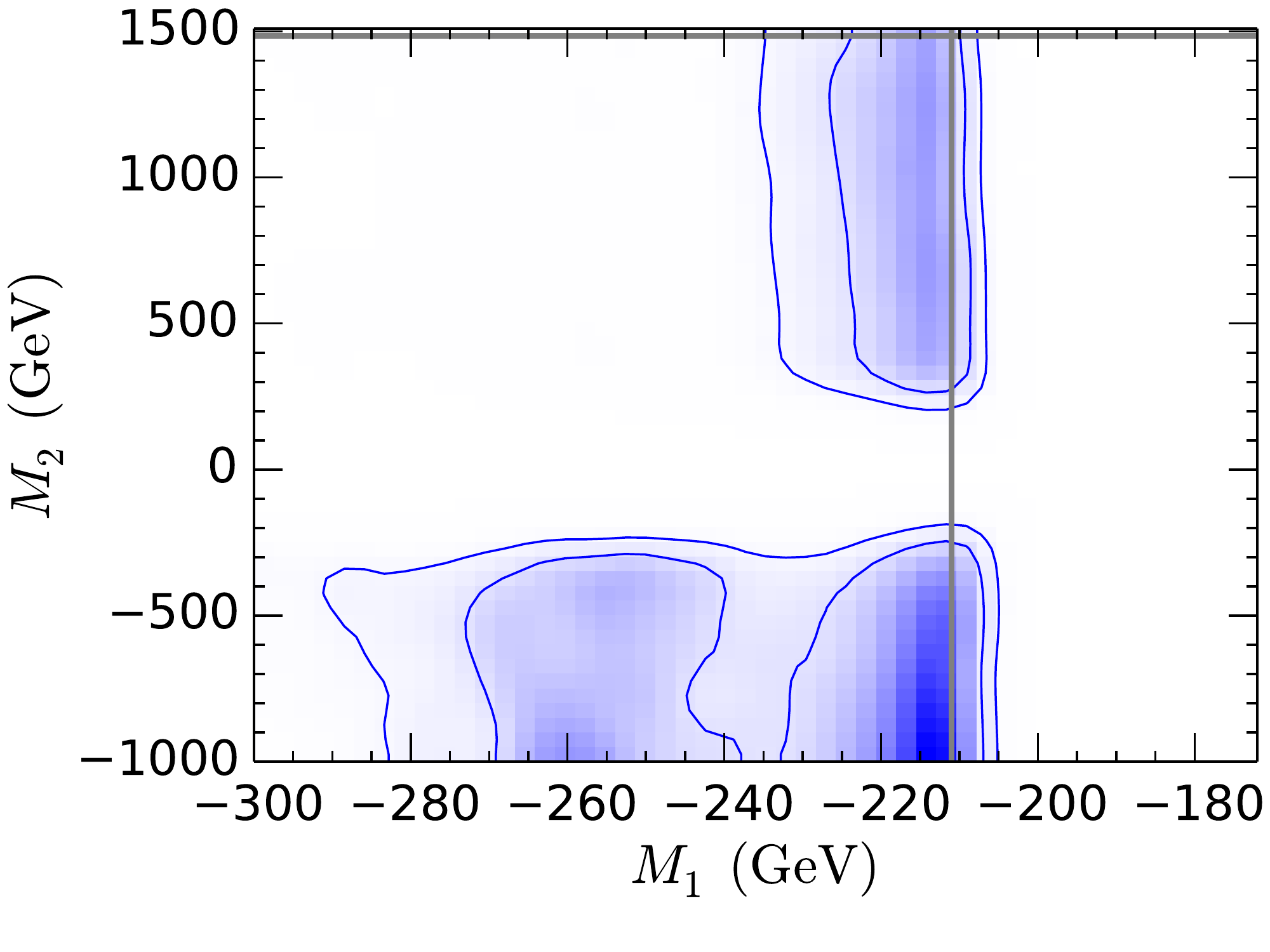}
  \includegraphics[height=0.37\textwidth, angle=0]{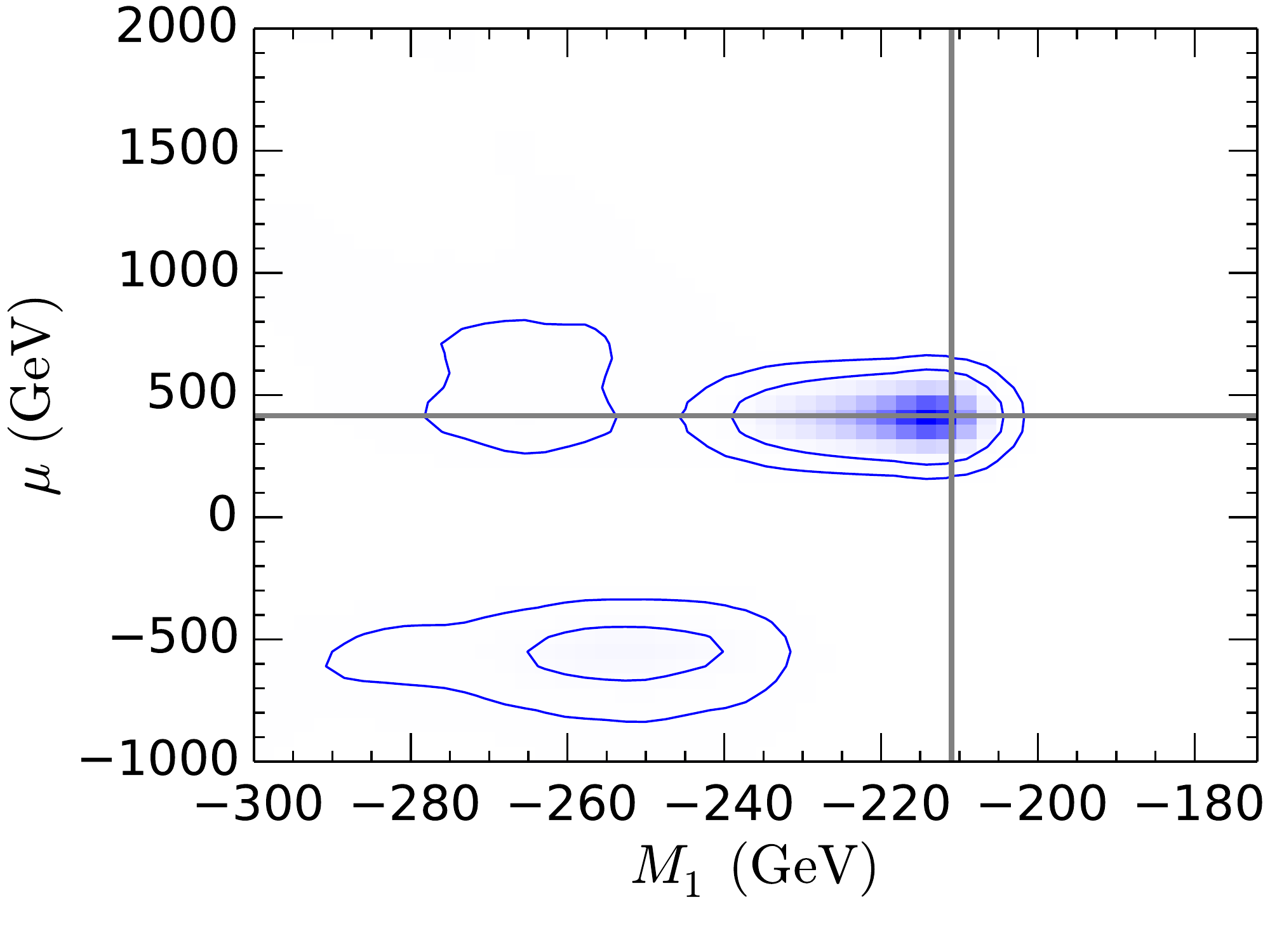}
  \includegraphics[height=0.37\textwidth, angle=0]{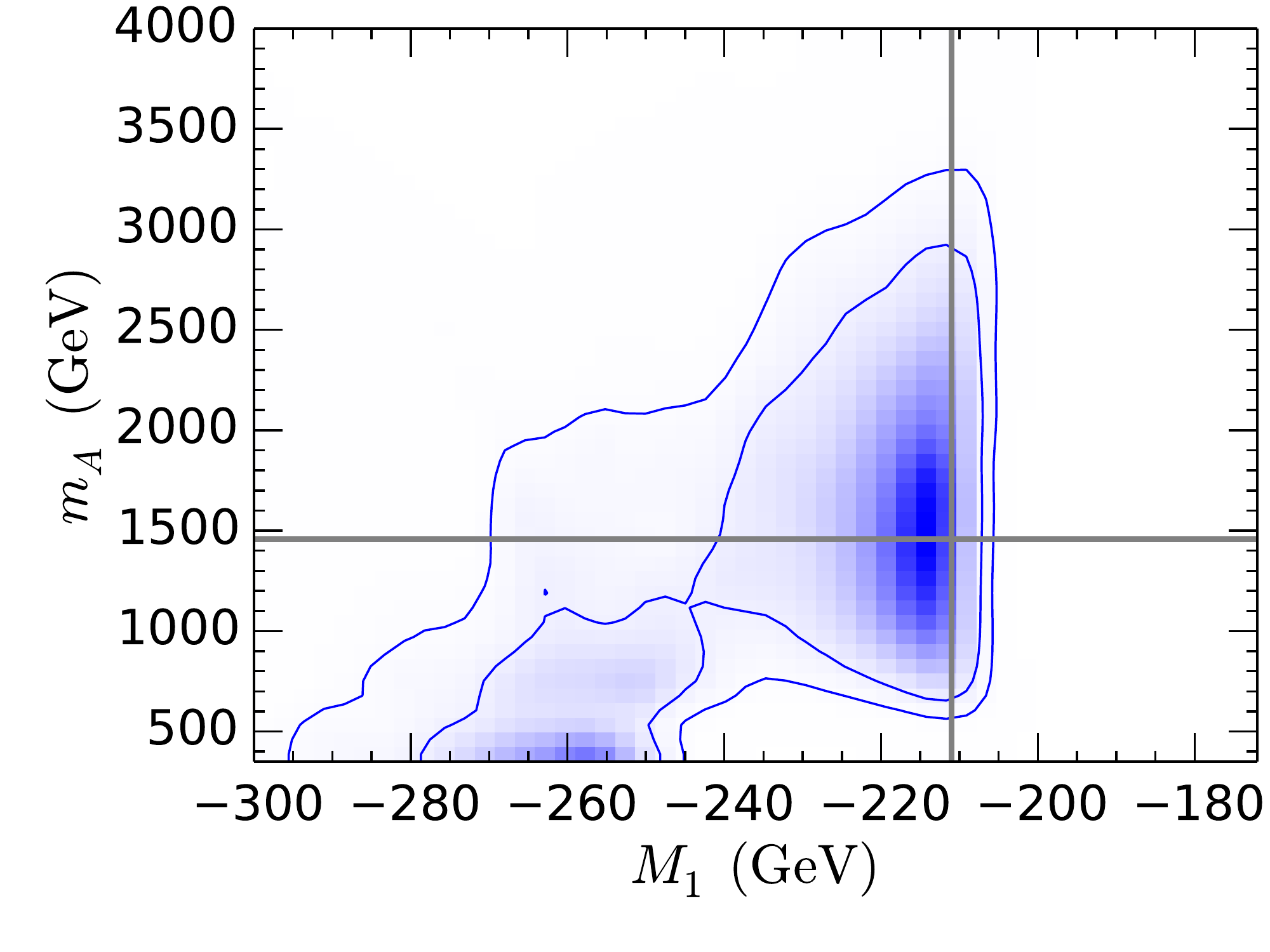}
  \includegraphics[height=0.37\textwidth, angle=0]{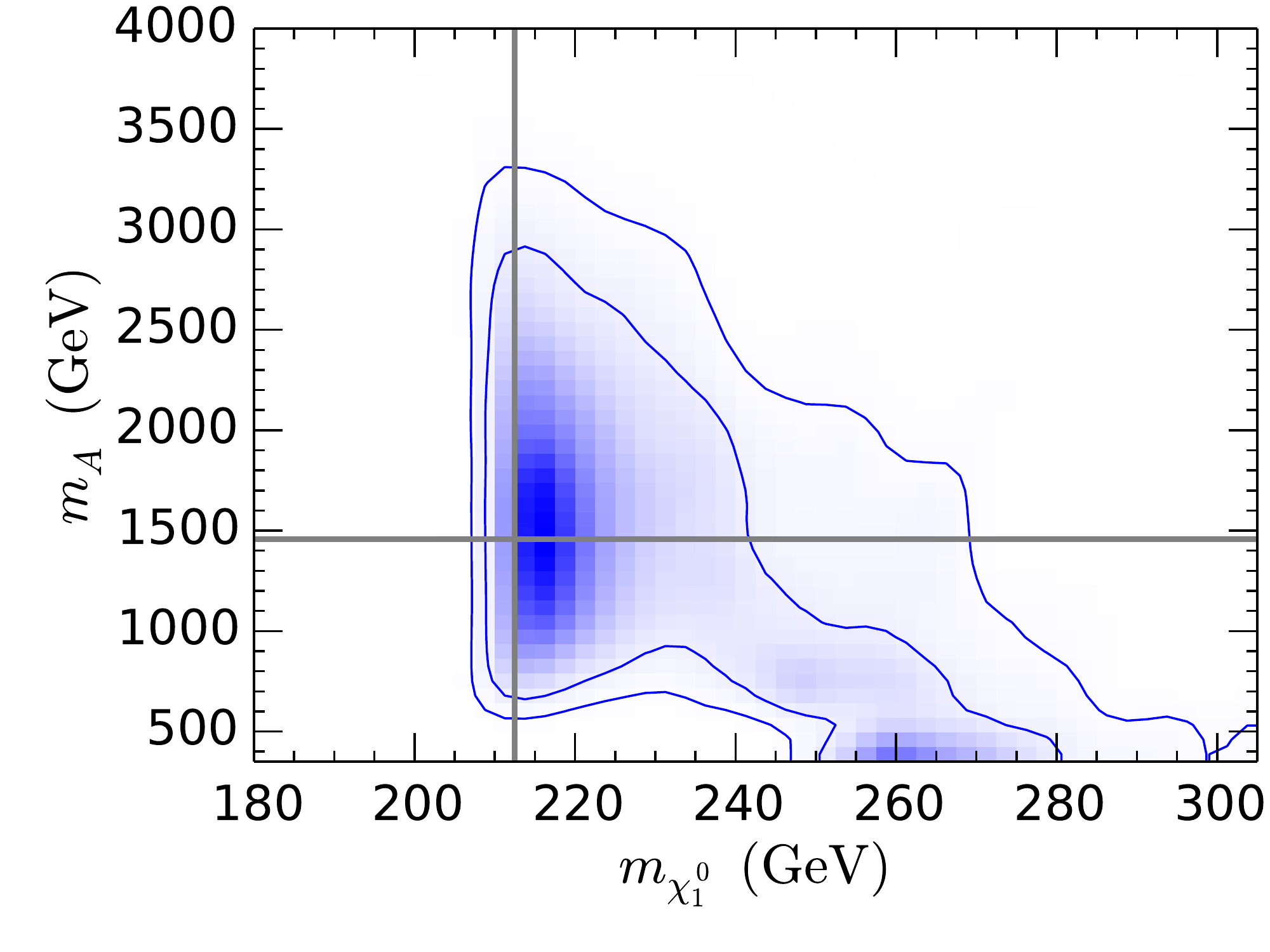}
  \includegraphics[height=0.37\textwidth, angle=0]{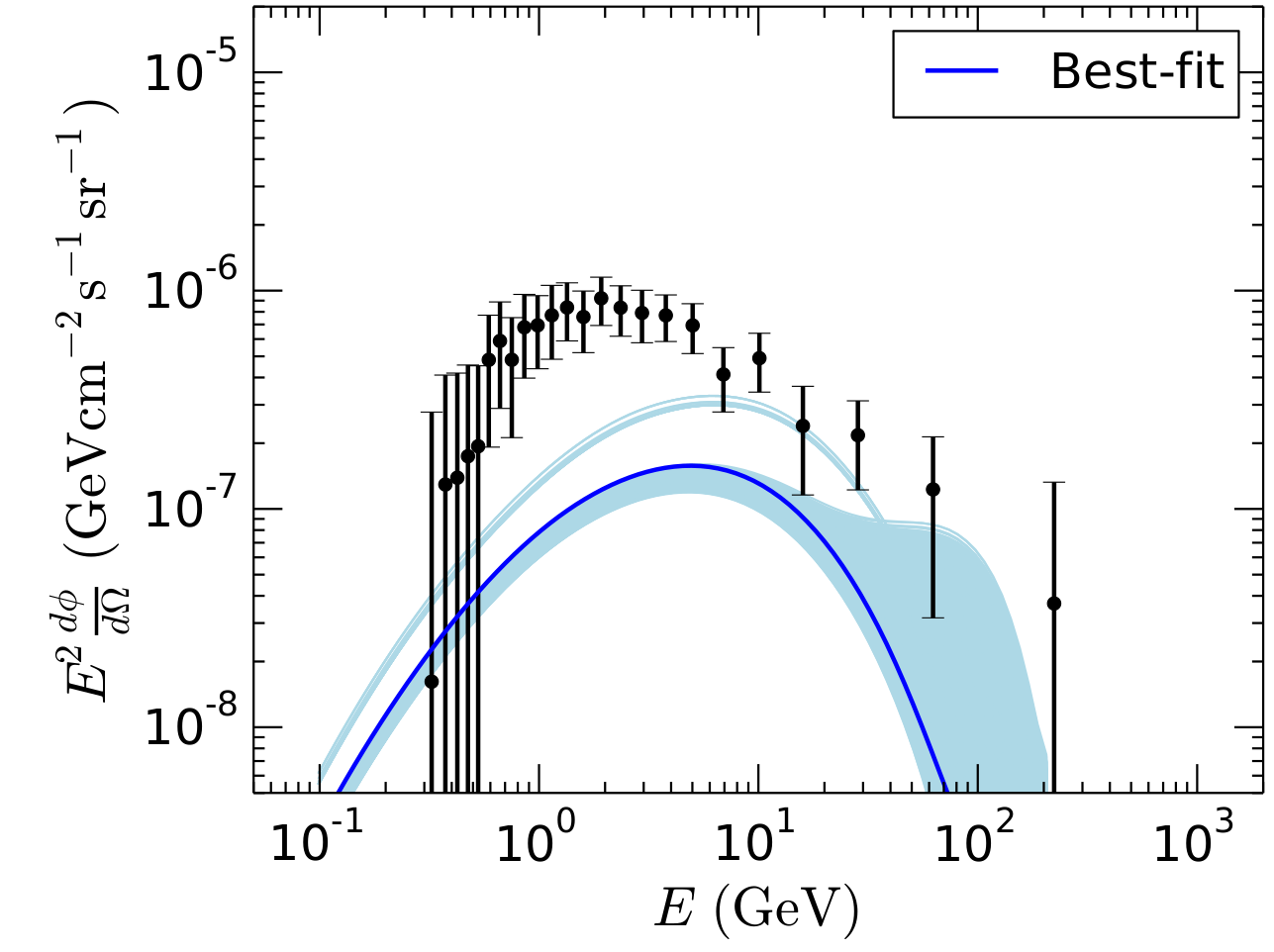}
  \includegraphics[height=0.37\textwidth, angle=0]{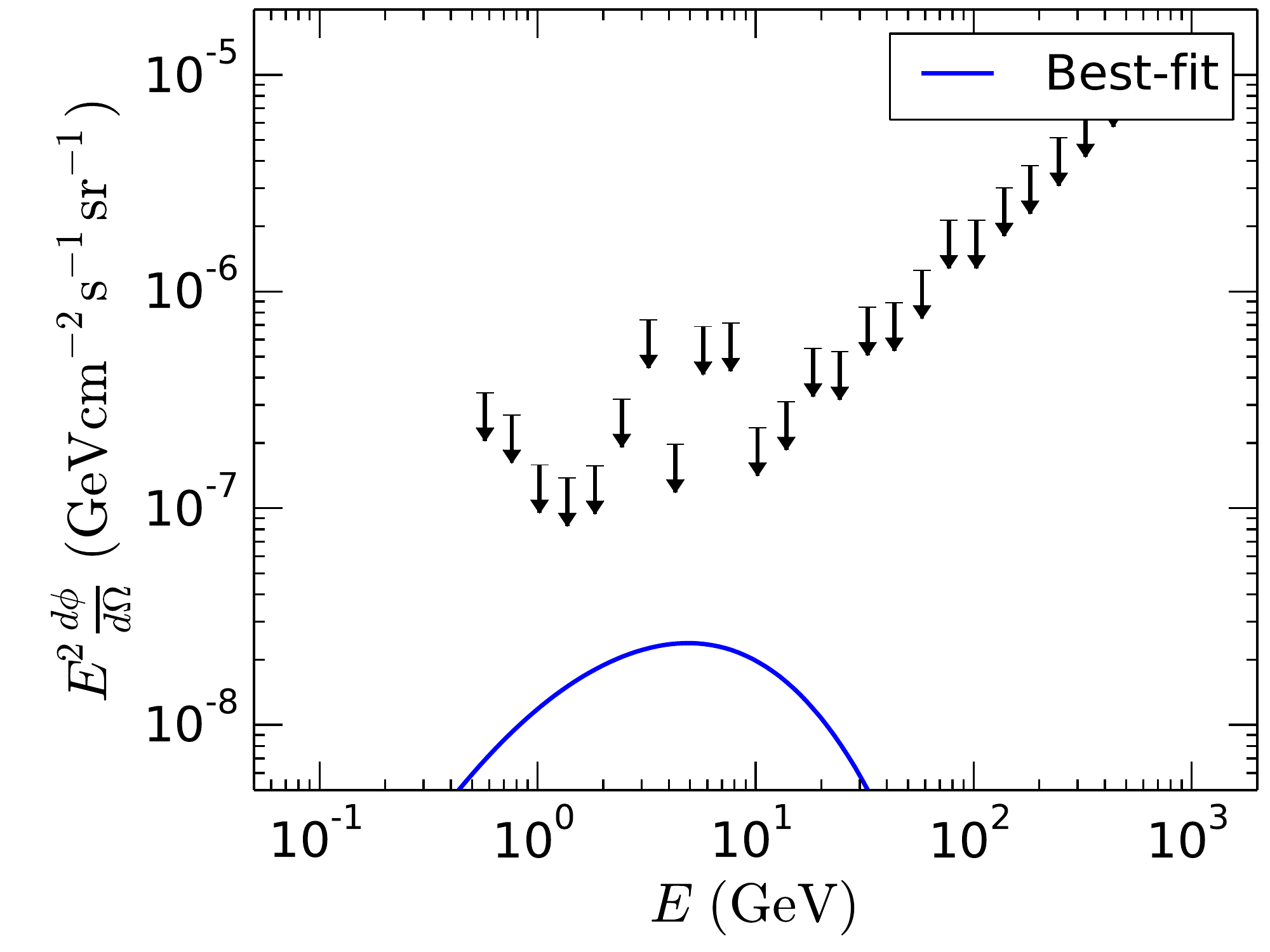}
  
  \caption{1$\sigma$ and 2$\sigma$ contours plots
    in the plane of $M_1 - M_2$ (top left), $M_1 - \mu$ (top right), $M_1 - m_A$ (middle left), $m_{\chi^0_1} - m_A$ (middle right) for Case 4b. Solid gray lines indicate the best fit values. Bottom left:
    2$\sigma$ bands of GC excess spectrum (light blue region)
    correspond to this case along with Fermi-LAT GC excess data and
    error bars (diagonal part of the covariance matrix). Deep blue
    line is the spectrum for best-fit points. Bottom right:
    Reticulum II $\gamma$-ray spectrum for the best-fit point (blue
    curve) along with the upper-limit on flux from Pass 8 analysis.}
\label{4b}
\end{figure*}

For $\tan\beta$= 5 and 20, the $t\bar{t}$ channel dominates in
DM annihilation in the GC as well as dwarf spheroidal galaxies.
The tau-pair final state is more competitive for   $\tan\beta$= 50,
where, however, the fit is far too wayward to present here.
The relevant plots are shown in Figs. \ref{4a} and \ref{4b}.
We also list in Table \ref{Table_chisq} the values of $\chi^2$/DOF for all
the cases discussed above.

The above discussion elicits the following points. 
First, {\em the viability of any type of MSSM spectrum is crucially
governed by the lower bound on the relic density together with the
shape of the GC $\gamma$-ray spectrum, together with the latest
direct search results}. Once the constraints from the above are adhered to, 
observations related to Reticulum II do not restrict any scenario significantly. 
And lastly, even the best of all
the fits to MSSM listed above lead to somewhat poor $\chi^2$/DOF. More 
will be said on this in section \ref{alternative_analysis}.

\begin{table}[h]
\begin{center}
\begin{tabular}{|c|c|c|}
\hline 
Case no & $\chi^{2}_{min}$/DOF \\ 
\hline
1a & 51.3/24  \\ 
 
1b & 65.1/24   \\ 

1c & 65.2/24  \\ 

2a & 65.7/24  \\ 
 
2b & 61.2/24  \\ 
 
2c & 62.2/24  \\ 

3a & 85.2/24  \\ 

3b & 97.2/24  \\ 

3c & 87.5/24  \\ 

4a & 85.9/24  \\ 
 
4b & 80.6/24   \\

\hline

\end{tabular}
\caption{values of $\chi^{2}_{min}$/DOF for various cases.}
\label{Table_chisq}
\end{center}
\end{table}


\section{Radio signal from Coma cluster} \label{radio}

DM annihilation inside the galaxies and galaxy clusters can produce
high-energy secondary charged particles (mostly electron and positron)
in the final state. Upon interaction with the ambient magnetic field,
these relativistic particles produce synchrotron radiation which can
lead to radio signals \cite{Storm:2016bfw, Hardcastle:2013bva,
  Natarajan:2015hma}.  One example of these is the Coma cluster, which
has been extensively observed in various frequency ranges
\cite{Colafrancesco:2005ji, Beck:2015rna}. 
In this section, we compute the radio signal for the DM models we have studied in the earlier sections and compare with the observational data. Note that the comparison with the radio data carried out here does not have any consequences for the constraints obtained on the MSSM parameter space.
The radio data points used
in these studies have been obtained from \cite{Thierbach:2002rs}.

Multi-frequency fits for radio emission from the Coma cluster, in
conjunction with the GC and Reticulum II data, have already been
reported in the literature \cite{Beck:2017wxu}. However, such studies
have by and large treated the DM mass ($m_{\chi^0_1}$) and
$\left\langle \sigma v \right\rangle$ as independent parameters,
adjusting the latter to saturate the radio data from the Coma cluster.
The dynamics of neutralino pair-annihilation, to which one is beholden
at any point in the MSSM parameter space, is not taken into
consideration there.

On the other  hand, we have started by fitting the MSSM parameters in 
details from the GC data where the background is known better. The shape
of the $\gamma$-ray spectrum is also used in detail.The channels
of annihilation and the net $\left\langle \sigma v \right\rangle$ is
computed rather than adjusted for any point in the MSSM parameter space.
The phenomenological constraints on  MSSM parameters (including the
observed Higgs mass), the lower limit on the relic density for MSSM being
the only dark matter source, and the direct search constraints on
the spectrum are all taken into account. 
The regions in the parameter space thus available are used to
compute the DM annihilation rates and the consequent radio flux from
the Coma Cluster.

The synchrotron flux from the Coma cluster depends on the $e^{\pm}$ source
function \cite{Colafrancesco:2005ji, Beck:2015rna}:
\begin{equation} \label{eq:radio_flux}
Q_e(r,E) = \left\langle \sigma v \right\rangle \frac{dN_e}{dE}
\mathcal{N}_{pairs}(r),
\end{equation}
where $\left\langle \sigma v \right\rangle$ is the same DM annihilation
rate we used previously for $\gamma$-ray spectrum calculation
(eq. \ref{eq:flux1} and \ref{eq:flux2}) and $dN_e/ dE$ is the
number of $e^{\pm}$ produced per annihilation per unit
energy. The quantity $\mathcal{N}_{pairs}(r)$ denotes the number of DM particle
pairs per unit volume squared at radial distance $r$ of the halo around
the cluster. The quantity depends on the DM density profile of the
halo (denoted by $\rho(r)$) and the contributions from sub-halos 
distributed inside the main halo
\cite{Colafrancesco:2005ji, Beck:2015rna}. For the calculation of
$\mathcal{N}_{pairs}(r)$ we follow the steps exactly as described in
\cite{Colafrancesco:2005ji}.

Using the above source function, the full calculation of radio
flux density spectrum $S(\nu)$ as a function of emitted radio
frequency $\nu$ is done subsequently, again following
\cite{Colafrancesco:2005ji} and \cite{radio_flux}. However, in view of
the various uncertainties, we have used two sets of values of the
parameters controlling the DM profile, and also the magnetic field, which are described below.

\begin{itemize}

\item {\bf Model A:}
In the first case, following reference \cite{Colafrancesco:2005ji}, we
have used the N04 profile, with $\alpha = 0.17$, and a homogeneous
magnetic field of magnitude 1.2 $\mu G$. 
\begin{equation}
\rho_{N04}(r) = \rho_s \exp\left[ -\frac{2}{\alpha}\left( \left( \frac{r}{r_s}\right) ^{\alpha} - 1 \right) \right],      \qquad \qquad B_{\mu} = 1.2 \mu G
\end{equation} 
Here $\rho_s$ is the characteristic density of the halo and $r_s$ is the cale radius of the profile.

\item {\bf Model B:} We show the difference in
prediction by switching over to the NFW profile in the second case,
with slope $\gamma = 1$ \cite{Navarro:1995iw}
\begin{equation}
\rho_{NFW}(r) = \frac{\rho_s}{(\frac{r}{r_s})(1 + \frac{r}{r_s})^2}
\end{equation}
 and a radial magnetic
field distribution motivated by observations of Faraday rotation for Coma, as in \cite{Beck:2015rna, Faraday_rotation} 
\begin{equation}
B_{\mu}(r) = B_0 \left(  1 + \left( \frac{r}{r_s} \right)^2    \right)^{-0.56}, \qquad  \qquad B_0 = 4.7 \mu G 
\end{equation}

\end{itemize}

The scale radius is taken to be the same in both cases \cite{Colafrancesco:2005ji,
  Beck:2015rna}. Our code is calibrated by exactly matching the fluxes
given in reference \cite{Colafrancesco:2005ji}.

\begin{figure}[h]
\begin{center}
\includegraphics[height=0.37\textwidth, angle=0]{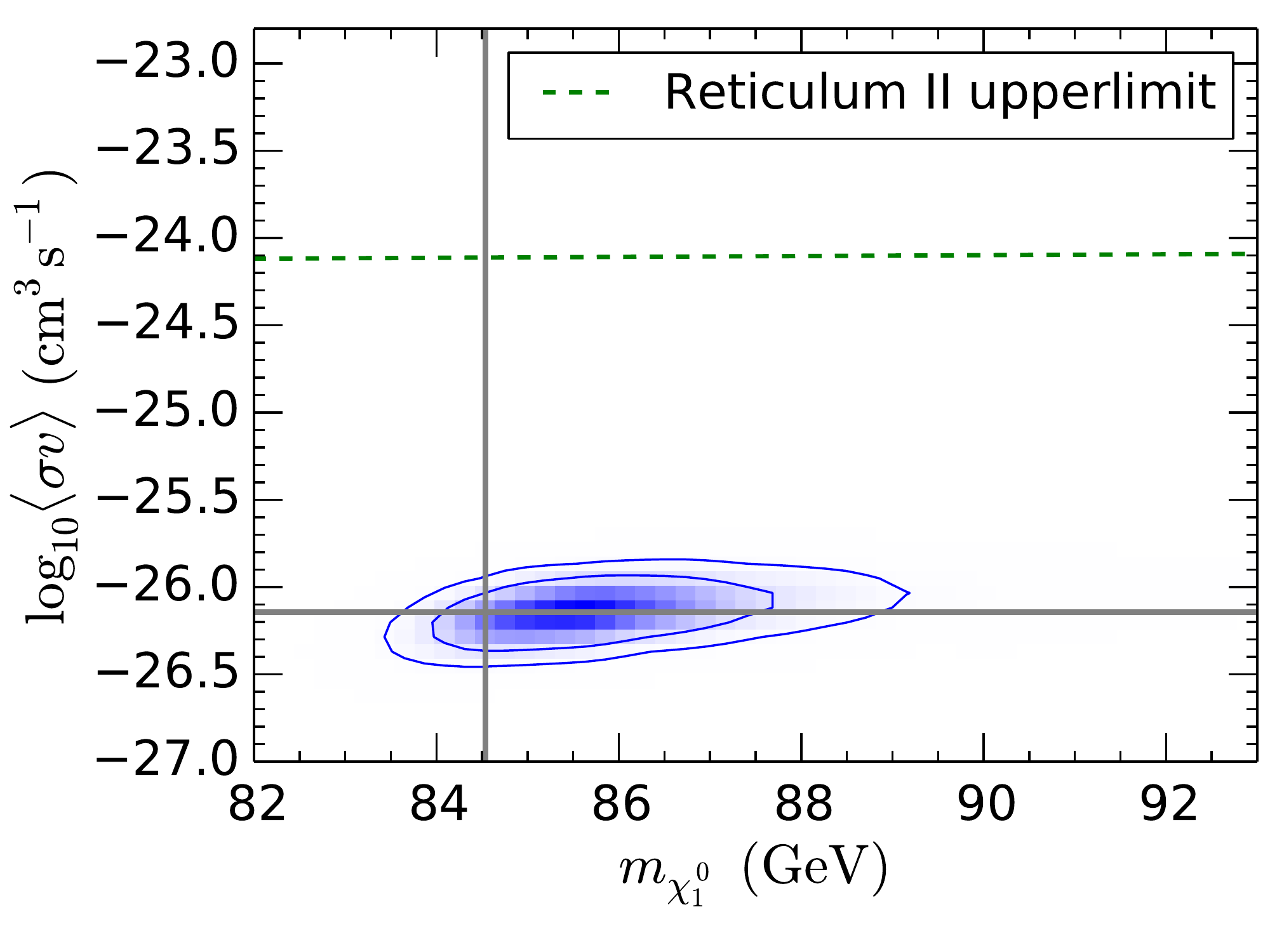}
\caption{1$\sigma$ and 2$\sigma$ (blue) contours plots in the plane of
  dark matter annihilation rate ($\left\langle \sigma v
  \right\rangle$) vs dark matter mass ($m_{\chi^0_1}$) for Case
  1a. Reticulum II upper limit (from Pass 8 analysis) also has been
  shown with green dashed line.}
\label{sv_1a}
\end{center}	
\end{figure}

The quantity $\mathcal{N}_{pairs}(r)$ (which plays a role similar to
the J-factor in eqs. \ref{eq:flux1} and \ref{eq:flux2}) is inversely
proportional to dark matter mass squared
\cite{Colafrancesco:2005ji}. Consequently the flux normalisation
(following equation \ref{eq:radio_flux}) is proportional to
$\frac{\left\langle \sigma v \right\rangle}{m_{\chi_1^0}^2}$.  Among
all the previously described cases, Case 1a, the best from the
standpoint of $\gamma$-ray data, has the largest $\frac{\left\langle
  \sigma v \right\rangle}{m_{\chi_1^0}^2}$. This case is thus expected
to yield the highest radio flux. The constraints in the $\left\langle
\sigma v \right\rangle$ - $m_{\chi_1^0}$ plane obtained from the
likelihood analysis for this case is shown in Fig. \ref{sv_1a}. We can
see that the allowed range in dark matter mass is quite small and has
value around $\sim 85$ GeV. The annihilation cross section peaks close
to $10^{-26}$ cm$^3$ s$^{-1}$, which is significantly smaller than the
upper limit inferred from the Reticulum II data.

\begin{figure}[h]
\begin{center}
\includegraphics[height=0.37\textwidth, angle=0]{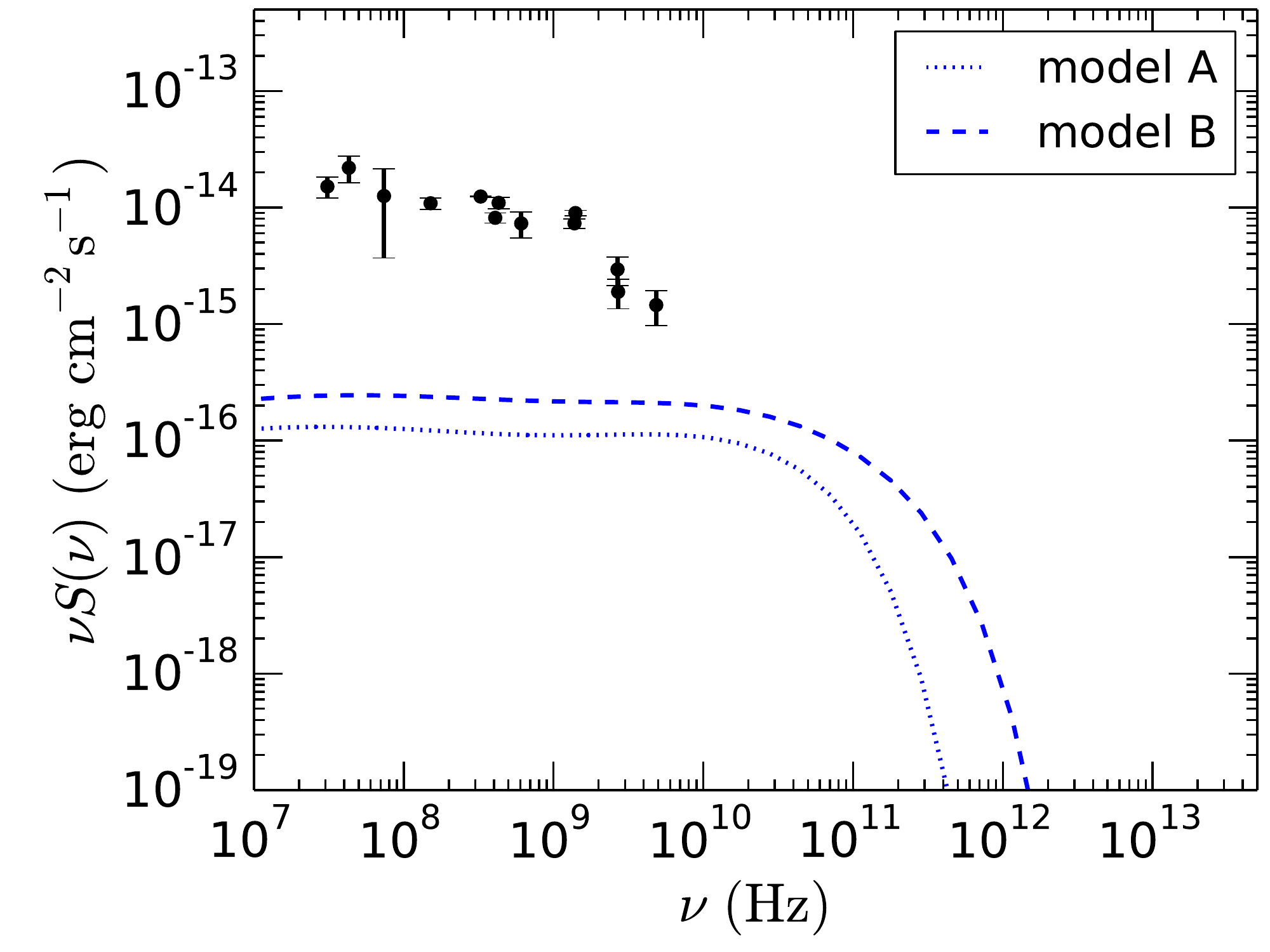}
\caption{Radio flux density spectrum for the Coma cluster for
  benchmark 1a for two astrophysical parameter sets. The blue dotted and dashed
  lines, marked as model A and model B, correspond respectively to the
  N04 and NFW profiles, and the corresponding magnetic fields. The
  radio data are taken from \cite{Thierbach:2002rs}}
\label{coma}
\end{center}	
\end{figure}

In Fig. \ref{coma}, we have shown the
comparison of the observed radio data from Coma cluster
\cite{Thierbach:2002rs} with the calculated flux density spectrum
($S(\nu)$) \cite{Colafrancesco:2005ji, Beck:2015rna} for the best-fit
point corresponding to Case 1a (ensuring consistency with M31 data and the
Reticulum upper limits),
for the two parameter sets discussed above.

It is clear that, even with the relative enhancement of the radio flux
in the second case (model B), the best fit curve from the $\gamma$-ray data
accounts for at most 20\% of this radio signal for $\approx
5$~GHz. The shortfall is considerably more conspicuous for lower
frequencies, and also with the first set of model parameters. One possible way of increasing the signal would be to use a higher value of the magnetic field (which in this case turns out to be $\sim 20 \mu$G), however, such values would be inconsistent with the Faraday rotation measures \cite{ Faraday_rotation}.

This, however, is not necessarily an inconsistency, since the
conventional astrophysical sources of $e^{\pm}$ in the Coma cluster
can contribute significantly to the radio emission.  There can of
course be a difference in conclusions due to a still different choice
of parameters in the DM profile.  As far as the GC profile is
concerned, our adopted choice corresponds to $J$-values on the high
side, the reduction of which results in worsening of the $\gamma$-ray
spectral fit. On the whole, the viability of any explanation of the
Coma cluster radio data in terms of DM annihilation will be better
known if one can extract the DM profile and the magnetic field from
other observation. Similarly, the observations of astrophysical
objects which have high DM content and where the astrophysical
processes giving rise to the relativistic $e^{\pm}$ are suppressed,
are likely to improve one's understanding in this direction.

On the whole, our estimate of the radio synchrotron flux in based on a
rigorous consideration of MSSM dynamics. As already stated, there is
thus no inconsistency between an $m_{\chi^0_1}$ used and the
corresponding $\left\langle \sigma v \right\rangle$, both being
also consistent with all phenomenological constraints. If at all the
$\gamma$-ray data are found explicable in terms of the MSSM, the
corresponding radio prediction can be tested by (a) extracting the DM
profile and magnetic field information independently, and (b)
observing other objects such as dwarf spheroidal galaxies where
astrophysical backgrounds are expected to be subdued.

\section{Likelihood analyses: comparison with direct search constraints
not initially imposed}  \label{alternative_analysis}

Let us finally ask ourselves the following question: how satisfactory
are the fits (and 2$\sigma$ regions) reported above from the
standpoint of MSSM? For an answer, let us remind ourselves of the
situation in the context of the Higgs boson of the standard
electroweak model. Before the actual discovery took place, leading to
the conclusion $m_h \approx 125$ GeV, global fits of precision data
had yielded $m_h \simeq$ 95 GeV.  This region had been ruled out from
direct searches for the Higgs at the Large Electron-Positron (LEP)
collider. In addition, Tevatron results further ruled out the region
$m_h \approx 158 - 174$ GeV. However, the region where the particle
has actually been detected fell outside the bands forbidden above but
well within the 2$\sigma$ region of the initial fit, sometimes called
the `standard fit'.  On the other hand, there has been the so-called
`complete fit' where the direct search constraints have been imposed
at the very outset, setting $\chi^2$ to high values in the
disallowed/disfavoured regions. The minimum of $\chi^2$ obtained
thereby was again within the 2$\sigma$ region of the standard
fit. This implied an overlap between the favoured regions of both
kinds of fit, and the particle, with interactions closely resembling
the SM Higgs, has been discovered in this overlap region. One may thus
conclude that {\it the SM is a good fit for the data on a Higgs boson
  having a mass of about 125 GeV}. The $\chi^2_{min}$ and also
$p$-values (defined as the probability $P(\chi^2 > \chi^2_{min})$)
including the LHC data support such a conclusion \cite{Baak:2011ze,
  Flacher:2008zq}. The $p$-value goes up from 0.21 in the standard fit to 0.23 in the complete fit.

\begin{figure*}[h]
\centering
  \centering
  \includegraphics[height=0.37\textwidth, angle=0]{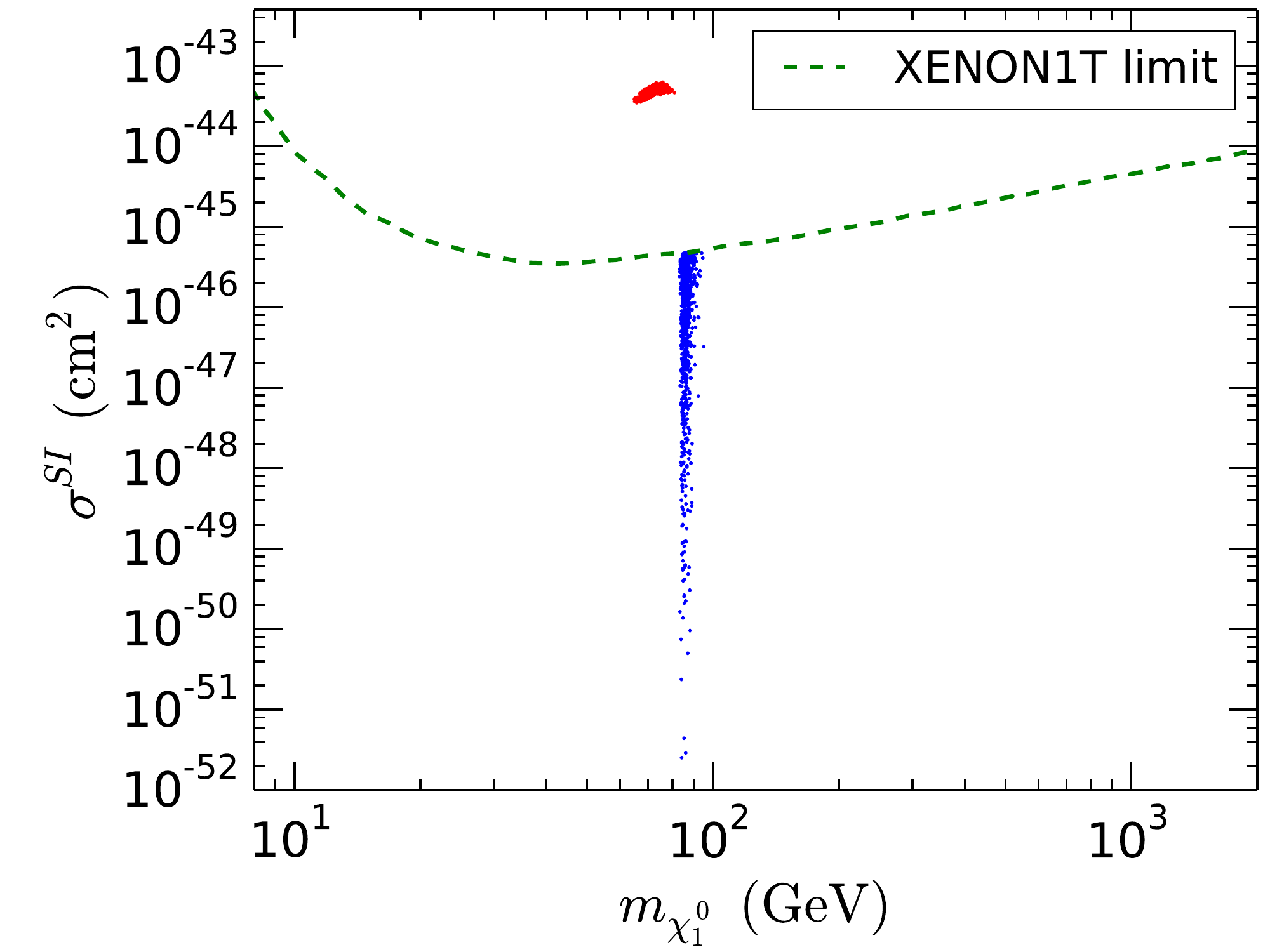}
  \includegraphics[height=0.37\textwidth, angle=0]{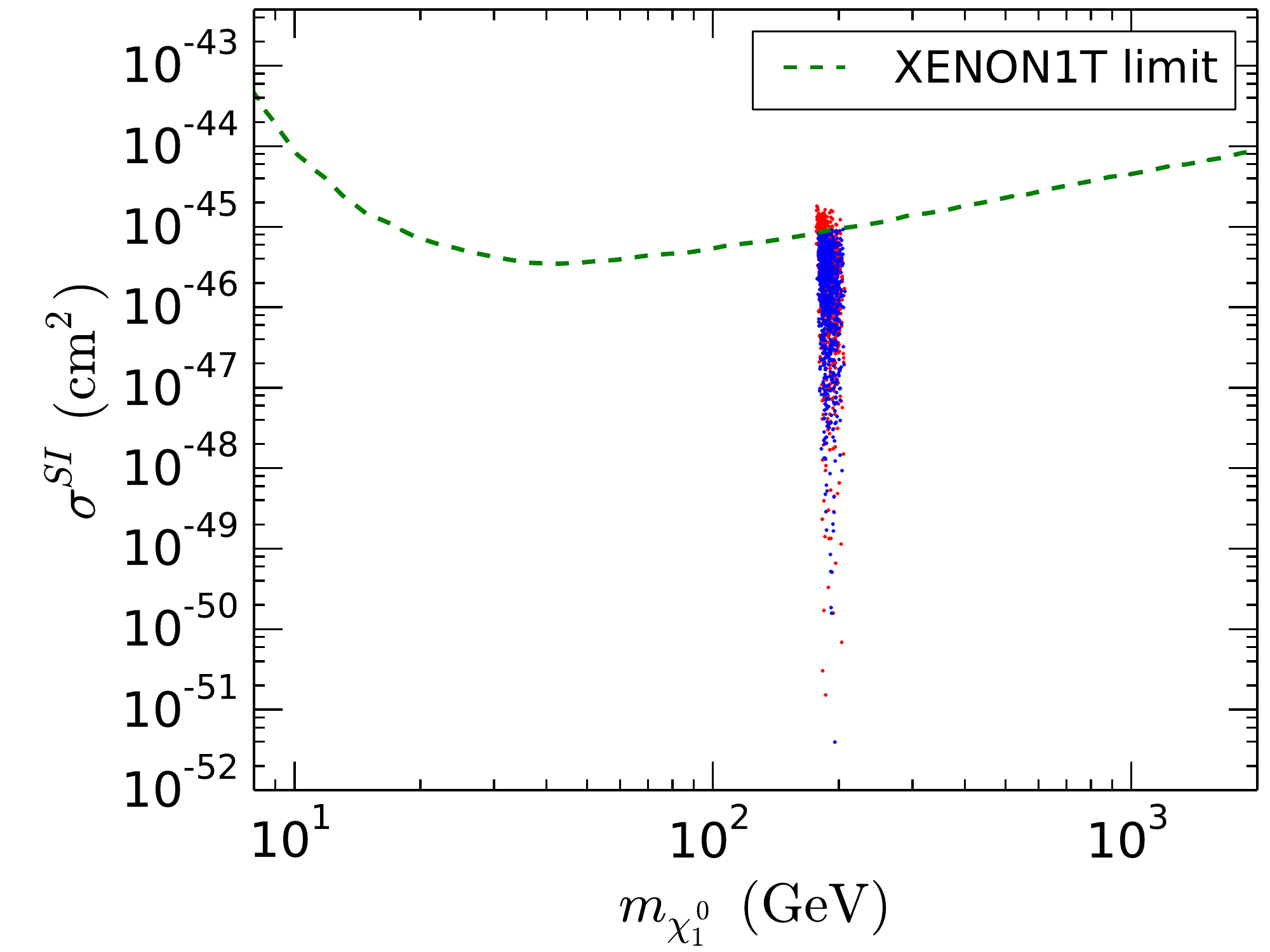}
  
  \caption{Left: $\chi_1^0$-neucleon spin-independent cross section as
    a function of $\chi_1^0$ mass. 2$\sigma$ upper limit on the cross
    section from XENON1T experiment has been shown by green dashed
    line. The blue region corresponds to the MSSM parameter space at
    95.6\% C.L. for the case shown in Fig. \ref{1a} (Case 1a in
    complete analysis). The red area is corresponds to the 95.6\%
    C.L. region of parameter space for Case 1a in 'fit without direct
    search' analysis. Right: same for Case 2b.}
\label{direct_search}
\end{figure*}

Let us, in comparison, examine the scenario in MSSM that fits the
$\gamma$-ray data best, which is our Case 1a. When the `complete fit'
in this case is made, the 2$\sigma$-contours in the marginalised plots
mark out a region of the MSSM parameter space.  Let us remember that
this fitting procedure imposes the direct search constraints at the
beginning. Now, analogously to the Higgs case, one may obtain the
2$\sigma$ regions about the best fit based on all results excepting
those related to direct DM search. {\it This entire region becomes
  disallowed when subjected to the latest direct search limits, as
  shown in Fig. \ref{direct_search}. This implies that the favoured
  regions in the complete fit fall outside the 95\% C.L. contours
  which one obtains via an impartial analysis of the $\gamma$-ray data
  in terms of the MSSM}.

\begin{table}[h]
\begin{center}
\begin{tabular}{|c|c|c|c|}
\hline 
Case no & $\chi^{2}_{min}$/DOF & $p$-value \\ 
\hline
1a(complete fit) & 51.3/24 & $1 \times 10^{-3}$\\ 
 
1a(fit without direct search) & 40.1/24 & $2 \times 10^{-2}$ \\

1a(fit without direct search and relic density lower limit) & 39.1/24 & $3 \times 10^{-2}$ \\ 
\hline
2b(complete fit) & 61.2/24 & $1 \times 10^{-4}$ \\ 
 
2b(fit without direct search) & 61.1/24 & $1 \times 10^{-4}$  \\ 
\hline
\end{tabular}
\caption{Comparison of the quality of fitting between two types of
  analysis for Cases 1a and 2b}
\label{Table_pvalue}
\end{center}
\end{table}

This can be understood from Table \ref{Table_pvalue} where we present
the values of $\chi^2_{min}$/DOF for both the `complete fit' and the
`fit without direct search'.  The numbers clearly show how
$\chi^2_{min}$ deteriorates significantly in the complete fit, while
that in the fit without direct search, too, is far from
satisfactory. The difference in the values of the $\chi^2_{min}$ for
two cases is $\sim 11$, which explains why the region allowed in the
fit without direct search becomes disallowed when the direct search
limits are imposed.  We also show the $p$-values of the two analyses
in the same table, where a higher $p$-value indicates that the
corresponding MSSM spectrum provides a better fit to the data. Again,
the $p$-value, being $2 \times 10^{-2}$ for the fit without direct search in Case
1a, becomes $1 \times 10^{-3}$ for the complete fit. For comparison, we have also shown the corresponding
$\chi^2_{min}$/DOF and the $p$-values for the two kinds of fit for
Case 2b which performs second best among our benchmarks. For this
benchmark, the 2$\sigma$ region from the fit without direct search is
largely retained in the complete fit, as shown in
Fig. \ref{direct_search}. However, the values of $\chi^2_{min}$/DOF,
similar for the two fits, are around 2.54. The $p$-values, too, are
around $10^{-4}$ for both cases. The main reason behind the
unsatisfactory fit is that the neutralino mass required for
annihilation the $t\bar{t}$ channel is so high that the $\gamma$-ray
peak shifts to hight frequencies. Accordingly, the deficits in the
low-frequency bins increase, and, because of the fact that the errors
there are small, these bins lead to high values of
$\chi^2_{min}$. MSSM parameters that can offset this by scaling up the
annihilation rates lead to the violation of the lower limit on $\Omega
h^2$, something that we interpret as going beyond MSSM.

The candidatures of the remaining benchmarks are distinctly
worse. {\it All these observations indicate that the MSSM, with the
  lightest neutralino as the only source of cold dark matter in our
  universe, is a rather unsatisfactory explanation of the observed
  $\gamma$-ray data from outer space.} This conclusion can change if
(a) future observations lead to drastic revision of the data used by
us, including shapes of distributions, or (b) a robust alternate
explanation, possibly based on undetected astrophysical sources, is
found for the extra-terrestrial data sets, or (c) much better fits for
the MSSM emerge from observation of numerous other stellar objects,
compared to which the presently used data get `weighted out'.

\section{Summary and conclusions} \label{summary}

We have analysed extra-terrestrial $\gamma$-ray data with special
emphasis on the excess flux within a specific ROI around the GC, which
corresponds to $2\degree < |b| < 20\degree$ and $|l| < 20\degree$, $b$
and $l$ being the galactic latitude and longitude. The excess, mostly
in the energy band 1 - 10 GeV, is fitted for four representative kinds
of MSSM spectrum. The constraints from all low-energy and collider
data are used, along with the limits from direct DM search
experiments. Consistency with observations of Reticulum II and M31 is
also maintained in the analysis.  We also apply the requirement that
the lower limit (modulo theoretical uncertainties) on the relic
density be satisfied by MSSM contributions, since otherwise the
minimality of the model is lost. It is found that the best fit to the
data is offered by a scenario with $|M_1|$ constrained around 100 GeV,
$\mu$ in the range 110 - 125 GeV and $\tan\beta$ = 20 and $m_A$ in the
range 450 - 550 GeV. A light stop scenario corresponding to
$\tan\beta$ = 50 emerges as the second best candidate, where the
lighter stop mass is close to 300 GeV and the $\chi^0_1$ is around 180
GeV.  The lower limit on the relic density has a significant role in
curving out $2\sigma$ regions in the above parameter regions.

However, if the fits are carried out blindly to direct search
constraint, then the $2\sigma$ regions thus obtained are completely
gone, indicating that the region presented after the `complete fit' is
not even a part of the original outcome of an impartial analysis of
the GC $\gamma$-ray data. The situation is not so contradictory for
the light stop region, but it corresponds to a distinctly worse fit
that the first case, as is reflected by $\chi^2_{min}$ and the
$p$-values for the two cases.  Drawing a parallel with somewhat similar
situations in the search for the Higgs boson in the standard
electroweak model,  the MSSM is found to offer a somewhat unsatisfactory 
fit for GC $\gamma$-rays, unlike the standard model which fits Higgs-related
data rather well.

Before we end, let us re-emphasize an important point that has been
already mentioned, namely, that galactic centre excess can be
due to other astrophysical sources such as millisecond pulsars (MSPs). This, we re-iterate, is still
an open issue.  While one may try to interpret the available evidences
either in terms of DM annihilation as a possible explanation or
perhaps in terms of astrophysical objects like MSPs, we show that the most common and
economical DM explanation, namely, that in terms of the MSSM, is
fraught with difficulties. It should also be noted that not allowing
under-abundance is but one component of this demonstration, a crucial
component being the shape of the GC spectrum itself. This becomes
obvious when one looks at Table \ref{Table_pvalue} where the statistical
significance for the best MSSM scenario {\em even without the relic
  density lower bound constraint is displayed.} Even if
under-abundance is allowed, the relative statistical insignificance of
the MSSM explanation, as compared with the explanation of the Higgs
data in terms of the standard electroweak theory, is something to take
serious note of. Such unsatisfactory $p$-values as have been obtained by
us are again due to the difficulty in matching the shape, while
ensuring consistency with laboratory constraints on the MSSM, and the
inescapable dynamics of the spectra that survive such constraints.

 {\bf Acknowledgements:}\\ We thank Anirban Biswas, Utpal
 Chattopadhyay, Raj Gandhi, Subhadeep Mondal, Niladri Paul and Tim
 Tait for helpful comments.  The work of AK and BM was partially
 supported by funding available from the Department of Atomic Energy,
 Government of India, for the Regional Centre for Accelerator-based
 Particle Physics (RECAPP), Harish-Chandra Research Institute. TRC
 acknowledges the hospitality of RECAPP while the project was being
 formulated, while AK and BM thank the National Centre for Radio
 Astrophysics, Pune, for hospitality while work was in progress.




\end{document}